%% file: main.tex
\documentclass[11pt,letterpaper]{article}
\usepackage[T1]{fontenc}
\usepackage[utf8]{inputenc}
\usepackage{lmodern}
\usepackage[margin=1in]{geometry}
\usepackage{microtype}
\usepackage{mathtools,amssymb,amsthm}
\usepackage{thm-restate}
\makeatletter
\@ifundefined{newcounteralias}{}{%
  \renewcommand\thmt@autorefsetup{%
    \@xa\def\csname\thmt@envname autorefname\@xa\endcsname\@xa{\thmt@thmname}%
  }%
}
\makeatother
\usepackage[numbers]{natbib}
\usepackage{xcolor}
\usepackage{tikz}
\usetikzlibrary{decorations.pathreplacing}
\usetikzlibrary{arrows.meta}
\usepackage[hidelinks]{hyperref}
\usepackage[nameinlink,noabbrev,capitalise]{cleveref}
\usepackage{enumitem}
\usepackage{placeins}
\usepackage{tcolorbox}

\usepackage{braket}

\newcommand{\norm}[1]{\lVert #1\rVert}

\tcbuselibrary{breakable}
\newtheorem{theorem}{Theorem}[section]
\newtheorem{lemma}[theorem]{Lemma}

\newtheorem{corollary}[theorem]{Corollary}
\newtheorem{claim}[theorem]{Claim}
\crefname{claim}{claim}{claims}
\Crefname{claim}{Claim}{Claims}

\theoremstyle{definition}
\newtheorem{definition}[theorem]{Definition}
\theoremstyle{plain}
\numberwithin{equation}{section}

\theoremstyle{remark}
\newtheorem{remark}[theorem]{Remark}

\newcommand{\bits}{\{0,1\}}
\newcommand{\PARITY}{\operatorname{PARITY}}

\hypersetup{
  pdftitle={Beyond the n/2 Threshold: Quantum Query Complexity for Computing Hamming Weight and Symmetric Functions}
}
\title{Beyond the $n/2$ Threshold: Quantum Query Complexity for Computing Hamming Weight and Symmetric Functions}
\author{
Yuan Deng\footnote{Google Research. \texttt{dengyuan@google.com}}
\and
Zihan Hao\footnote{UC San Diego.
\texttt{z1hao@ucsd.edu}}
\and
Honghao Lin\footnote{Google Research. \texttt{honghaol3010@gmail.com}}
\and
Qipeng Liu\footnote{UC San Diego.
\texttt{qipengliu@ucsd.edu}}
\and
Vahab Mirrokni\footnote{Google Research. \texttt{mirrokni@google.com}}
\and
Jianan Shao\footnote{UC San Diego.
\texttt{jis140@ucsd.edu}}
\and
David P. Woodruff\footnote{Google Research and Carnegie Mellon University. \texttt{dpwoodru@gmail.com }}
}
\date{}
\begin{document}
\maketitle

\begin{abstract}
The oracle-interrogation algorithm [van Dam, FOCS~'98] computes every
$n$-bit Boolean function with bounded error using $n/2+O(\sqrt n)$ quantum
queries. In contrast, parity can be computed exactly using $\lceil n/2\rceil$ queries,
but whether an additive overhead beyond $n/2$ is necessary for a
single Boolean output has remained open.

In this work, we construct an explicit family of symmetric Boolean
functions requiring $n/2+\Omega(n^{1/3})$ quantum queries at a fixed
positive constant error. We also give an algorithm that recovers the
exact Hamming weight using $n/2+O(n^{1/3})$ quantum queries with bounded
error, and therefore computes every symmetric Boolean function within
the same query bound. Together, these results determine the optimal
additive order for Hamming weight recovery and for the worst case over
symmetric Boolean functions at sufficiently small constant error. 
\end{abstract}

\input{intro}

\input{matching_walk_technical_overview}

\input{lower_bound_technical_overview}

\input{prelim}

\input{upper}

\input{lower}

\newpage
\input{aimethodology}

\input{bib}
\input{appendix}
\end{document}

%% file: intro.tex
\section{Introduction}
\label{sec:introduction}

How many quantum queries are necessary to compute a Boolean function of
$n$ input bits? The value $n/2$ is a natural threshold. Parity can be computed
exactly with $\lceil n/2\rceil$ quantum queries, and this bound is optimal even
with bounded error~\citep{BealsEtAl2001}. 
In the other direction, van Dam's oracle-interrogation
algorithm recovers the entire input with any fixed error using
$n/2+O(\sqrt n)$ queries~\citep{VanDam1998}. Farhi, Goldstone, Gutmann, and
Sipser proved that this square-root overhead is optimal for recovering the
entire input with bounded error~\citep{FarhiEtAl1999}. Consequently, the same
query upper bound holds for computing any Boolean function with constant error.

For symmetric Boolean functions, Paturi characterized the approximate degree
up to constant factors~\citep{Paturi1992}.
The threshold $n/2$ is also the ceiling of
the standard approximate-degree lower bound. Every $n$-bit Boolean function
has an exact multilinear polynomial representation of degree at most $n$; the standard  polynomial method suggests that quantum query lower bound is at least half the approximate
degree~\citep[Lemmas~4.1--4.2]{BealsEtAl2001}. Thus approximate degree alone
cannot prove a query complexity greater than $n/2$.
Therefore, it is intriguing to ask:
\begin{quote}
{\it Is any additive overhead beyond $n/2$ unavoidable when the task
is only to compute a single Boolean output, and if so, what is its optimal
worst-case overhead?}
\end{quote}
The basic question of whether such a function exists is due to Shelby Kimmel
and appears in Aaronson's 2014 problem list~\cite{Aaronson2014}.

Ambainis, Ba\v{c}kurs, Smotrovs, and de Wolf gave a complementary lower bound:
almost all Boolean functions require
$n/2-o(n)$ quantum queries~\citep{AmbainisEtAl2013}. Their proof uses the
sum-of-squares structure of quantum acceptance probabilities. Together with
van Dam's upper bound, their result identifies $n/2$ as the major term for
almost all Boolean functions, while leaving the additive question unresolved.

In this paper, we answer the existence question affirmatively.
We show a symmetric function, which  requires strictly greater than $n/2$ quantum queries to be computed in worst-case.
A function is symmetric if its value depends only on the Hamming weight
$|x|=\sum_{i=1}^n x_i$. We give explicit symmetric functions requiring
$n/2+\Omega(n^{1/3})$ queries, together with an algorithm that computes
\emph{every} symmetric function (indeed, the hamming weight) using $n/2+O(n^{1/3})$ queries.
In particular, the results achieve the tightest bound for computing hamming weights
with a constant probability. This also points out an interesting distinction between 
recovering hamming weight and recovering the entire string.

\subsection{Our results}
\label{sec:main-results}

For a Boolean function $f:\bits^n\to\bits$, let $Q_\varepsilon(f)$ denote the smallest
\emph{worst-case query complexity} of a quantum algorithm that computes $f(x)$ with
error at most $\varepsilon$ on every input $x$.

Our upper bound is independent of the particular symmetric Boolean function.

\begin{theorem}[Recovering the Hamming weight]
\label{thm:weight-upper}
For every fixed $0<\varepsilon<1/2$, there are constants
$C_\varepsilon,B_\varepsilon>0$ and a quantum query algorithm
$\mathcal A$ such that, for every $n$ and every $x\in\bits^n$, $\mathcal A$
outputs $|x|$ with probability at least $1-\varepsilon$ and makes at most
\[
 \frac n2+C_\varepsilon n^{1/3}+B_\varepsilon
\]
queries. Consequently, the same upper bounds
hold for every symmetric function $f:\bits^n\to\bits$.
\end{theorem}

\medskip

The matching lower bound holds for an explicit family at a fixed
positive constant error.

\begin{theorem}[A cube-root additive lower bound]
\label{thm:weight-lower}
There are an absolute constant $0<\varepsilon_0<1/2$ and an explicit
family of symmetric total Boolean functions $F_n$ such that, for all sufficiently
large $n$,
\[
 Q_{\varepsilon_0}(F_n)
 \geq \frac n2+\Omega(n^{1/3}).
\]
\end{theorem}

Next, we introduce the family of functions that achieves the tight bound,
first for even $n$.
For $n=2m$ and an integer $R\geq0$,
define
\begin{align}
 j_m(u)&=\max\{u-m, (m-1)-u\},
 \label{eq:function-index}\\
 F_{2m,R}(x)&=
 \mathbf{1}\bigl[
 j_m(|x|)\geq R+1
 \text{ and }j_m(|x|)-R\text{ is odd}
 \bigr].
 \label{eq:function-family}
\end{align}
The function is zero on the central region
$m-R-1\leq |x|\leq m+R$ and alternates on either side.
For odd $n=2m+1$, use the same $j_m(|x|)$ and the same radius as for
length $2m$, extending the profile to the additional layer $|x|=2m+1$.
Fixing the last bit to zero recovers
the even-length function, so the lower bound changes relative to
$n/2$ by only an additive $1/2$.

At $R=0$, we use the simpler representative given by parity with a central threshold:
\begin{equation}
 H_{2m}(x)=\PARITY(x)\oplus\mathbf{1}[|x|\geq m].
 \label{eq:output-flip}
\end{equation}
It differs from $F_{2m,0}$ only by a constant output flip, which does not change
query complexity. The two adjacent
layers $m-1,m$ have the same output in $H_{2m}$; widening this exceptional
region gives the family in \cref{thm:weight-lower}.

\begin{figure}[t]
\centering
\begin{tikzpicture}[x=0.82cm,y=0.66cm,font=\small]
  \node[anchor=west,font=\bfseries] at (-0.25,1.95)
    {(a) $F_{2m,0}$: two central zero layers};
  \foreach \i in {0,...,11} {
    \draw[fill=white] ({\i+0.7},0) rectangle ++(0.76,0.76);
  }
  \foreach \i in {0,2,4,7,9,11} {
    \fill ({\i+0.7},0) rectangle ++(0.76,0.76);
  }
  \node at (0,0.38) {$\cdots$};
  \node at (13.4,0.38) {$\cdots$};
  \node at (6.08,1.10) {$m-1$};
  \node at (7.08,1.10) {$m$};
  \draw[dashed,gray] (6.58,-0.02)--(6.58,0.85);
  \node at (2.85,-0.42) {alternating};
  \node at (10.1,-0.42) {alternating};
  \draw[decorate,decoration={brace,mirror,amplitude=4pt}]
    (5.70,-0.18)--(7.46,-0.18);
  \node at (6.58,-0.85) {two equal central layers};

  \node[anchor=west,font=\bfseries] at (-0.25,-2.0)
    {(b) $F_{2m,R}$: a wider central zero region};
  \node[anchor=east] at (13.65,-2.0) {$R\geq1$ shown};
  \foreach \i in {0,...,5,7,8,9,10,11,12} {
    \draw[fill=white] ({\i+0.2},-3.9) rectangle ++(0.76,0.76);
  }
  \foreach \i in {1,3,9,11} {
    \fill ({\i+0.2},-3.9) rectangle ++(0.76,0.76);
  }
  \node at (-0.50,-3.52) {$\cdots$};
  \node at (6.58,-3.52) {$\cdots$};
  \node at (13.65,-3.52) {$\cdots$};
  \node at (4.58,-2.78) {$m-R-1$};
  \node at (8.58,-2.78) {$m+R$};
  \node at (1.95,-4.36) {alternating};
  \node at (11.25,-4.36) {alternating};
  \draw[decorate,decoration={brace,mirror,amplitude=4pt}]
    (4.20,-4.08)--(8.96,-4.08);
  \node at (6.58,-4.79) {$j_m(|x|)\leq R$};
  \begin{scope}[yshift=-0.35cm]
    \node at (6.58,-5.29) {$2R+2$ zero layers};
    \draw[->] (0.2,-5.84)--(8.95,-5.84);
    \node[anchor=west] at (9.05,-5.84) {increasing $|x|$};
    \fill (5.00,-6.80) rectangle ++(0.4,0.4);
    \node[anchor=west] at (5.55,-6.60) {$=1$};
    \draw (7.10,-6.80) rectangle ++(0.4,0.4);
    \node[anchor=west] at (7.65,-6.60) {$=0$};
  \end{scope}
\end{tikzpicture}
\caption{Each square represents one
hamming weight. In (a), $F_{2m,0}$ is zero on the two central layers
$m-1,m$ and alternates on either side. In (b), every layer from
$m-R-1$ through $m+R$ has output zero.}
\label{fig:functions}
\end{figure}
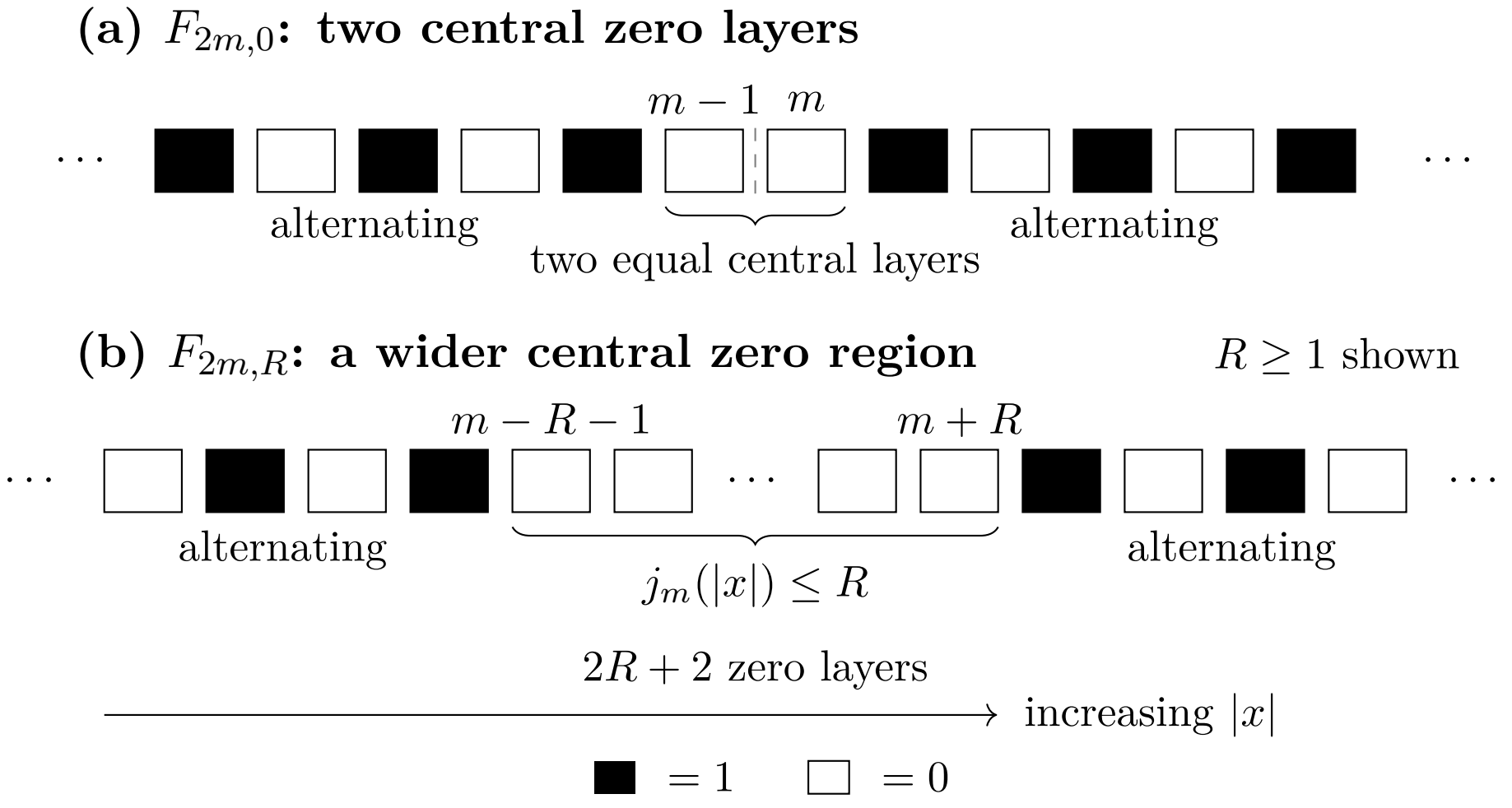

\begin{theorem}[One additional query]
\label{thm:one-query}
For every fixed $0\leq\varepsilon<1/2$ and all sufficiently large $m$, the
$R=0$ member of the family satisfies
\[
 Q_\varepsilon(F_{2m,0})=Q_\varepsilon(H_{2m})\geq m+1.
\]
More quantitatively, every algorithm for either function making at most $m$ queries has worst-case
success probability at most $1/2+O(1/\log m)$.
\end{theorem}

Thus the simplest function in the family already exceeds half the input length; which is already not known before. The even input length makes the additional query independent of
rounding. For the cube-root gap, the explicit choice
$F_n=F_{n,R(n)}$ uses
\[
 R(n)=\Theta(n^{2/3}).
\]
The central region therefore has $2R(n)+2$ layers.
\Cref{fig:functions} shows how widening the region preserves the alternating
tails while changing the additive lower bound.

\medskip

As an immediate corollary, we show that computing $|x|$ has worst-case quantum query complexity $n/2+\Omega(n^{1/3})$.
\begin{theorem}[Tight bounds for computing hamming weights]
\label{thm:weight-tight}
For every fixed $0<\varepsilon\le\varepsilon_0$, recovering $|x|$
with success probability at least $1-\varepsilon$ on every input has
worst-case query complexity $n/2+\Theta(n^{1/3})$.
\end{theorem}

For total Boolean functions without the symmetry restriction, our results
leave a gap between an $\Omega(n^{1/3})$ lower bound and an $O(\sqrt n)$
upper bound on the largest additive excess beyond $n/2$, at constant error $\varepsilon$.
Does there exist a family $f_n:\bits^n\to\bits$ for which
$Q_{\varepsilon}(f_n)\geq n/2+\Omega(\sqrt n)$, or can van Dam's universal upper
bound be improved to $n/2+o(\sqrt n)$ for single-bit outputs~\citep{VanDam1998}?
Such a family would necessarily be nonsymmetric and would match the order
of the additive overhead required for recovering the entire
input~\citep{FarhiEtAl1999}. Our lower-bound argument exploits symmetry;
extending it beyond this setting is a natural direction.

%% file: matching_walk_technical_overview.tex
\begingroup
\newcommand{\Finish}{\mathcal{F}}
\setlength{\emergencystretch}{1.5em}
\allowdisplaybreaks[2]

\section{Technical Overview}

\subsection{Upper Bound}

\label{sec:matching-walk-overview}

We first describe a bounded-error quantum algorithm that outputs the exact Hamming
weight of every input $x\in\{0,1\}^n$, using
at most $n/2+O(n^{1/3})$ queries.
Consequently, the same bound holds for every total symmetric Boolean function.
The algorithm on a high level, estimates the density, handles extremely biased inputs by quantum search, and otherwise uses a coherent walk to find almost as many disjoint unequal pairs as possible. A final quantum search recovers the remaining weight.

We use the phase bit oracle $O_x\ket{i,a}=(-1)^{a \cdot x_i} \ket{i,a}$. Write
\[
n_0=|\{i:x_i=0\}|,\quad n_1=|\{i:x_i=1\}|,\quad
b=n_0-n_1,\quad d=|b|,\quad w=\min\{n_0,n_1\}.
\]
Thus $w$ is the minority count and $d$ is the absolute bias.

\subsubsection{Algorithm framework}
\label{sec:ov-framework}

We first state two subroutines that can be derived immediately from amplitude amplification and Grover's search.
\begin{itemize}
    \item Procedure I (Quantum Counting~\cite{BHMT2002}): Given any input $y \in \{0,1\}^r$ of exactly $q$ ones, amplitude amplification/Grover's search allows one to estimate $\hat{q}$ within $E$ error, except with probability at most $1/100$, using $O(r/E)$ quantum queries.
    \item Procedure II (Input Recovering for Small Hamming Weights~\cite{deGraafDeWolf2002}): Given an input of $y \in \{0,1\}^r$ of at most $q$ ones, a Grover's algorithm can find all the ones with probability $1$ with at most $O(\sqrt{rq})$ quantum queries.

    The high level idea is to run Grover's search to recover one of these $q$ ones, remove it and recursively find another among the rest $r-1$ coordinates. This costs
    \[
    O\left(\sqrt{r/q} + \sqrt{r/(q-1)} + \cdots + \sqrt{r}\right).
    \]
\end{itemize}

\begin{tcolorbox}[
  breakable,
  colback=white,
  colframe=black,
  colbacktitle=white,
  coltitle=black,
  fonttitle=\bfseries,
  title={Algorithm: Computing the Hamming Weight},
  boxrule=0.5pt,
  titlerule=0.5pt,
  arc=0pt,
  boxsep=0pt,
  toptitle=4pt,bottomtitle=4pt,
  left=8pt,right=8pt,top=6pt,bottom=6pt
]
Fix a sufficiently small absolute constant $\alpha>0$. The algorithm has
three stages.
\begin{enumerate}[label=\textup{(\roman*)},leftmargin=*,itemsep=5pt,topsep=5pt]
\item \textbf{Estimate the density and handle a very small minority.}
Set
\begin{equation}
\label{eq:ov-estimate}
k=\lceil n^{1/3}\rceil,\qquad E=k^2.
\end{equation}
Use Procedure I to estimate the number of ones, and thus estimate the bias $\hat{b}$ with
$|\widehat b-b|\le E$ except with probability $1/100$, using
$O(n/E)=O(k)$ queries.

On successful estimation (with probability at least $99\%$),
\begin{equation}
\label{eq:ov-D}
d - E \le |\hat{b}| \le d+E.
\end{equation}
If $|\hat{b}| \ge(1-\alpha)n$, we know that there are either many zeros or many ones\footnote{In this case, we can flip all the bits and make it have very few ones.}. In this case, we call the Procedure II to learn the entire hamming weight. 

Its cost is $O(\sqrt\alpha\,n)$, which is below $n/4$ after fixing $\alpha$ sufficiently
small. 

\item \textbf{Walk toward a nearly maximum matching.}
Otherwise, on successful estimation, both bit values occur on a linear
number of coordinates.
Prepare a superposition of matchings whose sizes lie between $\ell$ and
$2\ell$, for a suitable width $\ell$. 

Apply a unitary $W$ (which can be constructed by one query to the input) a
predetermined number $t$ of times, and then measure the matching $M$.
The initial superposition is chosen so that the walk moves
toward larger matchings with a higher chance. With high probability, the measured size $j=|M|$ satisfies $w-j=O(n^{1/3})$.

\item \textbf{Finish the residual input.}
Remove the $j$ unequal pairs and apply Procedure II to the residual
minority, complementing the oracle if necessary. We use a known upper
bound on the residual minority count, specified below. Return $j$
plus the residual Hamming weight, accounting for any complementation.
\end{enumerate}
\end{tcolorbox}
Below we choose the initial state, the stopping time $t$, and the
residual-search parameters so that steps (ii) and (iii), including
state preparation, together use at most $n/2+O(n^{1/3})$ quantum queries.

\subsubsection{Matching states and the walk}
\label{sec:ov-states}

A matching $M$ is an unordered set of disjoint two-element subsets of $[n]$.
The register $\ket M$ stores the matching without recording the order in which its pairs were inserted. Put
$z_i=(-1)^{x_i}$ and define
\begin{equation}
\label{eq:ov-radial}
\begin{split}
h_M(x)&=\prod_{\substack{e = (u, v) \in M\\ u < v}}(z_{u}-z_{v}),\\
\ket{j;x}&=\frac1{\sqrt{\Lambda_j}}
\sum_{\substack{M\text{ matching}\\ |M|=j}}h_M(x)\ket M,
\qquad
\Lambda_j=\frac{4^j(n_0)_j(n_1)_j}{j!}.
\end{split}
\end{equation}
Here $(a)_j=a(a-1)\cdots(a-j+1)$, with $(a)_0=1$.
$\Lambda_j$ is the normalization factor.
If an equal pair $e = (u, v) \in M$ is in the matching, we have $z_u - z_v = 0$ and thus $h_M(x)=0$. Every unequal matching has the same amplitude
magnitude, but the signs can be different. There are $(n_0)_j(n_1)_j/j!$ such
matchings, which gives the normalization. These states exist for
$0\le j\le w$ (i.e., the maximum number of matchings), and $\ket{0;x}$ is the empty matching.

The notation $j$ describes the matching size; the physical register stores
the full matching. The label $x$ records the input dependence of its
amplitudes. Add a separate direction qubit with basis states
$\ket{\uparrow}$ and $\ket{\downarrow}$, labeling directions toward larger
and smaller matchings, respectively. The joint states are
\[
\ket{j;x}\ket{\uparrow},\qquad
\ket{j;x}\ket{\downarrow}.
\]
On the span of these states, the algorithm becomes a walk on the line
$0,1,\ldots,w$ with two direction labels at each position.

The core idea is to transport amplitude from smaller $j$ to larger $j$,
so that measuring the full matching reveals many disjoint unequal pairs.
Before introducing the unitary that moves amplitude, we first recall the parity sampling algorithm by~\cite{AmbainisIraidsSmotrovs2013}.

\subsubsection{One-query parity sampling}
\label{sec:ov-parity}

Consider a known remaining set of $r\ge1$ coordinates, numbered locally
$1,\ldots,r$ in increasing order of their original indices. Prepare the uniform index state, make one phase query, and
apply an input-independent isometry $A_r$:
\[
\frac1{\sqrt r}\sum_i\ket i
\ \xrightarrow{\text{one query}}\
\frac1{\sqrt r}\sum_i z_i\ket i
\ \xrightarrow{A_r}\
\frac{\sum_i z_i}{r}\ket{\mathrm{fail}}
+\frac1r\sum_{u<v}(z_u-z_v)\ket{u,v}.
\]
The transformation is the unequal-pair sampler used in exact symmetric
function algorithms~\cite{AmbainisIraidsSmotrovs2013}. Explicitly, the isometry does the following:
\begin{equation}
\label{eq:ov-A}
A_r\ket i=\frac1{\sqrt r}
\left(\ket{\mathrm{fail}}+
\sum_{i<v}\ket{i,v}-\sum_{u<i}\ket{u,i}\right).
\end{equation}

If we measure a pair $(u,v)$, its parity is certainly one:
$x_u\oplus x_v=1$. Thus, we know one of them is $1$, and the other is $0$.
Each unequal pair has probability $4/r^2$.
If the residual contains $n'_0$ zeros and $n'_1$ ones, the total success probability is
\begin{equation}
\label{eq:ov-parity-prob}
\frac{4n'_0 n'_1}{r^2}=1-\left(\frac{n'_0-n'_1}{r}\right)^2.
\end{equation}
Thus the sampler produces a useful pair with certainty on a balanced residual,
and less often on a biased residual.

\subsubsection{Constructing and using the walk}
\label{sec:ov-walk}

\paragraph{Coherent insertion and the walk operator}

At a high level, we use parity sampling to move amplitude from
$\ket{j;x}$ to $\ket{j+1;x}$. On a biased input, the sampler can fail
even when $j=0$. 
Removing $j$ unequal pairs leaves $r=n-2j$ coordinates and preserves
the signed bias $b$. For $0\le j<w$, define the amplitudes
\begin{equation}
\label{eq:ov-coeff}
\mu_j=\frac b{n-2j},\qquad
\beta_j=\sqrt{1-\mu_j^2}.
\end{equation}
Using one query to $O_x$, we construct a unitary $W$ with the action
\begin{equation}
\label{eq:ov-W}
\boxed{\begin{aligned}
W\ket{j;x}\ket{\uparrow}
&=\beta_j\ket{j+1;x}\ket{\uparrow}+\mu_j\ket{j;x}\ket{\downarrow},\\
W\ket{j;x}\ket{\downarrow}
&=\beta_{j-1}\ket{j-1;x}\ket{\downarrow}-\mu_{j-1}\ket{j;x}\ket{\uparrow}.
\end{aligned}}
\end{equation}
The $\beta$ terms move the matching size in the indicated direction; the
$\mu$ terms keep the size fixed and reverse the arrow. As a sanity check, when $b=0$, an upward state gains
one pair at every step until it reaches the maximum matching size.

We now describe the upward action. Fix a matching $M$ of size $j$
consisting of unequal pairs, and let $S(M)$ be its unmatched coordinates.
Write $A_M$ for the isometry in~\eqref{eq:ov-A} with indices in $S(M)$.
Starting from $\ket M\ket{\uparrow}$, prepare the uniform index state,
make one phase query, and apply $A_M$:
\[
\begin{aligned}
\ket M\ket{\uparrow}
&\longmapsto \frac1{\sqrt r}\sum_{i\in S(M)}\ket{M,i}\\
&\xrightarrow{\text{one phase query}}
\frac1{\sqrt r}\sum_{i\in S(M)}z_i\ket{M,i}\\
&\xrightarrow{A_M}
\mu_j\ket{M,\mathrm{fail}}
+\frac1r\sum_{\substack{u<v\\u,v\in S(M)}}
(z_u-z_v)\ket{M,\{u,v\}}.
\end{aligned}
\]
Then coherently insert $e=\{u,v\}$ into $M$ while retaining $e$:
\[
\ket{M,e}\longmapsto\ket{M\cup\{e\},e}.
\]
This relabeling is reversible, since the larger matching and the marked
edge determine $M$; the failure branch is left unchanged.
Now apply these operations to the full superposition $\ket{j;x}$.
For each resulting matching $N=M\cup\{e\}$,
\[
h_M(x)(z_u-z_v)=h_N(x).
\]
Thus all choices of the last inserted edge of $N$ have the same signed
amplitude. Grouping them gives the state
\[
\mu_j\ket{j;x}\ket{\mathrm{fail}}
+\frac1{r\sqrt{\Lambda_j}}
\sum_{\substack{N\text{ matching}\\|N|=j+1}}
h_N(x)\ket N\sum_{e\in N}\ket e.
\]
For each $N$, the last-edge register is now uniform over its $j+1$ edges.
Since this superposition is determined by $N$, an input-independent
unitary can erase the last edge and set the direction:
\[
\begin{aligned}
\ket N\left(\frac1{\sqrt{j+1}}\sum_{e\in N}\ket e\right)
&\longmapsto\ket N\ket{\uparrow},\\
\ket M\ket{\mathrm{fail}}
&\longmapsto\ket M\ket{\downarrow}.
\end{aligned}
\]
This provides the intuition for the
upward action in~\eqref{eq:ov-W}; the reversible construction follows similarly.

\paragraph{The initial state preparation}

Using the bias estimate from the first stage, define the ``upper bound'' for the absolute bias:
\[
D=|\hat{b}|+E.
\]
Then $D\ge E=k^2$ always, and $d\le D\le d+2E$ on successful estimation.
Choose a sufficiently large fixed constant $L$ and set $\ell=\lceil L\sqrt D\rceil$.
Choose a nonnegative smooth amplitude $f_0$, supported strictly
between $\ell$ and $2\ell$, symmetric about $3\ell/2$, and normalized so
that $\sum_j|f_0(j)|^2=1$.
Write $\mathbf{i}$ for the imaginary unit, satisfying $\mathbf{i}^2=-1$.
We prepare the initial state
\begin{equation}
\label{eq:ov-packet}
\ket{\Psi_0}=
\sum_j \mathbf{i}^j f_0(j)
\left(c_j\ket{j;x}\ket{\uparrow}+\mathbf{i} s_j\ket{j;x}\ket{\downarrow}\right).
\end{equation}
Here $f_0$ controls the distribution over matching sizes. The factors $\mathbf{i}^j$
and $\mathbf{i}$, together with the chosen direction amplitudes, produce
a state whose amplitudes shift toward larger matchings. We choose $c_j, s_j$ as follows:
\begin{equation}
\label{eq:ov-coin}
\begin{pmatrix}\beta_j&\mu_j\\ \mu_j&-\beta_j\end{pmatrix}
\binom{c_j}{s_j}=\binom{c_j}{s_j}.
\end{equation}

We prepare this state in two stages: first build the superposition over
matching sizes, then set the phases and the direction qubit. The step of setting phases and direction qubits are easy, so we focus on the first step. Introduce a
target register $T$ and a matching register $M$, initially in the state
\[
\left(\sum_j f_0(j)\ket j_T\right)\ket{\varnothing}_M.
\]
The register $T$ records the number of pairs that each branch should
eventually contain. Up to a small
preparation error, our goal is the transformation
\[
\sum_j f_0(j)\ket j_T\ket{\varnothing}_M
\longmapsto
\sum_j f_0(j)\ket j_T\ket{j;x}_M.
\]

Run $2\ell$ rounds. In round $s$, insert one unequal pair on branches
whose target satisfies $j\ge s$; branches with $j<s$ have already
reached their target and remain unchanged. Using the walk operator, there is a $\mu_j$ amplitude that corresponds to an unsuccessful movement, which is at most a constant since $|\hat{b}| \leq (1-\alpha) n$. Thus, by amplitude amplification~\cite{BHMT2002} and making $O(\ell)$ queries, one can prepare this state. By coherently erase the register $j$ and generate the relative phases as well as the direction qubits, we can prepare the starting state.

\paragraph{Why the state moves upward}

We explain how the amplitudes shift toward larger matching
sizes after one application of $W$. First suppose the amplitudes are supported away
from the endpoints, with $\mu,\beta,c,s$ constant throughout this range. Let $\ket{\Psi_f}$ denote
the state in~\eqref{eq:ov-packet} with $f$ in place of $f_0$. With these coefficients fixed, applying~\eqref{eq:ov-W}
we have:
\begin{equation}
\label{eq:ov-step-expanded}
\begin{aligned}
W\ket{\Psi_f}
=-\mathbf{i}\sum_j\mathbf{i}^j\ket{j;x}\Bigl(
[\beta c f(j-1)+\mu s f(j)]\ket{\uparrow}
+\mathbf{i}[\mu c f(j)-\beta s f(j+1)]\ket{\downarrow}\Bigr).
\end{aligned}
\end{equation}
The direction condition~\eqref{eq:ov-coin} gives
$\beta c+\mu s=c$ and $\mu c-\beta s=s$. We choose $f$ to be smooth so that $f(j\pm1)\approx f(j)\pm f'(j)$,
\[
\begin{aligned}
\beta c f(j-1)+\mu s f(j)
&\approx c\bigl(f(j)-\beta f'(j)\bigr),\\
\mu c f(j)-\beta s f(j+1)
&\approx s\bigl(f(j)-\beta f'(j)\bigr).
\end{aligned}
\]
Setting $\widetilde f(j)=f(j-\beta)$, we have the resulting state is
\begin{equation}
\label{eq:ov-frozen}
W\ket{\Psi_f}
\approx-\mathbf{i}\sum_j\mathbf{i}^j\widetilde f(j)\ket{j;x}
\bigl(c\ket{\uparrow}+\mathbf{i}s\ket{\downarrow}\bigr)=-\mathbf{i}\ket{\Psi_{\widetilde f}}.
\end{equation}
Thus the amplitudes shift toward larger matching sizes by about $\beta$ levels per query.

\paragraph{Stopping before the boundary 
and success probability.}
The above coarse analysis describes the walk as moving the amplitudes by $\beta_j$ per query.
However, it does not describe what happens when the amplitudes reach or get close to the endpoint $w$. At the endpoint, the exact walk
satisfies
\[
W\ket{w;x}\ket{\uparrow}=\mu_w\ket{w;x}\ket{\downarrow},
\qquad |\mu_w|=1.
\]
Thus after reaching $w$ and applying the walk operator again, the amplitudes can start moving toward smaller matchings. Thus, we will stop before it happens.

Each application of $W$ costs one query. The speed is $\beta_j$, so the time still needed to
move from level $j$ to $w$ is roughly
\begin{equation}
\label{eq:ov-travel-time}
\int_j^w\frac{du}{\beta_u}
=\frac12\sqrt{(n-2j)^2-b^2}
=\sqrt{(n_0-j)(n_1-j)}.
\end{equation}
We use this estimation to choose a shorter run so that not many amplitudes hit the boundary $w$.

In the formal analysis, 
we run the walk for the conservative time that is smaller than what it would take to reach $w$. This will leave a buffer before it reaches the boundary.
On successful estimation, replacing $|b|$ by its known upper bound $D$
gives the following stopping time:
\begin{equation}
\label{eq:ov-stopping-time}
\boxed{
t=\frac12\sqrt{(n-6\ell)^2-D^2}.
}
\end{equation}
For the actual number of steps, round $t$ down to an even integer,
which changes it by less than two.

The full walk analysis needs to establish the following guarantee:
conditional on successful estimation, state preparation followed by
the walk produces, with probability at least $99\%$, a matching whose
size $j$ satisfies
\[
\sqrt{(n_0-j)(n_1-j)}=O(D/k).
\]
We show that this implies $w-j=O(k)=O(n^{1/3})$, so only a small number of pairs
remain to reach a maximum matching. 

After removing the matching, $n_0-j$ zeros and $n_1-j$ ones remain. We use Procedure II to find all the remaining ones/zeros (depending which is the minority).
Since the residual length is at most twice the larger count, the square
root of its product with the minority count is at most
$\sqrt{2(n_0-j)(n_1-j)}=O(D/k)$.
Thus Procedure II finishes in $O(D/k)$ queries.

\paragraph{The total query bound.}
The initial estimate costs $O(k)$ queries, preparation costs
$O(\ell)$, the walk costs at most $t$, and finishing costs $O(D/k)$.
Since $L$ is fixed, $D\ge k^2$, and $\ell=O(\sqrt D)$, both $k$ and
$\ell$ are $O(D/k)$. The total query count is therefore
\begin{align*}
Q_{\rm total}
&\le t+O(k+\ell+D/k)\\
&\le \frac12\sqrt{n^2-D^2}+O(D/k)\\
&\le \frac n2-\frac{D^2}{4n}+O(D/k)\\
&\le \frac n2+O(n/k^2)
=\frac n2+O(n^{1/3}).
\end{align*}
When the finishing cost is larger, the walk is shorter. The saving
$D^2/(4n)$ pays for the extra $O(D/k)$ queries, apart from at most
$O(n/k^2)=O(n^{1/3})$ queries in the worst case.

\endgroup

%% file: lower_bound_technical_overview.tex
\begingroup
\setlength{\emergencystretch}{1.5em}
\allowdisplaybreaks[2]

\subsection{Lower Bounds}
\label{sec:lower-bound-overview}

We now explain the lower bounds for the functions introduced
in~\eqref{eq:function-family}. Write $n=2m$. We first show why the
parity--threshold function $H_{2m}$ requires at least $m+1$ queries for
constant errors and all sufficiently large $m$. We then widen its central constant region
to obtain $F_{2m,R}$ with query complexity
$m+\Omega(m^{1/3})$ for a fixed positive error.

\subsubsection{Signed weights and quantum acceptance probabilities}

We recall the sum-of-squares (SOS) approach to quantum query
lower bounds used by Ambainis et al.~\cite{AmbainisEtAl2013}, presenting
it here through signed weights.
If an algorithm uses $T$ queries, the state before measurement can be represented as 
\begin{align*}
    |\phi_x\rangle = \sum_i \alpha_i(x) |i\rangle, 
\end{align*}
where each $\alpha_i(\cdot)$ is a polynomial of degree at most $T$.
Its probability of outputting any outcome has
the form, by Born's rule, 
\begin{equation}
\label{eq:lbov-sos}
\left|\langle \psi|\phi_x\rangle \right|^2 = \left| \sum_i \alpha_i(x) \langle\psi|i\rangle \right|^2,
\end{equation}
For a fixed measurement basis $|\psi\rangle$, each overlap is a polynomial $p_\nu(x)$ of degree at most $T$.
Summing over the measurement outcomes that output one gives
$a(x)=\sum_\nu |p_\nu(x)|^2$.
Keeping this sum-of-squares structure gives more information than the
degree bound on $a$ alone~\citep{BealsEtAl2001,AmbainisEtAl2013}.

For real weights $\lambda(x)$ that are not all zero, define
\[
L_\lambda(F)=\sum_{x\in\bits^{2m}}\lambda(x)F(x).
\]
We are going to construct weights $\lambda:\{0,1\}^n\to\mathbb{R}$ such that
\begin{equation}
\label{eq:lbov-positive}
L_\lambda(p^2)\ge0
\qquad\text{for every real polynomial }p\text{ of degree at most }T.
\end{equation}
This also gives $L_\lambda(a)\ge0$ since $a$ is an SOS.

Let $P$ be the sum of the positive weights, i.e., 
\begin{align*}
P = \sum_{x: \lambda(x) > 0} \lambda(x).   
\end{align*}
Let $N$ be the total 
magnitude of
the negative weights, i.e., 
\begin{align*}
N = \sum_{x: \lambda(x) < 0} (-\lambda(x)). 
\end{align*}
Finally let $A=P+N=\sum_x|\lambda(x)|$ be the sum of absolute weights. 
Now we construct a function $f$ based on $\lambda$:
\begin{align*}
    f(x) = \begin{cases}
        0 & \text{ if } \lambda(x) \geq 0\\
        1 & \text{ if } \lambda(x) < 0
    \end{cases}.
\end{align*}
An algorithm with error
at most $\varepsilon$ would satisfy $a(x) \leq \varepsilon$ if $f(x) = 0$ and $a(x) \geq 1 - \varepsilon$ if $f(x) = 1$. Thus,
\begin{align*}
\label{eq:lbov-error-principle}
0\le L_\lambda(a) & = \sum_{x} \lambda(x) a(x) \\
& = \sum_{x: \lambda(x) \geq 0} \lambda(x) a(x) - \sum_{x: \lambda(x) < 0} (-\lambda(x)) a(x) \\
&\le\varepsilon P-(1-\varepsilon)N
=\varepsilon A-N.
\end{align*}
Thus every $T$-query algorithm has error at least $\varepsilon \geq N/A$. In particular,
if $N/A>\varepsilon$, the target requires at least $T+1$ queries. Writing
$S=P-N$ for the signed sum gives the useful identity
\[
\frac NA=\frac12\left(1-\frac SA\right).
\]
Therefore, to prove a lower bound of $T+1$ queries at error
$\varepsilon$, our goal is to find signed weights $\lambda(\cdot)$ whose
associated $L_\lambda(p^2)$ is nonnegative on every square $p^2$ with
$\deg p\le T$, while satisfying $N/A>\varepsilon$. The weights $\lambda(\cdot)$ will naturally induce a function $f(\cdot)$.

\paragraph{Using symmetry.}
We use a symmetry reduction similar to that in~\cite{LeeEtAl2016},
drawing on earlier work on symmetric polynomial decompositions and
SOS positivity~\cite{Srinivasan2011,KurpiszEtAl2016}.
Since $\lambda$ can arbitrarily depend on $x$ but we only care about $f$ being a symmetric function, we naturally force $\lambda$ to be symmetric as well: 
$\lambda(x)=v(|x|-m)$. For an integer $g\ge0$, decomposing polynomials into components that are
orthogonal on each Hamming layer reduces the positivity requirement~\eqref{eq:lbov-positive} at
$T=m+g$ to the one-variable constraints for all integers $0 \leq r \leq m$,
\begin{equation}
\label{eq:lbov-blocks}
\sum_{k=-r}^{r}\binom{2r}{r+k}v(k)q(k)^2\ge0,
\qquad
\text{for every real polynomial }q
\text{ with }\deg q\le\min\{r+g,2r\}.
\end{equation}

\subsubsection{One additional query: a single pole}

We start with the signed weights used in Grigoriev's knapsack
lower bound~\cite{Grigoriev2001}, in the form given in~\cite{KurpiszEtAl2016}.
The signed weights in~\eqref{eq:lbov-single-weights} and the
evaluation identity in~\eqref{eq:lbov-single-residue} are already established in prior work.
Here we give a new proof of the identity based on the residue theorem and use the weights to
prove the $m+1$ quantum query lower bound for $H_{2m}$. 
The new proof inspires the extension to the lower bounds $m+g$ for $g \gg 1$.

For the function $H_{2m}$ in~\eqref{eq:output-flip}, use the following function with a single pole at $-1/2$:
\begin{equation}
\label{eq:lbov-single-weights}
v(k)=\frac{(-1)^k}{k+\tfrac12},
\qquad k=|x|-m.
\end{equation}
The signs alternate on both sides of the middle (i.e., smaller than $-1$ or bigger than $0$), but the two central
layers $k=-1,0$ both have positive weight. The negative weights therefore
describe $F_{2m,0}$, or equivalently $H_{2m}$ up to a bit flip.

There are two things to show: (1) positivity on degree-$m$ squares,
as in~\eqref{eq:lbov-positive}, and (2) that the negative-weight
fraction $N/A$ approaches $1/2$.
Since the goal is to show a lower bound $m+1$, We show~\eqref{eq:lbov-blocks} with $g=0$. Fix any integer $0\le r\le m$.
Put
$\Phi_r(z)=\prod_{k=-r}^{r}(z-k)$. For every real polynomial $q$ of
degree at most $r$, 
the residue theorem gives
\begin{equation}
\label{eq:lbov-single-residue}
\sum_{k=-r}^{r}\binom{2r}{r+k}
\frac{(-1)^k}{k+\tfrac12}q(k)^2
=\frac{(2r)!}{|\Phi_r(-\tfrac12)|}\,q(-\tfrac12)^2
\ge0.
\end{equation}
To see the reason, consider the function
\[
G_r(z)=
\frac{(-1)^r(2r)!\,q(z)^2}
{\Phi_r(z)(z+\tfrac12)}.
\]
The only points where $G_r$ can have nonzero residues are $-r,-r+1,\ldots,r$ and $-\tfrac12$.

Consider the derivative of $\Phi_r(\cdot)$:
\[
\Phi_r'(k)
=\prod_{\substack{h=-r\\h\ne k}}^r(k-h)
=(-1)^{r-k}(r+k)!(r-k)!.
\]
Therefore the residue at $k$ is exactly the corresponding
summand on the left of~\eqref{eq:lbov-single-residue}:
\[
\begin{aligned}
\operatorname{Res}_{z=k}G_r(z)
&=\lim_{z\to k}(z-k)G_r(z)\\
&=\frac{(-1)^r(2r)!\,q(k)^2}
{\Phi_r'(k)(k+\tfrac12)}\\
&=\binom{2r}{r+k}\frac{(-1)^k}{k+\tfrac12}q(k)^2.
\end{aligned}
\]
The only other possible pole is $-1/2$. Exactly $r+1$
factors in $\Phi_r(-\tfrac12)$ are negative, so
$\Phi_r(-\tfrac12)=(-1)^{r+1}|\Phi_r(-\tfrac12)|$. Its residue is thus
nonpositive:
\[
\operatorname{Res}_{z=-1/2}G_r(z)
=\frac{(-1)^r(2r)!\,q(-\tfrac12)^2}{\Phi_r(-\tfrac12)}
=-\frac{(2r)!}{|\Phi_r(-\tfrac12)|}\,q(-\tfrac12)^2\le0.
\]
Finally, the numerator of $G_r$ has degree at most $2r$, while the
denominator has degree $2r+2$. Hence $G_r(z)=O(z^{-2})$, so there is
no residue at infinity. By the residue theorem, the integral of $G_r$
around a counterclockwise circle enclosing all finite poles equals
$2\pi i$ times the sum of their residues; this integral is $O(R^{-1})$
as the radius $R\to\infty$, so the sum is zero.
The sum of the residues at the integer points
is therefore the negative of the residue at $-1/2$, proving
\eqref{eq:lbov-single-residue}.
Applying~\eqref{eq:lbov-blocks} with $g=0$ proves positivity on
all degree-$m$ squares.

\vspace{1em}

It remains to estimate the fraction of negative weight. For
$1\le j\le m$, pair the layer $k=-j$, consisting of inputs with
$|x|=m-j$, with the layer
$k=j$, consisting of inputs with $|x|=m+j$. These layers have the same
number of inputs:
\[
B_j:=\binom{2m}{m-j}=\binom{2m}{m+j},\qquad 0\le j\le m.
\]
Thus $B_j$ is the multiplicity of each weight in the pair, and the two
layers together contain $2B_j$ inputs. The first few pairs (when
present) and the general pair are as follows:
\[
\renewcommand{\arraystretch}{1.5}
\begin{array}{c|c|c|c}
\text{Layers }(-j,j)&v(-j)&v(j)&\text{Multiplicity of each}\\
\hline
(-1,1)&2&-2/3&B_1\\
(-2,2)&-2/3&2/5&B_2\\
\vdots&\vdots&\vdots&\vdots\\
(-j,j)&\dfrac{(-1)^{j+1}}{j-\tfrac12}
&\dfrac{(-1)^j}{j+\tfrac12}&B_j
\end{array}
\]
For every pair of layers $(-j,j)$, the contributions have the same
multiplicity and opposite signs, with magnitudes that become close as
$j$ grows. A calculation shows that the signed total $S_m$ is small
compared with the absolute sum $A_m$. More specifically,
\begin{equation}
\label{eq:lbov-single-error}
\frac{N_m}{A_m}
=\frac12-\frac{S_m}{2A_m}
\ge\frac12-O(1/\log m).
\end{equation}
This proves~\cref{thm:one-query}: every
$m$-query algorithm has worst-case success probability at most
$1/2+O(1/\log m)$. Hence every fixed error below $1/2$ requires at
least $m+1$ queries for sufficiently large $m$.

\subsubsection{More poles allow a larger query lower bound}

To rule out algorithms using $m+g$ queries, the constraints
in~\eqref{eq:lbov-blocks} must allow $q$ to have degree up to $r+g$.
With only one pole, $G_r$ need not have the $O(z^{-2})$ decay that
made the sum of its residues zero; which is crucial for proving the positivity. The residue theorem still
applies, but the earlier positivity identity no longer follows.
Each extra degree of $q$ adds two degrees to $q^2$, so we need to increase the degree to $2r+2g+2$.

We first describe the signed weights without specifying the
locations of these poles. Write $k=|x|-m$. The main proof chooses an integer
$R$ with the same parity as $g$, and $2g+1$ distinct half-integer poles
$p_1,\ldots,p_{2g+1}$ lying between $0$ and $R$. Set
\begin{equation}
\label{eq:lbov-many-poles}
h_g(k)=\frac{(-1)^{g+k}}
{\prod_{i=1}^{2g+1}(k-p_i)}.
\end{equation}
The poles avoid all integer inputs. Their exact locations are given in the main body (in~\eqref{eq:lb-poles}).\footnote{The main proof denotes the integer roots by $\rho\in\mathcal P$; the corresponding poles in our notation are $\rho-\tfrac12$.}

For the weights $v=h_g$, consider the following function (similar to that in the previous proof)
\[
G_{r,g}(z)=\frac{(-1)^{r+g}(2r)!\,q(z)^2}
{\Phi_r(z)\prod_{i=1}^{2g+1}(z-p_i)}.
\]
Its residues at integer points are exactly the summands
in~\eqref{eq:lbov-blocks}; the remaining residues are at the added
poles $p_i$. Since $zG_{r,g}(z)\to0$ as $z\to\infty$, the residue
theorem implies that the sum of all finite residues is zero. Therefore,
\begin{equation}
\label{eq:lbov-multipole-identity}
\begin{aligned}
\sum_{k=-r}^{r}\binom{2r}{r+k}h_g(k)q(k)^2
&=\sum_{k=-r}^{r}\operatorname{Res}_{z=k}G_{r,g}(z)\\
&=-\sum_{i=1}^{2g+1}\operatorname{Res}_{z=p_i}G_{r,g}(z).
\end{aligned}
\end{equation}

We have two remaining goals:
\begin{enumerate}[nosep]
\item Prove positivity in~\eqref{eq:lbov-blocks} for every required
summation range. For ranges containing all poles, the residue
identity for the unmodified weights $h_g$ tells us which signs we need:
\[
\underbrace{\sum_{k=-r}^{r}\binom{2r}{r+k}h_g(k)q(k)^2}
_{\text{we want this }\ge0}
=-\underbrace{\sum_{i=1}^{2g+1}
\operatorname{Res}_{z=p_i}G_{r,g}(z)}
_{\text{we want this }\le0}.
\]
The idea here is to make each term
$-\operatorname{Res}_{z=p_i}G_{r,g}(z)$ nonnegative.
For every $r\ge R$, all added poles lie in $[-r,r]$.
Although the numerator contains $(-1)^{r+g}$, its dependence on $r$
is canceled by the sign of $\Phi_r(p_i)$ in each residue. The pole
choice therefore makes every added-pole residue nonpositive
simultaneously for all such $r$.
For $r<R$, some poles lie outside the summation range, and their
residue signs may depend on $r$; the same sign argument need not give
positivity. We handle these shorter ranges by replacing central
negative weights by zero. This makes every summand in them
nonnegative and can only increase the sums for the larger ranges.
\item After replacing central negative weights by zero, retain a
constant fraction of negative weight in the total absolute weight.
\end{enumerate}

To handle the shorter ranges and obtain the central constant region of
our Boolean function, we set negative weights to zero at $-R\le k<R$ and leave all other unchanged:
\begin{equation}
\label{eq:lbov-clipping}
u_g(k)=
\begin{cases}
\max\{0,h_g(k)\},&-R\le k<R,\\
h_g(k),&\text{otherwise},
\end{cases}
\qquad
\lambda_g(x)=u_g(|x|-m).
\end{equation}
We keep $G_{r,g}$ unchanged and use it only for the weights $h_g$.
If $r\ge R$, every term $-\operatorname{Res}_{z=p_i}G_{r,g}(z)$
in~\eqref{eq:lbov-multipole-identity} is nonnegative, so the weighted
sum involving $h_g$ is nonnegative. Since $u_g(k)\ge h_g(k)$, the
corresponding sum involving $u_g$ can only be larger.
If $r<R$, every integer $k\in[-r,r]$ lies in $[-R,R)$, so
$u_g(k)\ge0$ and the sum is nonnegative term by term.
Thus~\eqref{eq:lbov-blocks} holds with $v=u_g$, and the symmetry
reduction gives~\eqref{eq:lbov-positive} for $\lambda_g$ at $T=m+g$.

For the specific pole choice in the main proof,
Lemmas~\ref{lem:lb-multipole-positive} and~\ref{lem:lb-multipole-mass}
establish the two properties needed for the query lower bound:
non-negativity on degree-$(m+g)$ squares and a constant fraction of
negative weight, provided $m\ge512g^3$. We leave those proofs to the
main body. Choosing $g$ to be a sufficiently small constant times
$m^{1/3}$ therefore gives, for an
absolute constant error $\varepsilon_0>0$,
\begin{equation}
\label{eq:lbov-result}
Q_{\varepsilon_0}(F_{2m,R})
\ge m+g+1
=m+\Omega(m^{1/3}),
\qquad R=\Theta(m^{2/3}).
\end{equation}
This is the construction proved in~\cref{thm:lb-cuberoot}.

\endgroup

%% file: prelim.tex
\section{Preliminaries}
\label{sec:preliminaries}

For a bit string $x\in\bits^r$, write $|x|$ for its Hamming weight.
We use the standard bit oracle
$O_x\ket{i,a}=\ket{i,a\oplus x_i}$. Preparing the answer register in
$\ket{-}$ implements the phase oracle
\[
O_x^{\mathrm{ph}}\ket i=(-1)^{x_i}\ket i,
\qquad
O_x\bigl(\ket i\ket{-}\bigr)
=\bigl(O_x^{\mathrm{ph}}\ket i\bigr)\ket{-}.
\]

\subsection{Amplitude amplification}
\label{sec:prelim-amplitude-amplification}

\begin{lemma}[Amplitude amplification~{\citep{BHMT2002}}]
\label{lem:prelim-amplitude-amplification}
Let $U$ be unitary, $\ket u$ a unit vector, and $\Pi$ an orthogonal
projector onto the success subspace. Write
\[
U\ket u=\sin\theta\ket{v_1}+\cos\theta\ket{v_0},
\qquad 0<\theta<\pi/2,
\]
where $\ket{v_1},\ket{v_0}$ are unit vectors with
$\Pi\ket{v_1}=\ket{v_1}$ and $\Pi\ket{v_0}=0$.
Let $R_u=I-2\ket u\bra u$ and
$R_\Pi=I-2\Pi$.
Then, for every
integer $m\ge0$,
\begin{equation}\label{eq:prelim-amplitude-amplification}
\bigl(-UR_uU^\dagger R_\Pi\bigr)^mU\ket u
=\sin((2m+1)\theta)\ket{v_1}
+\cos((2m+1)\theta)\ket{v_0}.
\end{equation}
This uses $2m+1$ calls to $U$ or $U^\dagger$.
\end{lemma}

\subsection{Two quantum algorithms}
\label{sec:q-algo}

\begin{lemma}[Approximate counting~{\citep[Theorem~13]{BHMT2002}}]
\label{lem:prelim-counting}

Given query access to $x\in\bits^r$, where $r\ge1$, and a known accuracy
$0<E\le r$, a quantum algorithm returns a real estimate $\widehat q$ of
$q=|x|$ such that
\[
\Pr\bigl[|\widehat q-q|\le E\bigr]\ge\frac{99}{100},
\]
using at most $O(r/E)$ queries on every input.
\end{lemma}

\begin{lemma}[Exact recovery of a sparse input~{\citep[Lemma~2]{deGraafDeWolf2002}}]
\label{lem:prelim-sparse-recovery}
Given query access to $x\in\bits^r$ and a known integer $0\le K\le r$,
there is a quantum algorithm that finds all positions containing a one
with certainty whenever $|x|\le K$. It uses at most $O(\sqrt{rK})$ queries
on every input.
\end{lemma}

The same procedure finds all zeros when an upper bound on their number
is known, by complementing the queried bits.

\subsection{Residues and the residue theorem}
\label{sec:prelim-residues}

\begin{definition}[Holomorphic functions and residues]
\label{def:prelim-residue}
A complex function is \emph{holomorphic} on an open set if it is complex
differentiable at every point of that set.
Suppose $F$ is holomorphic in a punctured neighborhood
$0<|z-a|<\eta$ of $a$, and has at most a pole at $a$. Write its Laurent
expansion as
\[
F(z)=\sum_{j=-m}^{\infty}c_j(z-a)^j
\]
for some integer $m\ge1$. Its \emph{residue} at $a$ is
$\operatorname{Res}_{z=a}F(z)=c_{-1}$, the coefficient of $(z-a)^{-1}$.
If $F$ is holomorphic at $a$, there are no negative-power terms and its
residue is zero.
\end{definition}

To compute the residues appearing later, we recall the formula for a simple
pole. A point $a$ is a \emph{simple pole} of $F$ if $(z-a)F(z)$ extends to a
holomorphic function near $a$ whose value at $a$ is nonzero.
If $F$ has at most a simple pole at $a$, then
\begin{equation}
\label{eq:prelim-simple-residue}
\operatorname{Res}_{z=a}F(z)=\lim_{z\to a}(z-a)F(z).
\end{equation}

\begin{theorem}[Residue theorem]
\label{thm:prelim-residue-theorem}
Let $\Gamma$ be a positively oriented (counterclockwise), piecewise smooth
simple closed contour. Suppose $F$ is holomorphic on an open neighborhood
of $\Gamma$ and its interior, except at finitely many poles
$a_1,\ldots,a_s$ strictly inside $\Gamma$. Then
\begin{equation}
\label{eq:prelim-residue-integral}
\oint_{\Gamma}F(z)\,dz
=2\pi i\sum_{j=1}^{s}\operatorname{Res}_{z=a_j}F(z).
\end{equation}
\end{theorem}

\subsection{Rational functions and partial fractions}
\label{sec:prelim-rational-functions}

A \emph{rational function} is a quotient $F(z)=P(z)/Q(z)$ of complex
polynomials, with $Q$ not identically zero. It is \emph{proper} if
$\deg P<\deg Q$. The following decomposition applies when the denominator
has no repeated roots.

\begin{lemma}[Partial fractions for simple roots]
\label{lem:prelim-partial-fractions}
Let $F=P/Q$ be proper, and write
\[
Q(z)=c\prod_{j=1}^{s}(z-\alpha_j),
\qquad c\ne0,
\]
where the roots $\alpha_1,\ldots,\alpha_s$ are distinct. Then, for
$z\notin\{\alpha_1,\ldots,\alpha_s\}$,
\begin{equation}
\label{eq:prelim-partial-fractions}
F(z)=\sum_{j=1}^{s}\frac{c_j}{z-\alpha_j},
\qquad
c_j=\operatorname{Res}_{z=\alpha_j}F(z)
=\frac{P(\alpha_j)}{Q'(\alpha_j)}.
\end{equation}
\end{lemma}

\begin{corollary}[Residue theorem for proper rational functions]
\label{cor:prelim-zero-residue-sum}
Let $F=P/Q$ be a proper rational function with finite poles
$\alpha_1,\ldots,\alpha_s$. Then
\begin{equation}
\label{eq:prelim-rational-residue-sum}
\sum_{j=1}^{s}\operatorname{Res}_{z=\alpha_j}F(z)
=\lim_{z\to\infty}zF(z).
\end{equation}
In particular, if $\deg Q\ge\deg P+2$, then
\begin{equation}
\label{eq:prelim-zero-residue-sum}
\sum_{j=1}^{s}\operatorname{Res}_{z=\alpha_j}F(z)=0.
\end{equation}
The poles need not be simple.
\end{corollary}

\begin{proof}
Since $F$ is proper, $L=\lim_{z\to\infty}zF(z)$ exists and
$F(z)=L/z+O(z^{-2})$. For a sufficiently large counterclockwise circle
$\Gamma_R=\{z:|z|=R\}$ enclosing all finite poles, the residue theorem gives
\[
2\pi i\sum_{j=1}^{s}\operatorname{Res}_{z=\alpha_j}F(z)
=\oint_{\Gamma_R}F(z)\,dz
=2\pi i L+O(R^{-1}).
\]
Letting $R\to\infty$ proves the identity. If $\deg Q\ge\deg P+2$, then
$L=0$.
\end{proof}

%% file: upper.tex
\newcommand{\R}{\mathbb R}
\newcommand{\sx}{\mathbf{x}}
\newcommand{\ii}{\mathbf{i}}
\numberwithin{equation}{section}

\section{The Algorithm}\label{sec:upper-algorithm}

\subsection{Problem setting}\label{sec:setting}
In this section, we present an algorithm that computes the Hamming weight
with bounded error. For an input $x\in\bits^n$, the algorithm
has query access to the phase oracle
and is supposed to output the Hamming weight
$n_1=|\{i:x_i=1\}|$. With the algorithm described below, we have the
following theorem.

\begin{theorem}[Recovering the Hamming weight]\label{thm:target}
For every fixed error $\varepsilon > 0$, there is a quantum query algorithm
$\mathcal A$ such that, for any $n$ and any $n$-bit input $x\in\bits^n$,
$\mathcal A$ outputs the Hamming weight $|x|$ with probability at least
$1-\varepsilon$ and has query complexity
\begin{equation}\label{eq:handwritten-query-bound}
T(\mathcal A)\le\frac n2+O_\varepsilon(n^{1/3}).
\end{equation}
\end{theorem}
The constant factor in $O_\varepsilon$ depends on $\varepsilon$.

For the input $x$, let $n_a=|\{i\in[n]:x_i=a\}|$ for
$a\in\{0,1\}$, so $n_0+n_1=n$ and $n_1=|x|$. Define
\[
b=n_0-n_1=n-2|x|,\qquad d=|b|,\qquad
w=\min\{n_0,n_1\}=\frac{n-d}{2}.
\]
Here $b$ is the signed bias, $d$ is the absolute bias, and $w$ is the
minority count. Each unequal pair contains one zero and one one, so
$w$ is also the maximum number of disjoint unequal pairs. Removing
such a pair decreases both counts by one and leaves the bias unchanged.

\subsection{Algorithm}\label{sec:description}
The algorithm breaks down into three
stages: coarse estimation of the bias, a fixed-time quantum walk that
recovers part of the weight, and exact counting of the remaining weight.
The second stage contributes the $n/2$ factor and requires the most analysis.
The first and third stages use the algorithms introduced in
\Cref{lem:prelim-counting,lem:prelim-sparse-recovery}. We relist them,
as well as the other existing ingredients needed for the algorithm.

\subsubsection{Ingredients for the algorithm}\label{sec:alg-ingredients}

\begin{itemize}[leftmargin=*]
\item \textbf{$\mathsf{ApproxCount}(\varepsilon,E)$.}
On input $x\in\bits^n$
, it outputs $\widehat{b}$. By \Cref{clm:board-approx-count}, $|\widehat{b}-b|\leq E$ with probability at least $1-\varepsilon$.

\item \textbf{$\mathsf{ExactCount}(\varepsilon,h)$.} On input $x\in\bits^r$, suppose $|x|(r-|x|)\leq h^2$, then by \Cref{clm:board-exact-count} it outputs the Hamming weight $|x|$ with probability at least $1-\varepsilon$.

\item \textbf{Quantum operations needed for the quantum walk}:

Let $\mathcal M=\operatorname{span}\{\ket M\}$ be the space of
matching states, where $\ket M$ represents a matching $M$ of indices
in $[n]$. Let
$\mathcal C=\operatorname{span}\{\ket\uparrow,\ket\downarrow\}$
be a one-qubit register determining the walk direction. We construct
the following operations:

\begin{itemize}[leftmargin=*]
\item Controlled sampler $K$:
\begin{equation*}
\begin{split}
K&=\sum_M\ket M\bra M\otimes K_{R(M)},\\
K_{R(M)}&=A_{R(M)}O_x^{\mathrm{ph}}A_{R(M)}^\dagger
+\bigl(I-A_{R(M)}A_{R(M)}^\dagger\bigr),
\end{split}
\end{equation*}
where $R(M)\subseteq[n]$ is the set of indices not in $M$,
$K_{R(M)},A_{R(M)}$ act on indices in $R(M)$, and $A_{R(M)}$ is
defined in \Cref{eq:handwritten-isometry}.
Constructing $K$ involves one query.

\item $O_x^{\mathrm{ph}}$: phase oracle query.

\item Isometry $P$, as defined in \Cref{eq:wb-pair-merging}.

\item Unitary $Q=P\cdot K\cdot P^\dagger$.

\item Coin flipper
$S=\ket\uparrow\bra\downarrow+\ket\downarrow\bra\uparrow$.

\item One-step walk unitary $W$:
\begin{equation}\label{eq:handwritten-walk}
W=S\cdot Q=SPKP^\dagger.
\end{equation}
\end{itemize}

\item \textbf{State preparation for the quantum walk}

Let $\{\ket{j;x}\}$ be a set of orthogonal matching states as defined
in \Cref{eq:wb-matching-state}, with match size $0\leq j\leq w$. Let $\ket{\chi_j}$ prepare $\ket{j;x}$ with a superposition of coins,
\begin{equation*}
\ket{\chi_j}
=\cos\frac{\kappa_j}{2}\ket{j;x}\ket\uparrow
+\sin\frac{\kappa_j}{2}\ket{j;x}\ket\downarrow,
\end{equation*}
as defined in \Cref{eq:wb-chi-state}, with
$\kappa_j=\arcsin\frac{b}{n-2j}$ on the packet support.
At the unused endpoint $d=0$, $j=w$, take $\kappa_w=\pi/2$,
consistent with the convention in \Cref{sec:wb-matching}.
Let $\phi$ be a fixed nonzero real smooth bump supported on $[1,2]$, let $\ell$ be a
parameter, and define the amplitude function $f_0$:
\begin{equation}\label{eq:handwritten-packet}
f_0(j)=\frac{c_\ell}{\sqrt\ell}\phi\left(\frac j\ell\right),
\end{equation}
where $c_\ell$ is a normalizing factor such that
$\sum_{j=\ell}^{2\ell}|f_0(j)|^2=1$.
We take $n$ sufficiently large that this normalization is defined.

Define $\ket{\Psi_0}$ as
\begin{equation}\label{eq:handwritten-unphased-packet}
\ket{\Psi_0}=\sum_{j=\ell}^{2\ell}f_0(j)\ket{\chi_j}.
\end{equation}
Define the phase rotation unitary $G$, with $\mathbf{i}$ being the
imaginary unit:
\begin{equation}\label{eq:handwritten-carrier}
\begin{split}
G\ket{j;x}\ket\uparrow
&=\mathbf{i}^j\ket{j;x}\ket\uparrow,\\
G\ket{j;x}\ket\downarrow
&=\mathbf{i}^{j+1}\ket{j;x}\ket\downarrow.
\end{split}
\end{equation}
Then define the initial state
\begin{equation}\label{eq:board-initial-packet}
\ket{\Phi_0}=G\ket{\Psi_0}
=\sum_{j=\ell}^{2\ell}\mathbf{i}^j f_0(j)
\left(\cos\frac{\kappa_j}{2}\ket{j;x}\ket\uparrow
+\mathbf{i}\sin\frac{\kappa_j}{2}\ket{j;x}\ket\downarrow\right),
\end{equation}
which can be prepared by the algorithm described in
\Cref{sec:preparation}.
\end{itemize}

\subsubsection{Algorithm}\label{sec:alg-procedure}

\begin{tcolorbox}[
  breakable,
  colback=white,
  colframe=black,
  colbacktitle=white,
  coltitle=black,
  fonttitle=\bfseries,
  title={Algorithm: $\mathcal A(x;\varepsilon)$},
  boxrule=0.5pt,
  titlerule=0.5pt,
  arc=0pt,
  boxsep=0pt,
  toptitle=4pt,bottomtitle=4pt,
  left=8pt,right=8pt,top=6pt,bottom=6pt
]
Set $\varepsilon'=\varepsilon/10$ and $E=n^{2/3}$, and let $L,C_0$
be large constants depending on $\varepsilon$, let $\alpha$
be a sufficiently small constant depending on $\varepsilon$.

\begin{enumerate}[leftmargin=*,label=\arabic*.]
\item \textbf{Coarse estimation.}
Run the approximate counting subroutine to obtain the estimate
\begin{equation}\label{eq:handwritten-coarse-stage}
\widehat b\leftarrow\mathsf{ApproxCount}(\varepsilon',E),
\qquad D=|\widehat b|+E,
\end{equation}
This choice gives $d\le D\le d+2E$ except with probability $\varepsilon'$.

\item \textbf{Quantum walk.}
Set $\ell=\lceil L\sqrt D\rceil$.
If $D\ge(1-\alpha)n$, skip the quantum walk.
Otherwise, prepare $\ket{\Phi_0}$ using
\Cref{clm:board-packet-preparation} with error $\varepsilon'$.
Set
\begin{equation}\label{eq:handwritten-schedule}
t_0=\frac12\sqrt{(n-6\ell)^2-D^2}.
\end{equation}
Without loss of generality, we take $t_0$ to be even.
Apply the one-step unitary $W$ for $t_0$ steps, obtaining
\begin{equation}\label{eq:handwritten-evolution}
\ket{\Phi_{t_0}}=W^{t_0}\ket{\Phi_0}.
\end{equation}
Measure the state, obtaining a matching $M$ of size $j$. Each edge
represents a $(0,1)$ pair in the input bits. Remove all indices in $M$
and obtain the remaining index subset $R\subseteq[n]$:
\begin{equation}\label{eq:handwritten-residual}
R=\{i:\text{for every }e\in M,\ i\notin e\},
\qquad |R|=r=n-2j.
\end{equation}
Extract the remaining input $x_R\in\bits^r$.

\item \textbf{Finish the remaining input.}
Set the promise bound
\begin{equation}\label{eq:handwritten-cleanup-budget}
h=\begin{cases}
C_0DE/n=C_0 D/n^{1/3},&\text{if }D<(1-\alpha)n,\\[3pt]
\sqrt{\alpha}n,&\text{otherwise}.
\end{cases}
\end{equation}
Run the exact sparse counting subroutine to compute
\begin{equation}\label{eq:handwritten-cleanup}
\widehat n_1(x_R)\leftarrow\mathsf{ExactCount}(\varepsilon',h).
\end{equation}
Output $j+\widehat n_1(x_R)$ as the final Hamming weight.
\end{enumerate}
\end{tcolorbox}

\subsection{Algorithm overall analysis}\label{sec:overall-analysis}

We first claim that each subroutine satisfies the desired properties, and derive the overall analysis.

\begin{claim}[Approximate counting]\label{clm:board-approx-count}
For any fixed error $\varepsilon>0$ and accuracy $E\le n$, the subroutine $\mathsf{ApproxCount}(\varepsilon,E)$ makes $O_\varepsilon(n/E)$ queries to the input $x\in\bits^n$. Let $\widehat b$ be its output, then with probability at least $1-\varepsilon$, $|\widehat b-b|\le E,$ where $b$ is the bias of $x$.

\end{claim}

\begin{proof}
By \Cref{lem:prelim-counting}, we generalize the error $1/100$ to a fixed error $\varepsilon$, we can find an estimator of $|x|$ with additive error at most $E/2$ in $O_\varepsilon(n/E)$ queries. Since $b=n-2|x|$, we can approximate $b$ with additive error at most $E$.
\end{proof}

\begin{claim}[Exact counting]\label{clm:board-exact-count}
For any fixed error $\varepsilon>0$ and known promise bound $h\ge0$,
the subroutine $\mathsf{ExactCount}(\varepsilon,h)$ makes at most
$O(h+1)$ queries on every input $y\in\bits^r$. Whenever
$\sqrt{|y|(r-|y|)}\le h$, it outputs the Hamming weight $|y|$
with probability at least $1-\varepsilon$.
\end{claim}

\begin{proof}
If $r=0$, return zero. Otherwise, let
$w_y=\min\{|y|,r-|y|\}$. Under the promise,
\begin{equation*}
rw_y\le2w_y(r-w_y)=2|y|(r-|y|)\le2h^2.
\end{equation*}
Set $K=\lfloor2h^2/r\rfloor$, so $w_y\le K$.
If $K\ge r/2$, read all $r$ bits and return their Hamming weight.
This case uses at most $2h$ queries, since $K\ge r/2$ implies $h\ge r/2$.

Otherwise, $2K+1\le r$. Query any $2K+1$ positions and let $b_*$
be their majority bit. At most $K$ positions in the entire input
contain the minority bit, so $b_*$ is also the majority bit of $y$.
Apply \Cref{lem:prelim-sparse-recovery} with bound $K$ to find all
positions whose value is $1-b_*$, complementing the queried bits
when $b_*=1$. If this finds $t$ positions, return $t$ when $b_*=0$
and $r-t$ when $b_*=1$. Under the promise, the recovery succeeds
with certainty and the output is $|y|$.

Thus the second case always uses
at most
\begin{equation*}
2K+1+O(\sqrt{rK})=O(\sqrt{rK}+1)=O(h+1)
\end{equation*}
queries, using $K\le r$ and $rK\le2h^2$.
When $K=0$, one query identifies the common bit under the promise,
and sparse recovery uses no queries.
\end{proof}

\begin{lemma}[Initial state preparation]\label{clm:board-packet-preparation}
Let $\varepsilon>0$ be a fixed error, $\phi$ be a smooth bump supported
on $[1,2]$, and $E,L,\alpha$ be defined in
\Cref{sec:alg-procedure}.
Let $\widehat b$ be an estimator of $b$ with the guarantee that
$|\widehat b-b|\le E$ and
$D=|\widehat b|+E<(1-\alpha)n$.
Set $\ell=\lceil L\sqrt D\rceil$. For sufficiently large $n$,
the initial state $\ket{\Phi_0}$ defined in
\Cref{eq:board-initial-packet} can be prepared using
$O(\ell)$ queries, with failure probability at most $\varepsilon$,
and trace distance at most $\varepsilon$ when successful.
\end{lemma}

We defer the proof of this lemma to the proof of \Cref{cor:preparation}.

\begin{lemma}[Quantum walk guarantee]\label{lem:board-walk-guarantee}
Following the algorithm, suppose $D<(1-\alpha)n$, $D\geq E$, and $D\ge d$.
Let $\ket{\Phi_{t_0}}=W^{t_0}\ket{\Phi_0}$, with $\ket{\Phi_0}$ as in
\Cref{eq:board-initial-packet}. Let $x_R$ be the remaining input after
measuring $\ket{\Phi_{t_0}}$, with length $r$ and Hamming weight $w_R$.
Then, with probability at least $1-\varepsilon$,
\begin{equation*}
\sqrt{w_R(r-w_R)}
\le C_1\left(\frac{D^2-d^2}{n}+\ell\right).
\end{equation*}
\end{lemma}

With all the above guarantees, we then prove \Cref{thm:target}.
\begin{proof}[Proof of \Cref{thm:target}]

We first bound the total error of the
algorithm:
\begin{equation*}
\begin{split}
\mathrm{err}(\mathrm{total})
&\le\mathrm{err}(\mathrm{step\ 1})
+\mathrm{err}(\mathrm{step\ 2})
+\mathrm{err}(\mathrm{step\ 3}),\\
\mathrm{err}(\mathrm{step\ 2})
&\le\mathrm{err}(\mathrm{prep})
+\mathrm{TD}(\mathrm{prep})+\mathrm{err}(\mathrm{walk}).
\end{split}
\end{equation*}
Here the error bounds for each stage are conditioned on the preceding
stages satisfying their guarantees.
The term $\mathrm{err}(\mathrm{prep})$ stands for the failure
probability when preparing $\ket{\Phi_0}$;
$\mathrm{TD}(\mathrm{prep})$ stands for the trace distance between
$\ket{\Phi_0}$ and the actual initial state when preparation succeeds,
which also bounds the trace distance between the actual post-walk state
and $\ket{\Phi_{t_0}}$.
The term $\mathrm{err}(\mathrm{walk})$ stands for the probability
that the matching measured from the ideal post-walk state does not
satisfy the guarantee in \Cref{cor:ea-cleanup}.
By \Cref{lem:ea-measurement},
$\mathrm{err}(\mathrm{walk})\le\mathrm{err}_{\mathrm{walk}}^2$;
here $\mathrm{err}_{\mathrm{walk}}$ is the norm error defined in
\Cref{sec:wb-hybrids}.

By the guarantees above, each of the five error contributions is at
most $\varepsilon'$. Therefore, the total error is at most
$5\varepsilon'\le\varepsilon$, since
$\varepsilon'=\varepsilon/10$.

Then, for the query complexity, when $D<(1-\alpha)n$, we have
\begin{equation*}
\begin{split}
T(\mathcal A)=T_1+T_2+T_3
&\le O\left(\frac nE\right)+t_0
+O\left(\frac{DE}{n}\right)+O(\ell)\\
&\le\frac12\sqrt{n^2-D^2}+O(n^{1/3})
+O\left(\frac{ED}{n}\right)+O(\sqrt D)\\
&\le\frac n2-\frac{D^2}{4n}
+O\left(\frac{D}{n^{1/3}}\right)+O(n^{1/3})\\
&\le\frac n2+O(n^{1/3}).
\end{split}
\end{equation*}
All hidden constants may depend on the fixed error $\varepsilon$.
The cleanup term uses the fixed query budget in
\Cref{eq:handwritten-cleanup-budget,clm:board-exact-count}.
Here we used $\sqrt D=O(D/n^{1/3})$, since $D\ge E=n^{2/3}$,
and $\ell=\lceil L\sqrt D\rceil=O(\sqrt D)$.
For the last inequality,
$-D^2/(4n)+CD/n^{1/3}\le C^2n^{1/3}$ for every fixed $C>0$.

When $D\ge(1-\alpha)n$, the preparation and walk are skipped.
On successful coarse estimation,
$d\ge D-2E\ge(1-\alpha)n-2E\ge(1-2\alpha)n$
for sufficiently large $n$.
Taking $\alpha<1/2$, this gives
$n_0n_1=(n^2-d^2)/4\le\alpha n^2$, so
$h=\sqrt\alpha\,n$ is a valid cleanup promise.
Let $C$ be a constant such that the query bound in
\Cref{clm:board-exact-count} is at most $C(h+1)$.
Choose $\alpha$ sufficiently small and $n$ sufficiently large that
$C(\sqrt\alpha\,n+1)\le n/3$. Then
\begin{equation*}
T(\mathcal A)=T_1+0+T_3
\le O\left(\frac nE\right)+\frac n3
=\frac n3+O(n^{1/3}).
\end{equation*}
The query bounds hold even on failure outcomes because the subroutines
have fixed query caps. The finitely many smaller input lengths can be
handled by reading all bits and increasing the constant in the final
$O(n^{1/3})$ term.
\end{proof}

\subsection{Quantum walk analysis}\label{sec:walk-analysis}

The main idea of the quantum walk is to obtain more unequal
pairs, i.e.\ index pairs $(i,j)$ such that $x_i\ne x_j$, which contribute Hamming weight  by $1$ without
querying $x_i,x_j$ individually. Such pairs can be sampled with a
one-query sampler~\cite{AmbainisIraidsSmotrovs2013}, and we integrate it into a multi-step quantum
walk of gathering them into a matching, so that the progress accumulates
faster than classically repeating the sampling process. Moreover, the
evolution of weight distribution over matching sizes during the quantum
walk can be approximated with a continuous-time evolution, which has
the nice property of preserving boundedness of the distribution. Thus,
by a series of hybrid arguments, we can show the weight distribution of
the post-walk state is close to a continuously evolved distribution,
which allows us to guarantee that by setting the walk step properly,
most weight is supported within a bounded interval of big enough
matching size. In fact, it is guaranteed that the remaining unmatched
minority bits are few enough that we can move on to the next step of
exactly counting them using \Cref{clm:board-exact-count}.

The analysis is divided into 6 steps: we will first define the walk operator and how to implement it using one query, then preparing the initial state, extending the
weight distribution to continuous space, approximating the walk with
a continuous-time evolution, bounding the approximation error with a
set of hybrid arguments, guaranteeing the post-walk measurement outcome.

In the following analysis, for input $x\in\{0,1\}^n$, we denote $n_0$
as the number of $0$s in $x$, and $n_1$ as the number of $1$s,
$w=\min\{n_0,n_1\}$ as the minority weight. Denote $b=n_0-n_1$ as
the bias, $d=|b|$ as the absolute bias. For $i\in[n]$, let
$z_i=(-1)^{x_i}$. For a
subset $R\subseteq[n]$, denote $x_R\in\{0,1\}^{|R|}$ as the
subsequence of $x$ at the indices in $R$.

\subsubsection{Operations in each step}\label{sec:wb-matching}

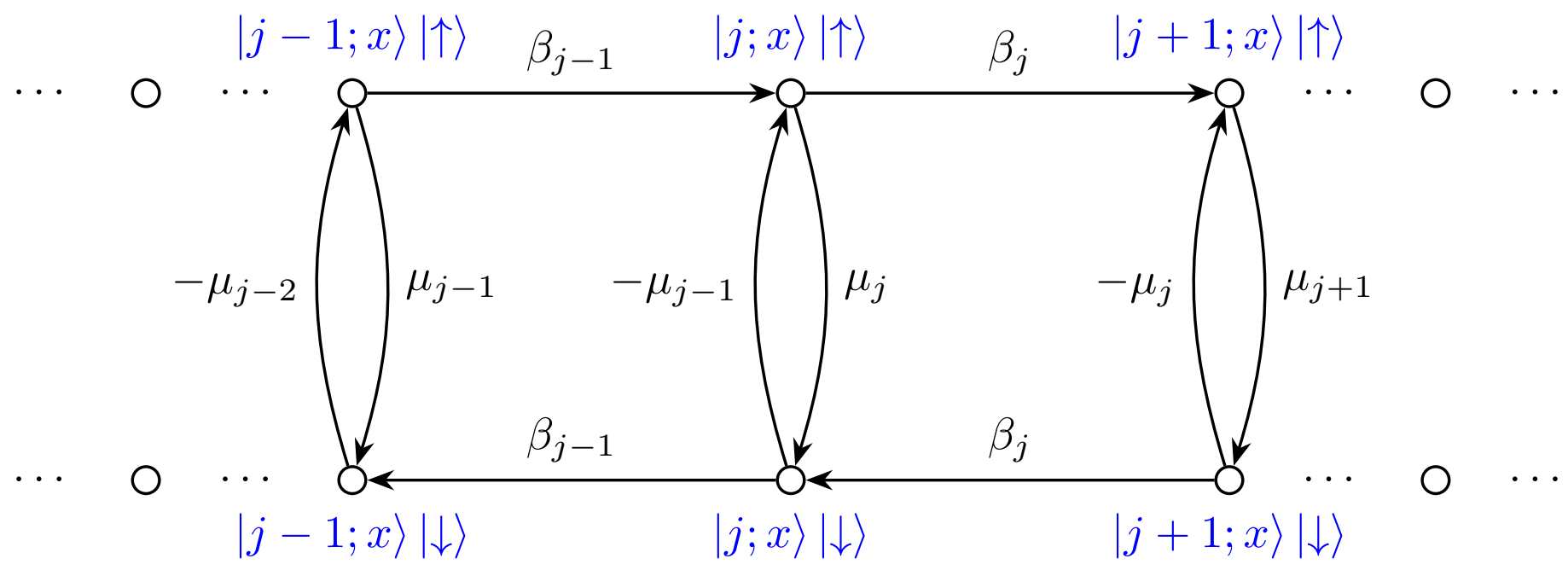
\begin{figure}[htbp]
\centering
\begin{tikzpicture}[
  >=Stealth, line width=.65pt, font=\small,
  vertex/.style={circle,draw,fill=white,inner sep=2.1pt},
  every label/.style={text=blue},
  every edge/.style={draw,->}]
\node[vertex,label=above:{$\ket{j-1;x}\ket{\uparrow}$}] (ul) at (0,3) {};
\node[vertex,label=above:{$\ket{j;x}\ket{\uparrow}$}] (um) at (3.4,3) {};
\node[vertex,label=above:{$\ket{j+1;x}\ket{\uparrow}$}] (ur) at (6.8,3) {};
\node[vertex,label=below:{$\ket{j-1;x}\ket{\downarrow}$}] (dl) at (0,0) {};
\node[vertex,label=below:{$\ket{j;x}\ket{\downarrow}$}] (dm) at (3.4,0) {};
\node[vertex,label=below:{$\ket{j+1;x}\ket{\downarrow}$}] (dr) at (6.8,0) {};
\foreach \row in {0,3} {
  \foreach \column in {-1.6,8.4} {
    \node[vertex] at (\column,\row) {};
  }
  \foreach \column in {-2.4,-.8,7.6,9.2} {
    \node at (\column,\row) {$\cdots$};
  }
}
\path (ul) edge node[above] {$\beta_{j-1}$} (um)
      (um) edge node[above] {$\beta_j$} (ur)
      (dr) edge node[above] {$\beta_j$} (dm)
      (dm) edge node[above] {$\beta_{j-1}$} (dl);
\path (ul) edge[bend left=17] node[right] {$\mu_{j-1}$} (dl)
      (dl) edge[bend left=17] node[left] {$-\mu_{j-2}$} (ul)
      (um) edge[bend left=17] node[right] {$\mu_j$} (dm)
      (dm) edge[bend left=17] node[left] {$-\mu_{j-1}$} (um)
      (ur) edge[bend left=17] node[right] {$\mu_{j+1}$} (dr)
      (dr) edge[bend left=17] node[left] {$-\mu_j$} (ur);
\end{tikzpicture}

\caption{Quantum walk transitions on the matching basis.}
\label{fig:matching-walk-transitions}
\end{figure}

Following the high-level idea above, the quantum walk is defined on a
graph (Fig.~\ref{fig:matching-walk-transitions}) with each vertex representing a basis state $\ket{j;x}$
corresponding to a matching of size $j$, along with a direction state
$\ket{\uparrow}$ or $\ket{\downarrow}$. We first define $\mathcal M$
as the general matching space spanned by the basis $\{\ket M\}$,
where $M$ is a matching of indices in $[n]$, i.e.\ a set of disjoint
pairs. Then, for $0\le j\le w$, we define
\begin{equation}\label{eq:wb-matching-state}
\ket{j;x}=\frac1{\sqrt{\Lambda_j}}
\sum_{\substack{M\text{ matching}\\|M|=j}}h_M(x)\ket M,
\end{equation}
\begin{equation*}
\begin{split}
\text{with   }h_M(x)=\prod_{\substack{(u,v)\in M\\u<v}}(z_u-z_v),\ \ 
\Lambda_j=4^j\cdot\#\{\text{unequal matchings of size }j\}
=\frac{4^j(n_0)_j(n_1)_j}{j!},
\end{split}
\end{equation*}
where $(a)_j=a(a-1)\cdots(a-j+1)$, and $(a)_0=1$. Notice that if
$M$ ever contains an equal pair, i.e.\ a pair $(u,v)$ with $x_u=x_v$,
then $z_u-z_v=0$, $h_M(x)=0$. Therefore $\ket{j;x}$ is a
superposition of all the size $j$ unequal matchings, with signs
indicated by $h_M(x)$, and $\Lambda_j$ is a normalization factor.
We also use one qubit to represent
$\mathcal C=\operatorname{span}\{\ket{\uparrow},\ket{\downarrow}\}$,
as the coin to decide the moving direction of the next step.

\paragraph{The one-query sampler~\cite{AmbainisIraidsSmotrovs2013}.}
We use the one-query sampler to coherently sample a new unequal pair
each time, and it is applied on indices in $R(M)\subseteq[n]$, the
indices not in $M$. The whole controlled sampler $K$ is defined as
\begin{equation*}
K=\sum_M\ket M\bra M\otimes K_{R(M)},
\end{equation*}
where we abbreviate $K_{R(M)}$ as $K_r$, with
$r=|R(M)|=n-2|M|$, and similarly write $A_r=A_{R(M)}$
after relabeling the residual indices by $[r]$. $K_r$ is defined as:
\begin{equation*}
K_r=A_r O_x^{\mathrm{ph}}A_r^\dagger+(I-A_rA_r^\dagger),
\end{equation*}
Here $O_x^{\mathrm{ph}}$ is a phase query on $x_{R(M)}$,
$O_x^{\mathrm{ph}}\ket i=z_i\ket i$, and
$A_r:\mathcal V\to\mathcal E$ is an isometry mapping the index space
$\mathcal V=\operatorname{span}\{\ket u:u\in[r]\}$ to the edge space
$\mathcal E=\operatorname{span}\bigl(
\{\ket{u,v}:u<v,\ u,v\in[r]\}\cup\{\ket{\mathsf{fail}}\}\bigr)$
that includes an extra flag $\ket{\mathsf{fail}}$. $A_r$ is defined as
\begin{equation}\label{eq:handwritten-isometry}
A_r=\ket{\mathsf{fail}}\left(\frac1{\sqrt r}\sum_{u\in[r]}\bra u\right)
+\frac1{\sqrt r}\sum_{u<v}\ket{u,v}(\bra u-\bra v),
\end{equation}
To see how $K_r$ samples an unequal pair, we initialize with
$\ket{\mathsf{fail}}$,
\begin{equation*}
\begin{split}
\ket{\mathsf{fail}}
&\xrightarrow{A_r^\dagger}\frac1{\sqrt r}\sum_{u\in[r]}\ket u
\xrightarrow{O_x^{\mathrm{ph}}}\frac1{\sqrt r}\sum_{u\in[r]}z_u\ket u\\
&\xrightarrow{A_r}\frac1r\sum_u z_u\ket{\mathsf{fail}}
+\frac1r\sum_{u<v}(z_u-z_v)\ket{u,v}\\
&\phantom{\xrightarrow{A_r}}=\frac b r\ket{\mathsf{fail}}
+\frac1r\sum_{u<v}(z_u-z_v)\ket{u,v},
\end{split}
\end{equation*}
while whenever we measure and obtain a pair $(u,v)$, that indicates
$z_u-z_v\ne0$, $x_u\ne x_v$.

\paragraph{Absorbing a pair into a matching.}
We define an isometry $P$ on the span of the following input vectors
to merge the pair into the matching,
\begin{equation}\label{eq:wb-pair-merging}
\begin{split}
P\left(\frac1{\sqrt{|M|}}\sum_{e\in M}\ket{M\setminus e}\ket e\right)
&=\ket M\ket{\downarrow},\\
P\ket M\ket{\mathsf{fail}}&=\ket M\ket{\uparrow}.
\end{split}
\end{equation}
We define it this way, instead of $P\ket M\ket e=\ket{M\cup e}$,
so that $\ket{M\setminus e_1}\ket{e_1}$,
$\ket{M\setminus e_2}\ket{e_2}$ do not map to $\ket M$ together,
and $P$ can be an isometry. Then define a unitary $Q$ that applies to
$\mathcal M\otimes\mathcal C$,
\begin{equation*}
Q=P\cdot K\cdot P^\dagger,
\end{equation*}
so that we have
\begin{equation*}
Q\ket{j;x}\ket{\uparrow}
=\frac b{n-2j}\ket{j;x}\ket{\uparrow}
+\sqrt{1-\left(\frac b{n-2j}\right)^2}\ket{j+1;x}\ket{\downarrow},
\end{equation*}
with the intuition that $P^\dagger$ prepares state $\ket{\mathsf{fail}}$
while the match size is $j$; after $K$ we either remain at
$\ket{\mathsf{fail}}$, which sends us back to $\ket{j;x}\ket{\uparrow}$,
or sample a random unequal pair not
in the matching, which will be added into the matching by $P$.

We define parameters $\beta_j,\mu_j$ for $0\le j\le w$ as
\begin{equation*}
\mu_j=\frac b{n-2j},\qquad\beta_j=\sqrt{1-\mu_j^2},
\end{equation*}
with special boundary case $\mu_{-1}=-1$, $\beta_{-1}=0$.
Then, the action of $Q$ yields
\begin{equation}\label{eq:wb-Q}
\begin{split}
Q\ket{j;x}\ket{\uparrow}
&=\mu_j\ket{j;x}\ket{\uparrow}
+\beta_j\ket{j+1;x}\ket{\downarrow},\\
Q\ket{j+1;x}\ket{\downarrow}
&=\beta_j\ket{j;x}\ket{\uparrow}
-\mu_j\ket{j+1;x}\ket{\downarrow}.
\end{split}
\end{equation}

\paragraph{The complete one-step unitary $W$.}
Observe that $Q$ preserves the subspace
$\operatorname{span}\{\ket{j;x}\ket{\uparrow},\allowbreak
\ket{j+1;x}\ket{\downarrow}\}$
for $0\le j\le w-1$, which is not what we want as the walk is stuck
locally. To form an all-connected path, we define a coin-flipping
unitary $S$, then combine them together to obtain the complete
one-step unitary $W$,
\begin{equation*}
S=\ket{\uparrow}\bra{\downarrow}+\ket{\downarrow}\bra{\uparrow}.
\end{equation*}
\begin{equation*}
W=S\cdot Q=SP K P^\dagger,
\end{equation*}
while $W$ still only needs one phase query. Therefore, we form
complete quantum walk dynamics.

By representing
$\{\ket{j;x}\ket{\uparrow},\ket{j;x}\ket{\downarrow}\}_j$
with vertices, we have Fig.~\ref{fig:matching-walk-transitions}, where arrows represent amplitude
transfer, described by
\begin{equation*}
\begin{split}
W\ket{j;x}\ket{\uparrow}
&=\beta_j\ket{j+1;x}\ket{\uparrow}
+\mu_j\ket{j;x}\ket{\downarrow},\\
W\ket{j;x}\ket{\downarrow}
&=-\mu_{j-1}\ket{j;x}\ket{\uparrow}
+\beta_{j-1}\ket{j-1;x}\ket{\downarrow}.
\end{split}
\end{equation*}

\paragraph{The analysis one-step unitary $V$.}
Although the actual unitary applied during each step of the walk is
$W$, it contains a fast-varying phase rotation factor, which is
distracting when we only care about the weight on each basis.
Therefore, we extract this phase rotation factor by defining $G$ as
\begin{equation*}
\begin{split}
G\ket{j;x}\ket{\uparrow}&=\mathbf{i}^j\ket{j;x}\ket{\uparrow},\\
G\ket{j;x}\ket{\downarrow}&=\mathbf{i}^{j+1}\ket{j;x}\ket{\downarrow},
\end{split}
\end{equation*}
where $\mathbf{i}$ is the imaginary unit. Then we obtain $V$ as the
unitary we analyze,
\begin{equation*}
V=\mathbf{i}\cdot G^\dagger WG,
\end{equation*}
Thus the state after $t$ steps, with the initialization in
\Cref{eq:handwritten-evolution}, can be represented as
\begin{equation}\label{eq:wb-phase-evolution}
W^t\ket{\Phi_0}=(-\mathbf{i})^tG V^t\ket{\Psi_0}.
\end{equation}
The phase unitary $G$ and the global phase do not influence the
matching distribution, so we can analyze $V^t\ket{\Psi_0}$.

To view the transformation of $V$ on the weight distribution of
matching size clearly, we denote $u_{j,\uparrow},u_{j,\downarrow}$
to be the amplitudes of $\ket{j;x}\ket{\uparrow},\ket{j;x}\ket{\downarrow}$,
and $u$ to be the amplitude distribution. We then view $V$ as a
transformation on the amplitude distribution,
\begin{equation*}
\begin{split}
(Vu)_{j,\uparrow}
&=\beta_{j-1}\cdot u_{j-1,\uparrow}+\mu_{j-1}\cdot u_{j,\downarrow},\\
(Vu)_{j,\downarrow}
&=\mu_j\cdot u_{j,\uparrow}-\beta_j\cdot u_{j+1,\downarrow}.
\end{split}
\end{equation*}

Taking the zeroth-order approximation of $V$, we have
\begin{equation*}
\begin{split}
(Vu)_j
=\binom{(Vu)_{j,\uparrow}}{(Vu)_{j,\downarrow}}
\approx\binom{\beta_j\cdot u_{j,\uparrow}+\mu_j\cdot u_{j,\downarrow}}
{\mu_j\cdot u_{j,\uparrow}-\beta_j\cdot u_{j,\downarrow}}
=\begin{pmatrix}\beta_j&\mu_j\\\mu_j&-\beta_j\end{pmatrix}
\cdot u_j=:C_ju_j,
\end{split}
\end{equation*}
where $C=\bigoplus_jC_j$ is defined in this way to be the zeroth-order
approximation of $V$,
let $\chi_j,\eta_j$ be eigenvectors of $C_j$, with
$C_j\chi_j=\chi_j$, $C_j\eta_j=-\eta_j$,
\begin{equation*}
\chi_j=\binom{\cos\frac{\kappa_j}{2}}{\sin\frac{\kappa_j}{2}},
\qquad
\eta_j=\binom{-\sin\frac{\kappa_j}{2}}{\cos\frac{\kappa_j}{2}},
\end{equation*}
with $\kappa_j=\arcsin\mu_j$.
We therefore denote $\ket{\chi_j}$ as
\begin{equation}\label{eq:wb-chi-state}
\ket{\chi_j}
=\cos\frac{\kappa_j}{2}\ket{j;x}\ket{\uparrow}
+\sin\frac{\kappa_j}{2}\ket{j;x}\ket{\downarrow},
\end{equation}
We set $\{\ket{\chi_j}\}$ as the support of the unphased initial state
$\ket{\Psi_0}$, and we will show in further analysis that the weight
approximately stays on $\{\ket{\chi_j}\}$ during the evolution by $V$.

\subsubsection{Preparing the starting superposition}\label{sec:preparation}
We then show how to prepare the initial state using
amplitude amplification.

By \Cref{eq:wb-Q}, the one-query operator $Q$ preserves the link
spanned by $\ket{s;x,\uparrow}$ and $\ket{s+1;x,\downarrow}$, and
\begin{equation*}
Q\ket{s;x,\uparrow}
=\mu_s\ket{s;x,\uparrow}+\beta_s\ket{s+1;x,\downarrow}.
\end{equation*}

The following lemma shows how to prepare $\ket{s+1;x,\downarrow}$ from
$\ket{s;x,\uparrow}$ when the transition amplitude is bounded below.
\begin{lemma}[One transition]\label{lem:amplified-insertion}
Fix $0<c\le1$ and $0\le\eta\le c/4$. Suppose $\beta_s$ and a known
estimate $\widehat\beta_s$ belong to $[c,1]$, with
$|\widehat\beta_s-\beta_s|\le\eta$.
There is a unitary $U_s$ using $O_c(1)$ queries with a real amplitude
$a_s\ge0$ satisfying
\[
\begin{aligned}
a_s&:=\bra{s+1;x,\downarrow,1}U_s\ket{s;x,\uparrow,0},\\
&1-O_c(\eta^2)\le a_s\le1.
\end{aligned}
\]
The last register denotes an ancillary bit.
\end{lemma}
\begin{proof}
Choose a sufficiently large fixed odd integer $N=2m+1$ so that
$\sin\theta_0\le c/2$, where $\theta_0=\pi/(2N)$.
Append an ancillary bit that is $1$ with probability
$\rho_s=\sin^2\theta_0/\widehat\beta_s^{\,2}\le1$.
Implement the reflections by
\[
\begin{aligned}
R_u&=I-2\bigl(I_{\mathcal M}\otimes
\ket{\uparrow,0}\bra{\uparrow,0}\bigr),\\
R_\Pi&=I-2\bigl(I_{\mathcal M}\otimes
\ket{\downarrow,1}\bra{\downarrow,1}\bigr).
\end{aligned}
\]
Here $I_{\mathcal M}$ is the identity on the matching register; both
require no queries. Applying \Cref{lem:prelim-amplitude-amplification}
to $Q$ and the bit preparation on this radial link for $m$ rounds gives
a unitary $U_s$ using $N$ queries with amplitude
\[
a_s=\sin\!\left(N\arcsin(\sqrt{\rho_s}\,\beta_s)\right)
=\sin\!\left(\frac\pi2+O_c(\eta)\right)=1-O_c(\eta^2).
\]
The bound $\eta\le c/4$ keeps the amplified angle in $(0,\pi)$,
so $a_s>0$.
\end{proof}

\begin{corollary}[Preparing the initial superposition]\label{cor:preparation}
Use $b,E,D$ from \Cref{sec:setting,eq:handwritten-coarse-stage},
put $k=n/E$, and let $\ell=\lceil L\sqrt D\rceil$, with $L>0$ fixed. If
$|\widehat b-b|\le E$ and $D<(1-\alpha)n$, for fixed $0<\alpha<1$
and sufficiently large $n$,
the state $\ket{\Phi_0}$ in \Cref{eq:board-initial-packet} can be prepared using at
most $O_\alpha(\ell)$ queries, with failure probability $O_\alpha(\ell/k^2)$.
Conditioned on acceptance, the normalized output state is at trace
distance $O_\alpha(\ell/k^2+1/k)$ from $\ket{\Phi_0}$.
\end{corollary}
\begin{proof}
For $s<2\ell$, the amplitude
$\beta_s=\sqrt{1-(b/(n-2s))^2}$ and its estimate obtained by replacing
$b$ with $\widehat b$ are bounded below by a constant depending on $\alpha$
and differ by $O_\alpha(1/k)$. Thus \Cref{lem:amplified-insertion} gives
$a_s=1-O_\alpha(k^{-2})$ per transition.
Prepare $\sum_jf_0(j)\ket j_T\ket{\varnothing,\uparrow}$.
In round $\nu=1,\ldots,2\ell$, apply the transition only on targets $j\ge\nu$;
all gates and amplification phases leave completed targets unchanged.
Measure only a common pass/fail flag, with inactive targets always
passing; a rejection aborts preparation. On acceptance, reset the port
and ancillary bit from $\ket{\downarrow,1}$ to $\ket{\uparrow,0}$ on
active targets, leaving completed targets unchanged.
The coefficient at target $j$ is multiplied by
$\prod_{s<j}a_s=1-O_\alpha(\ell/k^2)$ uniformly. Thus the rejection
probability is $O_\alpha(\ell/k^2)$, and the $2\ell$ rounds use
$O_\alpha(\ell)$ queries in total.

Erase $T$ using the matching size. Conditioned on acceptance, the
normalized state is at norm distance $O_\alpha(\ell/k^2)$ from
\[
\sum_{j=\ell}^{2\ell}f_0(j)\ket{j;x}\ket{\uparrow}.
\]
Finally, apply the direction rotation in
\Cref{eq:wb-chi-state} with $b$ replaced by $\widehat b$.
This approximates the unphased state
$\ket{\Psi_0}=\sum_{j=\ell}^{2\ell}f_0(j)\ket{\chi_j}$.
The rotation requires no queries and adds $O_\alpha(1/k)$ to the norm
distance. Finally apply $G$, which requires no queries and preserves
this distance, to obtain the desired state $\ket{\Phi_0}=G\ket{\Psi_0}$.
The resulting norm bound $O_\alpha(\ell/k^2+1/k)$ also bounds the
trace distance between these pure states.
\end{proof}

\subsubsection{Extend to continuous space}\label{sec:wb-continuous-space}

One thing not satisfying enough about the action of $V$ on the
amplitude distribution is that $(Vu)_j$ not only depends on $u_j$
but depends on $u_{j-1},u_{j+1}$ as well, which prevents us from
directly analyzing the time evolution of the weight on some $j$.
To handle this, we first extend our initial weight distribution onto
``real number matching size'', then approximate the value of neighboring
positions in terms of its derivatives at the current point. We use
$\mathbf{x}$ as the continuous spatial coordinate as an extension of
$j$; we use boldface $\mathbf{x}$ to distinguish it from input $x$.

We start by extending the amplitude distribution $u$, which now becomes
a function $u:\mathbb R\to\mathbb R^2$;
\begin{equation*}
u(\mathbf{x})=\binom{u_\uparrow(\mathbf{x})}{u_\downarrow(\mathbf{x})}.
\end{equation*}
On the interior $0<\mathbf{x}<w$, extend the coefficients by
\begin{equation*}
\mu(\mathbf{x})=\frac b{n-2\mathbf{x}},\qquad
\beta(\mathbf{x})=\sqrt{1-\mu(\mathbf{x})^2},\qquad
\kappa(\mathbf{x})=\arcsin\mu(\mathbf{x}),
\end{equation*}
and extend $f_0$ to $\mathbb R$ and $\chi,\eta$ to this interior by
\begin{equation*}
f_0(\mathbf{x})=
\begin{cases}
\dfrac{c_\ell}{\sqrt\ell}\cdot\phi\!\left(\dfrac{\mathbf{x}}\ell\right),
&\mathbf{x}\in[\ell,2\ell],\\
0,&\text{otherwise},
\end{cases}
\end{equation*}
\begin{equation*}
\chi(\mathbf{x})=
\binom{\cos\frac{\kappa(\mathbf{x})}{2}}{\sin\frac{\kappa(\mathbf{x})}{2}},
\qquad
\eta(\mathbf{x})=
\binom{-\sin\frac{\kappa(\mathbf{x})}{2}}{\cos\frac{\kappa(\mathbf{x})}{2}},
\end{equation*}
with $c_\ell,\phi$ as in \Cref{eq:handwritten-packet}.
Recall that the evolution of $V$ starts with
$\ket{\Psi_0}=\sum_j f_0(j)\cdot\ket{\chi_j}$, then the extended initial continuous distribution is 
\begin{equation*}
u_0(\mathbf{x})=(\chi\cdot f_0)(\mathbf{x})
=\binom{f_0(\mathbf{x})\cdot\cos\frac{\kappa(\mathbf{x})}{2}}
{f_0(\mathbf{x})\cdot\sin\frac{\kappa(\mathbf{x})}{2}}.
\end{equation*}

We then extend the action of $V$ to a continuous distribution. For
$u:\mathbb R\to\mathbb R^2$, define
\begin{equation*}
\begin{split}
(Vu)(\mathbf{x})
=\binom{(Vu)_\uparrow(\mathbf{x})}{(Vu)_\downarrow(\mathbf{x})}
=\binom{\beta(\mathbf{x}-1)\cdot u_\uparrow(\mathbf{x}-1)
+\mu(\mathbf{x}-1)\cdot u_\downarrow(\mathbf{x})}
{\mu(\mathbf{x})\cdot u_\uparrow(\mathbf{x})
-\beta(\mathbf{x})\cdot u_\downarrow(\mathbf{x}+1)}.
\end{split}
\end{equation*}
Apply Taylor expansion, and let $\mathsf T_\pm h(\mathbf{x})$ denote
the residue such that
\begin{equation*}
h(\mathbf{x}\pm1)=h(\mathbf{x})\pm h'(\mathbf{x})
+\frac12h''(\mathbf{x})+\mathsf T_\pm h(\mathbf{x}),
\end{equation*}
Then, we have
\begin{equation*}
\begin{split}
(Vu)_\uparrow
&=\beta u_\uparrow+\mu u_\downarrow
-(\beta u_\uparrow)'-\mu'u_\downarrow+\frac12(\beta u_\uparrow)''+\frac12\mu''u_\downarrow
+\mathsf T_-(\beta u_\uparrow)+(\mathsf T_-\mu)u_\downarrow,\\
(Vu)_\downarrow
&=\mu u_\uparrow-\beta u_\downarrow-\beta u_\downarrow'
-\frac12\beta u_\downarrow''-\beta\mathsf T_+u_\downarrow.
\end{split}
\end{equation*}
Therefore, $V$ can be represented using coefficient matrices depending
on $\mathbf{x}$,
\begin{equation}\label{eq:wb-spatial-expansion}
V=C+D_1+D_2+R_3,
\end{equation}
where
\begin{equation}\label{eq:wb-spatial-operators}
C=
\begin{pmatrix}
\beta&\mu\\
\mu&-\beta
\end{pmatrix},\qquad
D_1u=\binom{-(\beta u_\uparrow)'-\mu'u_\downarrow}
{-\beta u_\downarrow'},
\end{equation}
\begin{equation*}
D_2u=\frac12\binom{(\beta u_\uparrow)''+\mu''u_\downarrow}
{-\beta u_\downarrow''},\qquad
R_3u=\binom{\mathsf T_-(\beta u_\uparrow)+(\mathsf T_-\mu)u_\downarrow}
{-\beta\mathsf T_+u_\downarrow}.
\end{equation*}
Here $C,D_1,D_2,R_3$ represent the zeroth-, first-, second-order terms
and the residue.

Specifically, consider amplitudes of the form
$u(\mathbf{x})=\chi(\mathbf{x})f(\mathbf{x})$, where
$f:\mathbb R\to\mathbb R$ is a scalar function.
For our choice of $\chi$, the up and down components are
\begin{equation*}
u_\uparrow(\mathbf{x})=f(\mathbf{x})\cos\frac{\kappa(\mathbf{x})}{2},
\qquad
u_\downarrow(\mathbf{x})=f(\mathbf{x})\sin\frac{\kappa(\mathbf{x})}{2}.
\end{equation*}
Differentiation therefore acts on both $f$ and the trigonometric
factors. Set $a=\tfrac12\kappa'$ and define the scalar differential
operator $\mathcal L$ by
\begin{equation*}
\mathcal Lf=-\beta\cdot\partial_{\mathbf{x}}f-\frac12\beta'f.
\end{equation*}
We now evaluate the zeroth- and first-order terms for this input.
Since $C\chi=\chi$, the zeroth-order term is $C(\chi f)=\chi f$.
For the first-order term, substitute the two components above into the
definition of $D_1$. Applying the product rule for differentiation,
using $\chi'=a\eta$, $\beta'=-2a\mu$, and $\mu'=2a\beta$, gives
\begin{equation*}
\begin{split}
D_1(\chi f)
&=\binom{-(\beta f\cos(\kappa/2))'-\mu'f\sin(\kappa/2)}
{-\beta(f\sin(\kappa/2))'}\\
&=-\beta\chi f'+a(\mu\chi-\eta)f\\
&=\chi\left(-\beta f'-\frac12\beta'f\right)-a\eta f
=\chi\mathcal Lf-a\eta f.
\end{split}
\end{equation*}
The same two-component vector is now written in the $\chi,\eta$
basis; the component along $\eta$ comes from the position dependence
of $\chi$. Combining these identities with $V=C+D_1+D_2+R_3$ gives
\begin{equation*}
\begin{split}
V(\chi f)&=C(\chi f)+D_1(\chi f)+(D_2+R_3)(\chi f)\\
&=\chi f+\chi\mathcal Lf-\eta af+(D_2+R_3)(\chi f).
\end{split}
\end{equation*}

If we apply $V$ twice, the first-order term will be fully supported on $\chi$, rather than $\eta$.
Since $C\eta=-\eta$, the first-order term of
$V^2\chi f$ is
\begin{equation}\label{eq:wb-two-step-first}
\begin{split}
(CD_1+D_1C)\chi f
&=C(\chi\mathcal Lf-\eta af)+(\chi\mathcal Lf-\eta af)
=2\chi\mathcal Lf.
\end{split}
\end{equation}

\subsubsection{Approximate with continuous-time evolution}
\label{sec:wb-continuous-time}

Following the action of $V$, the evolution of the amplitude distribution
$u$ can be denoted as
\begin{equation*}
u_t=V^tu_0=V^t(\chi f_0),
\end{equation*}
and the amplitude on $\chi$ can be denoted as
\begin{equation*}
f_t=\chi^\dagger u_t.
\end{equation*}
For amplitudes of the form $\chi f$, the first-order calculation
in \Cref{eq:wb-two-step-first} gives a change of $2\chi\mathcal Lf$
over two steps.

With these observations, we approximate the discrete evolution of
$f_t$ with a continuous-time evolution $\widetilde f(t,\mathbf{x})$,
satisfying
\begin{equation}\label{eq:wb-transport}
\begin{split}
\widetilde f(0,\mathbf{x})&=f_0(\mathbf{x}),\\
\partial_t\widetilde f
&=-\beta(\mathbf{x})\cdot\partial_{\mathbf{x}}\widetilde f
-\frac12\beta'(\mathbf{x})\cdot\widetilde f.
\end{split}
\end{equation}
We will show in \Cref{lem:ea-hybrid} that such an approximation only causes
fixed constant error at any even timestamp $t$; now we focus on
analyzing $\widetilde f$ instead.

The definition of $\widetilde f$ also gives
\begin{equation*}
\partial_t|\widetilde f|^2
=-\partial_{\mathbf{x}}\bigl(\beta|\widetilde f|^2\bigr),
\end{equation*}
so its squared amplitude moves with speed $\beta$.

We introduce the characteristic flow of the partial differential
equation, $X(t,\xi)\in[0,w]$, by
\begin{equation*}
\begin{split}
X(0,\xi)&=\xi,\\
\partial_tX(t,\xi)&=\beta(X(t,\xi)).
\end{split}
\end{equation*}
The flow is used for interior starting points and times before reaching
the boundary.

Intuitively, the time evolution $X(\cdot,\xi)$ represents the movement
of an object that starts at $\xi$ when $t=0$ and moves with velocity
$\beta(X)$ at position $X$. We can view $X$ as a moving reference
frame. The map $X(t,\cdot)$ takes an object's starting point at time
$0$ to its position at time $t$ following the reference frame.

We then view the evolution $\widetilde f$ under the reference frame
$X$. For fixed $\xi$, we may abbreviate $X(t,\xi)$ as $X(t)$ or $X$.
\begin{equation*}
\begin{split}
\frac{d}{dt}\widetilde f(t,X(t,\xi))
&=(\partial_t\widetilde f)(t,X)
+(\partial_{\mathbf{x}}\widetilde f)(t,X)\partial_tX\\
&=-\beta(X)\partial_{\mathbf{x}}\widetilde f(t,X)
-\frac12\beta'(X)\widetilde f(t,X)
+\partial_{\mathbf{x}}\widetilde f(t,X)\beta(X)\\
&=-\frac12\beta'(X)\widetilde f(t,X).
\end{split}
\end{equation*}
Thus we have
\begin{equation*}
\begin{split}
\widetilde f(t,X(t,\xi))
&=f_0(\xi)\exp\!\left(-\frac12\int_0^t
\beta'(X(s,\xi))\,ds\right)\\
&=f_0(\xi)\sqrt{\frac{\beta(\xi)}{\beta(X(t,\xi))}}.
\end{split}
\end{equation*}
This shows that the initial weight at $\xi$ is mapped to the weight at
$X(t,\xi)$ of $\widetilde f(t,\cdot)$, with a reweighting factor,
and if $\xi$ is unsupported initially, i.e.\ $f_0(\xi)=0$, then
$\widetilde f(t,\cdot)$ is also not supported at $X(t,\xi)$.
Therefore, a continuous interval $[\ell,2\ell]$ is mapped to
$[X(t,\ell),X(t,2\ell)]$, and $\widetilde f(t,\cdot)$ is exactly
supported on that when $\operatorname{supp}f_0=[\ell,2\ell]$.
For a general bump, the exact support statement is
\begin{equation}\label{eq:wb-flow-support}
\operatorname{supp}\widetilde f(t,\cdot)
=X\bigl(t,\operatorname{supp}f_0\bigr)
\subseteq[X(t,\ell),X(t,2\ell)].
\end{equation}

We then define the width of the support of $\widetilde f(t,\cdot)$
to be $W$, and the distance between the interval and the endpoint to be $\delta$,
\begin{equation}\label{eq:wb-width-gap}
\begin{split}
W(t)=&X(t,2\ell)-X(t,\ell),\\
\delta(t)=&w-X(t,2\ell).
\end{split}
\end{equation}

For a general bump, $W(t)$ is the length of this enclosing interval.
We can approximate them with the following claims.
In the estimates below, $A\lesssim B$ means $A\le cB$ for a
constant $c$ depending only on $\alpha$, and $A\asymp B$ means
both $A\lesssim B$ and $B\lesssim A$.

The branch condition $D<(1-\alpha)n$ and
$6\ell\le\alpha n/2$ give $\beta(\xi)\asymp1$ on $[\ell,3\ell]$.
In particular, $\beta(\xi)\ge v_\alpha>0$ for a constant
$v_\alpha$ depending only on $\alpha$.
We use the following conditions, and verify
below that they hold for all $0\le t\le t_0$:
\begin{equation}\label{eq:hd-width-regime}
\begin{split}
v_\alpha\le\beta(\xi)&\le1\qquad(\ell\le\xi\le2\ell),\\
\delta(t)&\ge\delta_\ell,
\qquad \delta_\ell:=\frac{\sqrt{d^2+4\ell^2}-d}{2}.
\end{split}
\end{equation}
\begin{claim}[Packet width]\label{clm:wb-width}
Under \Cref{eq:hd-width-regime},
$W(t)\asymp\ell\beta(X(t,2\ell))$.
\end{claim}
\begin{proof}
Differentiating the characteristic flow with respect to $\xi$ gives
\begin{equation*}
\partial_\xi X(t,\xi)=\frac{\beta(X(t,\xi))}{\beta(\xi)},
\qquad
W(t)=\int_\ell^{2\ell}\frac{\beta(X(t,\xi))}{\beta(\xi)}\,d\xi.
\end{equation*}
Using the remaining-time function $g$ defined in
\Cref{eq:hd-remaining-time}, put
\begin{equation*}
\begin{split}
\Delta&=g(w-\ell)-g(w-2\ell)
=\int_\ell^{2\ell}\frac{d\xi}{\beta(\xi)}
\asymp\ell,\\
s&=g(\delta(t))\ge\ell.
\end{split}
\end{equation*}
The remaining times across the interval lie in $[s,s+\Delta]$.
The speed as a function of remaining time $y$ is
$2y/\sqrt{d^2+4y^2}$, so
\begin{equation*}
1\le\frac{\beta(X(t,\xi))}{\beta(X(t,2\ell))}
\le\frac{s+\Delta}{s}\lesssim1
\qquad(\ell\le\xi\le2\ell).
\end{equation*}
Together with $\beta(\xi)\asymp1$, the integral formula for $W(t)$
proves the claim.
\end{proof}

\begin{claim}[Distance from the endpoint]\label{clm:wb-gap}
Under \Cref{eq:hd-width-regime}, $W(t)\lesssim\delta(t)$.
\end{claim}
\begin{proof}
Since $g(\delta_\ell)=\ell$, we have $g(\delta(t))\ge\ell$, and
\begin{equation*}
\frac{W(t)}{\delta(t)}
\asymp\frac{\ell\beta(X(t,2\ell))}{\delta(t)}
\lesssim\frac{\ell}{g(\delta(t))}\le1,
\end{equation*}
using \Cref{clm:wb-width} and
$\beta(w-q)=2\sqrt{q(q+d)}/(2q+d)$.
\end{proof}

Following the characteristic flow $X$, a weight on $X(t,\xi)$ will
move according to $\beta(X)$, and all further movement is determined
by $X$. Denote $\delta=w-X$ to be the remaining distance; we can then
determine its remaining time before moving to $w$, and denote it as
\begin{equation}\label{eq:hd-remaining-time}
g(\delta)=\sqrt{\delta(\delta+d)}
=\frac12\sqrt{(n-2X)^2-d^2}.
\end{equation}
Along the characteristic,
\begin{equation}\label{eq:hd-travel-derivative}
\begin{split}
\partial_tg(\delta)
&=g'(\delta)\partial_t\delta\\
&=-\frac{2\delta+d}{2\sqrt{\delta(\delta+d)}}
\cdot\frac{2\sqrt{\delta(\delta+d)}}{2\delta+d}=-1.
\end{split}
\end{equation}

Let $t^*=g(w-3\ell)=\frac12\sqrt{(n-6\ell)^2-d^2}$
denote the moving time from $3\ell$ to $w$.
Since $D\ge d$, the rounded time in \Cref{eq:handwritten-schedule}
satisfies $t_0\le t^*$. Thus, for $0\le t\le t_0$,
\Cref{eq:hd-travel-derivative} gives
\begin{equation*}
\begin{split}
g(\delta(t))
&=g(w-2\ell)-t\\
&\ge g(w-2\ell)-g(w-3\ell)
=\int_{2\ell}^{3\ell}\frac{d\xi}{\beta(\xi)}\ge\ell.
\end{split}
\end{equation*}
Because $g(\delta_\ell)=\ell$, the conditions in
\Cref{eq:hd-width-regime} hold throughout the walk.

\begin{claim}[Positions used in the approximation]\label{clm:ea-interior}
For sufficiently large fixed $L$ and $0\le t\le t_0$,
both $W(t)$ and $\delta(t)$ are $\Omega(L^2)$.
During any interval $[t,t+2]\subseteq[0,t_0]$, the packet and
all positions within distance four of it lie inside $(0,w)$.
On these positions, $\beta(\mathbf{x})\asymp\beta(X(t,2\ell))$
and $w-\mathbf{x}\asymp\delta(t)$.
\end{claim}
\begin{proof}
The identity $\delta_\ell(\delta_\ell+d)=\ell^2$ and
$\delta_\ell\le\ell$ imply
$\delta_\ell\ge\ell^2/(\ell+d)\gtrsim L^2$:
here $\ell\ge L^2$ and $d\le D\le\ell^2/L^2$.
The preceding bound gives $\delta(t)\ge\delta_\ell$.
Since $\beta(w-q)=2\sqrt{q(q+d)}/(2q+d)$ is increasing in $q$,
the integral formula in the proof of \Cref{clm:wb-width} gives
\begin{equation*}
W(t)\ge\ell\beta(X(t,2\ell))
\ge\ell\beta(w-\delta_\ell)\asymp\delta_\ell.
\end{equation*}
The left endpoint is at least $\ell$, and the right endpoint is
at distance at least $\delta_\ell$ from $w$.
Since $\partial_tX=\beta\le1$, evolving for two units of time and
enlarging the packet by four changes its positions by $O(1)$.
Taking $L$ sufficiently large, the lower bounds on
$\ell,\delta_\ell$ keep this enlargement inside $(0,w)$.
Across the packet, $w-\mathbf{x}\in[\delta(t),\delta(t)+W(t)]$,
so \Cref{clm:wb-gap} gives $w-\mathbf{x}\asymp\delta(t)$.
The same lower bounds keep this comparison valid after the
$O(1)$ displacements. Substitution in the displayed formula
for $\beta(w-q)$ then gives
$\beta(\mathbf{x})\asymp\beta(X(t,2\ell))$ throughout the
enlarged interval.
\end{proof}

\subsubsection{Hybrids between the discrete and continuous processes}
\label{sec:wb-hybrids}
We now bound the difference between the walk and the continuous
evolution defined above. Write $\widetilde f_t=\widetilde f(t,\cdot)$,
$\widetilde u_t=\chi\widetilde f_t$, and
$\beta_t=\beta(X(t,2\ell))$.

In this and the next subsection, for nonnegative quantities $A,B$,
we write $A\lesssim B$ if $A\le cB$ for a constant $c$, and
$A\asymp B$ if both $A\lesssim B$ and $B\lesssim A$.
Unless indicated otherwise, these constants may depend on $\alpha$
and the fixed function $\phi$, but not on $n,d,D,L$ or time.
A subscript, as in $\lesssim_q$ or $\lesssim_L$, allows dependence
on that additional parameter. We use the same convention for
$O(\cdot)$ and $\Theta(\cdot)$.

\begin{definition}[Discrete sampling]\label{def:ea-sampling}
For a continuous amplitude function $u$, define
\begin{equation}\label{eq:ea-sampling}
\ket{\mathcal I(u)}=\sum_{j=0}^{w}
\left(u_\uparrow(j)\ket{j;x}\ket\uparrow
+u_\downarrow(j)\ket{j;x}\ket\downarrow\right).
\end{equation}
\end{definition}
This vector need not be normalized. However,
$\ket{\mathcal I(\widetilde u_0)}=\ket{\Psi_0}$ is normalized by
\Cref{eq:handwritten-packet}. Let
$\ket{\Psi_t}=V^t\ket{\Psi_0}$ for integer $t$.
We will compare this state with $\ket{\mathcal I(\widetilde u_t)}$.
Let $\mathrm{err}_{\mathrm{walk}}$ denote the maximum of this
norm difference over even integers $0\le t\le t_0$.
For a continuous function, $\norm{f}^2$ denotes
$\int\norm{f(\mathbf{x})}^2\,d\mathbf{x}$; for a state vector it
denotes its usual squared norm.

We first record the derivative estimates used in the comparison.

\begin{claim}[Initial packet]\label{clm:ea-initial-derivatives}
For every fixed integer $q\ge0$,
$\norm{f_0^{(q)}}_\infty\lesssim_q\ell^{-q-1/2}$ and
$\norm{f_0}=\Theta(1)$.
\end{claim}
\begin{proof}
The sum defining $c_\ell^{-2}$ has at most $\ell+1$ nonzero
terms, each at most $\norm{\phi}_\infty^2/\ell$.
Since $\phi$ is continuous and nonzero, there is a fixed interval
inside $(1,2)$ on which $|\phi|$ is bounded below by a positive
constant. For sufficiently large $\ell$, this interval contains
at least a constant times $\ell$ points of the form $j/\ell$.
Thus $c_\ell=\Theta(1)$. The formula
$f_0^{(q)}(\mathbf{x})=c_\ell\ell^{-q-1/2}
\phi^{(q)}(\mathbf{x}/\ell)$ proves the derivative bound, and
$\norm{f_0}^2=c_\ell^2\int_1^2\phi(s)^2\,ds=\Theta(1)$.
\end{proof}

\begin{claim}[Coefficient derivatives]\label{clm:ea-coefficients}
For each fixed integer $q\ge1$ and $0<\mathbf{x}<w$,
\begin{equation}\label{eq:ea-coefficient-bound}
|\beta^{(q)}(\mathbf{x})|+|\mu^{(q)}(\mathbf{x})|
+\norm{\chi^{(q)}(\mathbf{x})}+\norm{\eta^{(q)}(\mathbf{x})}
\lesssim_q\beta(\mathbf{x})(w-\mathbf{x})^{-q}.
\end{equation}
For $q=0$, each of $|\beta|,|\mu|,\norm{\chi},\norm{\eta}$
is at most one.
\end{claim}

\begin{lemma}[Derivatives of the continuous packet]\label{lem:ea-regularity}
For every fixed integer $q\ge0$ and $0\le t\le t_0$,
$\norm{\partial_{\mathbf{x}}^q\widetilde f_t}_\infty
\lesssim_q W(t)^{-q-1/2}$ and
$\norm{\widetilde f_t}=\norm{f_0}=\Theta(1)$.
Moreover, for $0\le t\le t_0-2$,
\begin{equation}\label{eq:ea-shifted-sampling}
\sup_{\substack{t\le s\le t+2\\|z|\le2}}
\left(\sum_{j\in\mathbb Z}
|\partial_{\mathbf{x}}^q\widetilde f_s(j+z)|^2\right)^{1/2}
\lesssim_q W(t)^{-q}.
\end{equation}
\end{lemma}
The proofs of \Cref{clm:ea-coefficients,lem:ea-regularity} are in
\Cref{app:ea-regularity-proofs}. The second statement uses the
explicit solution along $X$, as well as
\Cref{clm:wb-width,clm:wb-gap}; preserving the norm alone would
not bound the derivatives.

\begin{lemma}[Two-step approximation]\label{lem:ea-local}
For every even integer $0\le t\le t_0-2$,
\begin{equation}\label{eq:ea-local}
\norm{V^2\ket{\mathcal I(\widetilde u_t)}
-\ket{\mathcal I(\widetilde u_{t+2})}}
\lesssim\beta_t\left(\frac1{W(t)^3}
+\frac1{\delta(t)W(t)}+\frac1{\delta(t)^2}\right).
\end{equation}
\end{lemma}

To prove \Cref{lem:ea-local}, we separate three error terms.
The definitions of $C,D_1,D_2,R_3$ are those in
\Cref{eq:wb-spatial-expansion,eq:wb-spatial-operators}.
For $u=\chi f$, define
\begin{equation}\label{eq:ea-error-definitions}
\begin{split}
E_2f&=\chi\left(\beta'f'+\tfrac12\beta''f\right)
+\eta\left(\mu'f'+\tfrac12\mu''f\right),\\
\mathcal R_3[u]&=(V+I)R_3u
+(V-C-D_1)D_1u+(V-C)D_2u,\\
\mathcal R_4(t)&=\frac12\chi\int_0^2
(2-s)^2\mathcal L^3\widetilde f_{t+s}\,ds.
\end{split}
\end{equation}
Expanding the two walk steps and the two units of continuous time gives
\begin{equation}\label{eq:ea-two-expansions}
\begin{split}
V^2\widetilde u_t
&=\chi\bigl(\widetilde f_t+2\mathcal L\widetilde f_t
+2\mathcal L^2\widetilde f_t\bigr)
+E_2\widetilde f_t+\mathcal R_3[\widetilde u_t],\\
\widetilde u_{t+2}
&=\chi\bigl(\widetilde f_t+2\mathcal L\widetilde f_t
+2\mathcal L^2\widetilde f_t\bigr)+\mathcal R_4(t).
\end{split}
\end{equation}
The first identity follows by keeping second-order terms in the
expansion of $V^2$; its calculation is included in the proofs of
\Cref{clm:ea-second-order-error,clm:ea-spatial-error}.
The second is Taylor's formula in
time, using $\partial_t\widetilde f=\mathcal L\widetilde f$.
In particular, the terms involving $\widetilde f_t''$ cancel in
the difference. We bound the remaining terms separately.

\begin{claim}[Second-order error]\label{clm:ea-second-order-error}
For $0\le t\le t_0-2$,
$\norm{\ket{\mathcal I(E_2\widetilde f_t)}}
\lesssim\beta_t\bigl(\delta(t)^{-1}W(t)^{-1}+\delta(t)^{-2}\bigr)$.
\end{claim}

\begin{claim}[Spatial remainder]\label{clm:ea-spatial-error}
For $0\le t\le t_0-2$,
$\norm{\ket{\mathcal I(\mathcal R_3[\widetilde u_t])}}
\lesssim\beta_t\bigl(W(t)^{-1}+\delta(t)^{-1}\bigr)^3$.
\end{claim}

\begin{claim}[Time remainder]\label{clm:ea-time-error}
For $0\le t\le t_0-2$,
$\norm{\ket{\mathcal I(\mathcal R_4(t))}}
\lesssim\beta_t^3\bigl(W(t)^{-1}+\delta(t)^{-1}\bigr)^3$.
\end{claim}
The proofs of these three claims are in
\Cref{app:ea-error-proofs}.

\begin{proof}[Proof of \Cref{lem:ea-local}]
By \Cref{clm:ea-interior}, the packet and all positions within two
walk steps of its support lie inside $(0,w)$.
Evaluating the continuous update at integers therefore agrees
with the discrete walk, so
$V^2\ket{\mathcal I(\widetilde u_t)}
=\ket{\mathcal I(V^2\widetilde u_t)}$.
Subtracting the two lines of \Cref{eq:ea-two-expansions} leaves
$E_2\widetilde f_t+\mathcal R_3[\widetilde u_t]-\mathcal R_4(t)$.
Apply the triangle inequality and
\Cref{clm:ea-second-order-error,clm:ea-spatial-error,clm:ea-time-error}.
Finally, $W(t),\delta(t)\ge1$ and $\beta_t\le1$, and
\begin{equation*}
\bigl(W(t)^{-1}+\delta(t)^{-1}\bigr)^3
\lesssim W(t)^{-3}+\delta(t)^{-1}W(t)^{-1}
+\delta(t)^{-2}.
\end{equation*}
These bounds give \Cref{eq:ea-local}.
\end{proof}

\begin{lemma}[Accumulated walk error]\label{lem:ea-hybrid}
For every even integer $0\le t\le t_0$,
$\norm{\ket{\Psi_t}-\ket{\mathcal I(\widetilde u_t)}}
\le\mathrm{err}_{\mathrm{walk}}$, where
\begin{equation}\label{eq:ea-global-L}
\mathrm{err}_{\mathrm{walk}}
\lesssim\frac1{L^2}+\frac{\log n}{L n^{1/3}}.
\end{equation}
Thus the error is at most any prescribed positive constant by taking
$L$ sufficiently large and then $n$ sufficiently large.
\end{lemma}
\begin{proof}
We prove the bound at $t_0$; stopping the same argument earlier
gives the bound at every even $t\le t_0$.
For even $s=0,2,\ldots,t_0$, consider the states
$\ket{H_s}=V^{t_0-s}\ket{\mathcal I(\widetilde u_s)}$.
Their first and last terms are $\ket{\Psi_{t_0}}$ and
$\ket{\mathcal I(\widetilde u_{t_0})}$.
Unitarity and the triangle inequality give
\begin{equation}\label{eq:ea-hybrid-sum}
\begin{split}
\norm{\ket{\Psi_{t_0}}-\ket{\mathcal I(\widetilde u_{t_0})}}
&\le\sum_{\substack{0\le s\le t_0-2\\s\text{ even}}}
\norm{V^2\ket{\mathcal I(\widetilde u_s)}
-\ket{\mathcal I(\widetilde u_{s+2})}}\\
&\lesssim\sum_{\substack{0\le s\le t_0-2\\s\text{ even}}}
\beta_s\left(W(s)^{-3}+\delta(s)^{-1}W(s)^{-1}
+\delta(s)^{-2}\right).
\end{split}
\end{equation}

We compare this sum with an integral in the remaining distance.
For $q>0$, put $B(q)=2\sqrt{q(q+d)}/(2q+d)$ and
\begin{equation*}
M(q)=
\ell\min\{1,\sqrt{q/d}\},
\qquad
H(q)=M(q)^{-3}+q^{-1}M(q)^{-1}+q^{-2}.
\end{equation*}
Here $B(\delta(s))=\beta_s$. The formula for $B$ and
\Cref{clm:wb-width} give $W(s)\asymp M(\delta(s))$.
Thus each summand is $O(B(\delta(s))H(\delta(s)))$.
The function $H$ is nonincreasing, and
$\partial_s\delta(s)=-B(\delta(s))$.
By \Cref{clm:ea-interior} and $B\le1$,
$\delta(v)\asymp\delta(s)$ for $s\le v\le s+2$.
The formula for $B$ then gives
$B(\delta(v))\asymp B(\delta(s))$ on the same interval.
Consequently,
\begin{equation*}
\begin{split}
\delta(s)-\delta(s+2)
&=\int_s^{s+2}B(\delta(v))\,dv\asymp B(\delta(s)),\\
B(\delta(s))H(\delta(s))
&\lesssim\int_{\delta(s+2)}^{\delta(s)}H(q)\,dq.
\end{split}
\end{equation*}
The intervals in the last integral are disjoint except at their
endpoints. Hence \Cref{eq:ea-hybrid-sum} is
$O(\int_{\delta_\ell}^{n}H(q)\,dq)$.

For $q\ge\max\{\delta_\ell,d\}$, $M(q)=\ell$, so the first
two terms of this integral are $O(n/\ell^3)$ and
$O(\log n/\ell)$, respectively.
If $\delta_\ell<d$, then
$\delta_\ell=\ell^2/(\delta_\ell+d)\asymp\ell^2/d$, and
\begin{equation*}
\begin{split}
\int_{\delta_\ell}^d M(q)^{-3}\,dq
&\lesssim\frac{d^{3/2}}{\ell^3\sqrt{\delta_\ell}}
\lesssim\frac{d^2}{\ell^4},\\
\int_{\delta_\ell}^d q^{-1}M(q)^{-1}\,dq
&\lesssim\frac{\sqrt d}{\ell\sqrt{\delta_\ell}}
\lesssim\frac d{\ell^2}.
\end{split}
\end{equation*}
These two integrals are absent when $\delta_\ell\ge d$, including
$d=0$. The last term is bounded by
\begin{equation*}
\int_{\delta_\ell}^{n}q^{-2}\,dq
\le\frac1{\delta_\ell}
=\frac{\delta_\ell+d}{\ell^2}
\le\frac d{\ell^2}+\frac1\ell.
\end{equation*}
We therefore obtain
\begin{equation}\label{eq:ea-global}
\mathrm{err}_{\mathrm{walk}}
\lesssim\frac n{\ell^3}+\frac{\log n}\ell
+\frac{d^2}{\ell^4}+\frac d{\ell^2}+\frac1\ell.
\end{equation}
Using $d\le D$, $D\ge E=n^{2/3}$, and $\ell\ge L\sqrt D$
bounds their sum by
$O(L^{-3}+L^{-4}+L^{-2}+\log n/(Ln^{1/3}))$.
Absorbing the smaller terms proves \Cref{eq:ea-global-L}.
\end{proof}

\subsubsection{Measuring the post-walk state}
\label{sec:wb-measurement}

We use the support of the continuous packet and
\Cref{lem:ea-hybrid} to show that the walk produces a sufficiently
large matching. The more precise remaining-time bound also gives
the promise needed for the final counting step.

\begin{claim}[Location of the continuous packet]\label{clm:ea-support-end}
Every $\mathbf{x}\in[X(t_0,\ell),X(t_0,2\ell)]$ satisfies
$g(w-\mathbf{x})\lesssim\ell+(D^2-d^2)/n$.
\end{claim}
\begin{proof}
By \Cref{eq:wb-flow-support}, it suffices to bound
$g(w-X(t_0,\ell))=g(w-\ell)-t_0$.
Put $\tau=\frac12\sqrt{(n-6\ell)^2-D^2}$, so
$\tau-t_0=O(1)$ accounts for rounding. The conditions
$D<(1-\alpha)n$ and $6\ell\le\alpha n/2$ give $\tau\asymp n$.
Recall that $t^*=g(w-3\ell)\ge\tau$.
Taking the difference of their squares gives
$t^*-\tau\lesssim(D^2-d^2)/n$.
Splitting the remaining time at $3\ell$, we obtain
\begin{equation*}
\begin{split}
g(w-\ell)-t_0
&=\int_\ell^{3\ell}\frac{d\xi}{\beta(\xi)}
+(t^*-\tau)+(\tau-t_0)\\
&\lesssim\ell+\frac{D^2-d^2}{n}+1
\lesssim\ell+\frac{D^2-d^2}{n}.
\end{split}
\end{equation*}
Here $\beta(\xi)\asymp1$ on $[\ell,3\ell]$, and the
constant rounding term is absorbed by $\ell\ge1$.
Finally, $g$ is increasing and each position in this interval is at least
$X(t_0,\ell)$, proving the claim.
\end{proof}

\begin{lemma}[Post-walk measurement]\label{lem:ea-measurement}
Measure the matching register of $\ket{\Phi_{t_0}}$ in
\Cref{eq:handwritten-evolution} and ignore the coin. With probability
at least $1-\mathrm{err}_{\mathrm{walk}}^2$, the resulting unequal
matching leaves a residual input satisfying
\begin{equation}\label{eq:ea-measured-cleanup}
\sqrt{|x_R|(|R|-|x_R|)}
\lesssim\ell+\frac{D^2-d^2}{n}.
\end{equation}
\end{lemma}
\begin{proof}
Let $\Pi$ project onto all computational basis states $\ket M$
for unequal matchings with sizes in
$[X(t_0,\ell),X(t_0,2\ell)]$, with either coin value.
The support bound in \Cref{eq:wb-flow-support} implies
$(I-\Pi)\ket{\mathcal I(\widetilde u_{t_0})}=0$.
Therefore \Cref{lem:ea-hybrid} gives
\begin{equation*}
\norm{(I-\Pi)\ket{\Psi_{t_0}}}
\le\norm{\ket{\Psi_{t_0}}-
\ket{\mathcal I(\widetilde u_{t_0})}}
\le\mathrm{err}_{\mathrm{walk}}.
\end{equation*}
The probability of a measurement outcome outside this interval is
at most the square of this norm. By \Cref{eq:wb-phase-evolution},
$\ket{\Phi_{t_0}}=(-\mathbf{i})^{t_0}G\ket{\Psi_{t_0}}$.
Since $G$ only changes phases, measuring the two states gives the
same matching distribution.

On the remaining outcomes, let $j$ be the size of the matching.
Removing $j$ unequal pairs leaves minority count $w-j$ and majority
count $w-j+d$. Consequently,
$\sqrt{|x_R|(|R|-|x_R|)}=\sqrt{(w-j)(w-j+d)}=g(w-j)$,
and \Cref{clm:ea-support-end} proves
\Cref{eq:ea-measured-cleanup}. This proof uses the unnormalized
sampled vector directly, so sampling introduces no additional
normalization error. Taking $\mathrm{err}_{\mathrm{walk}}^2\le
\varepsilon$ proves \Cref{lem:board-walk-guarantee}.
\end{proof}

\begin{corollary}[Large matching and the cleanup promise]\label{cor:ea-cleanup}
If also $D-d\le2E$, then on the event in
\Cref{lem:ea-measurement}, the matching size satisfies
$w-j\lesssim\ell$ and
$\sqrt{|x_R|(|R|-|x_R|)}\lesssim_L DE/n$.
Thus a sufficiently large $C_0$ makes the walk-branch promise in
\Cref{eq:handwritten-cleanup-budget} valid.
\end{corollary}
\begin{proof}
The additional assumption gives
$D^2-d^2=(D-d)(D+d)\lesssim DE$. Put $B=DE/n=D/n^{1/3}$.
Because $D\ge n^{2/3}$, we have $B\ge\sqrt D\ge1$ and
$\ell\lesssim_L\sqrt D\le B$.
Thus \Cref{eq:ea-measured-cleanup} is $O_L(B)$, proving the
cleanup promise.

For the matching-size bound, $D-d\lesssim E$ and $D\le n$
imply
\begin{equation*}
B^2=\frac{D^2}{n^{2/3}}
\lesssim D+\frac{Dd}{n^{2/3}}
\lesssim D+\sqrt D\,d
\le\ell(\ell+d).
\end{equation*}
Consequently, \Cref{eq:ea-measured-cleanup} gives
$g(w-j)\lesssim\ell+B\lesssim\sqrt{\ell(\ell+d)}=g(\ell)$.
If $q:=w-j\ge\ell$, then
$q/\ell\le g(q)^2/g(\ell)^2\lesssim1$, so $q\lesssim\ell$.
The same bound is immediate when $q<\ell$.
\end{proof}

%% file: lower.tex
\section{The Lower bound}
Let $n = 2m$ denote the input length. Write \(|x|=\sum_{i=1}^n x_i\), and let \(Q_\varepsilon(F)\) denote the
minimum number of queries required by a quantum algorithm that computes
\(F\) with error at most \(\varepsilon\) on every input.

In this section, we construct symmetric boolean functions requiring strictly more than $\frac{n}{2} = m$ quantum queries to achieve a bounded error guarantee. The first constructed boolean function corresponds to the pointwise value of Grigoriev's function{~\cite{Grigoriev2001}} which has one pole and can rule out $m$-query algorithms for any fixed error $\varepsilon < 1/2$. Then by adding more poles to the original function, we construct a new boolean function with stronger lower bound $m + \Omega(m^{1/3})$ for some universal constant error.

In particular, we prove the following two theorems. The explicit
symmetric Boolean functions $F_m^{(0)}$ and $F_{m,g}$ on
$\{0,1\}^{2m}$ are defined in
\cref{eq:lb-one-pole-function,eq:lb-final-function}, respectively.

\begin{restatable}[One additional query]{theorem}{lbOneExtraTheorem}
\label{thm:lb-one-extra}
For every fixed $0\le\varepsilon<1/2$ and all sufficiently large $m$,
\[
 Q_\varepsilon(F_m^{(0)})\ge m+1.
\]
We prove the following stronger statement: every algorithm using at most $m$ queries to
compute $F_m^{(0)}$ has worst-case error at least
$1/2-O(1/\log m)$.
\end{restatable}

\begin{restatable}[A cube-root additive lower bound]{theorem}{lbCuberootTheorem}
\label{thm:lb-cuberoot}
There is an absolute constant $0<\varepsilon_0<1/2$ such that, for
all integers $g\ge1$ and $m\ge512g^3$,
\[
 Q_{\varepsilon_0}(F_{m,g})\ge m+g+1.
\]
In particular, for every $m\ge512$, set
\[
 g_m=\left\lfloor(m/512)^{1/3}\right\rfloor,
 \qquad F_{2m}=F_{m,g_m}.
\]
Then
\[
 Q_{\varepsilon_0}(F_{2m})
 \ge m+g_m+1
 =m+\Omega(m^{1/3})
 =\frac n2+\Omega(n^{1/3}).
\]
\end{restatable}

We now give a common framework for establishing these lower bounds.
The acceptance probability of a \(T\)-query quantum algorithm on input \(x\) can be written in sum-of-squares form{~\citep{BealsEtAl2001}} as
\begin{equation}\label{eq:lb-sos}
 p(x)=\sum_\alpha q_\alpha(x)^2,
 \qquad \deg (q_\alpha)\le T, x \in \{0,1\}^n
\end{equation}
with real polynomials \(q_\alpha\).

If a quantum algorithm computes the Boolean function $F: \{0,1\}^n \to \{0,1\}$ with bounded error $\varepsilon$, the corresponding polynomial $p(x)$ representing the probability of outputting $1$ must satisfy $p(x) \le \varepsilon$ when $F(x) = 0$, and $p(x) \ge 1 - \varepsilon$ when $F(x) = 1$. This yields a system of $2^n$ linear inequalities. To establish a query lower bound, we assign signed weights to these $2^n$ inequalities according to the value of $F(x)$ and take their linear combination to derive a contradiction for small $T$.

For a real function \(\nu:\{0,1\}^n\to\mathbb R\), define
\[
 \nu_+(x)=\max\{\nu(x),0\},\qquad
 \nu_-(x)=\max\{-\nu(x),0\},
\]
and
\[
 V_+=\sum_{x \in \{0,1\}^n}\nu_+(x),\qquad V_-=\sum_{x \in \{0,1\}^n}\nu_-(x),\qquad
 A=V_++V_-,\qquad S=V_+-V_-.
\]

For function $f: \{0,1\}^n \mapsto \mathbb{R} $, define \[L_\nu(f) =\sum_{x \in \{0,1\}^n} \nu(x) f(x)\]

We say \(\nu\) is non-negative on degree-\(T\) squares if
\[
 L_\nu(q^2)\ge0
 \quad\text{for every real polynomial \(q\) with \(\deg q\le T\)}.
\]

\begin{lemma}\label{lem:lb-obstruction}
Suppose that \(\nu\ne0\) is non-negative on degree-\(T\) squares.
Let \(F:\{0,1\}^n\to\{0,1\}\) satisfy
\[
 F(x)=1\quad\text{if }\nu(x)<0,\qquad
 F(x)=0\quad\text{if }\nu(x)>0,
\]
with arbitrary values when \(\nu(x)=0\).
Every quantum algorithm using at most \(T\) queries to compute \(F\)
has worst-case error at least
\[
 \frac{V_-}{V_++V_-}.
\]
Consequently, if \(\varepsilon<V_-/(V_++V_-)\), then
\(Q_\varepsilon(F)\ge T+1\).
\end{lemma}
\begin{proof}
Let $\delta$ be the worst-case error of an algorithm using at most $T$
quantum queries, and let $p(x)$ be its probability of outputting $1$.
By \eqref{eq:lb-sos} and the assumption on $\nu$,we have
\[
L_\nu(p)=\sum_\alpha L_\nu(q_\alpha^2)\ge 0.
\]
When $\nu(x)>0$, we have $F(x)=0$ and hence $p(x)\le\delta$.
When $\nu(x)<0$, we have $F(x)=1$ and hence $p(x)\ge1-\delta$.
Therefore,
\[
\begin{aligned}
0\le L_\nu(p)
&= \sum_{\nu(x)>0}\nu(x)p(x)
 + \sum_{\nu(x)<0}\nu(x)p(x) \\
&\le \delta V_+-(1-\delta)V_- \\
&= \delta(V_++V_-)-V_-.
\end{aligned}
\]
Since $\nu\ne0$, we have $V_++V_->0$, so rearranging gives
\[
\delta\ge \frac{V_-}{V_++V_-}.
\]
Thus, if $\varepsilon<V_-/(V_++V_-)$, no algorithm using at most
$T$ queries can achieve error at most $\varepsilon$, so
$Q_\varepsilon(F)\ge T+1$.
\end{proof}

In order to prove lower bound, we need to find $\nu$ that is non-negative on high degree squares. And the negative mass fraction \(\frac{V_-}{V_+ + V_-}\) is bounded below by a positive constant. 

We construct \(\nu(x) \) according to the hamming weight of \(x\) and the main tool to prove non-negativity for such $\nu$ is the following lemma.
\begin{lemma}[Boolean-lattice reduction]
\label{lem:lb-reduction}
Let $m\ge1,g\ge0$ be integers and let
$\nu(x)=u(|x|-m)$ on $\{0,1\}^{2m}$.
Then $\nu$ is non-negative on degree-$(m+g)$ squares
if and only if
\[
 \sum_{k=-\ell}^{\ell}
 \binom{2\ell}{\ell+k}u(k)
 q\!\left(k\right)^2\ge0
\]
for every integer $0\le\ell\le m$ and every real
univariate polynomial $q$ with
$\deg q\le\min\{\ell+g,2\ell\}$.
\end{lemma}
The proof of \cref{lem:lb-reduction} uses the polynomial decomposition on boolean cube. The details are deferred to \Cref{proof:lb}.

\subsection{Warmup: an \texorpdfstring{$n/2+1$}{n/2+1} lower bound}
\label{sec:lb-warmup}

Use the one-pole signed weighting $\nu(x)=u(|x|-m)$, where
\[
u(k) = \frac{(-1)^k}{k + \frac{1}{2}}.
\]
Its negative weights define the symmetric Boolean function
\begin{equation}\label{eq:lb-one-pole-function}
 F_m^{(0)}(x)=
 \begin{cases}
  1,&u(|x|-m)<0,\\
  0,&u(|x|-m)>0.
 \end{cases}
\end{equation}
To apply \cref{lem:lb-obstruction} with $T=m$, we prove positivity
on degree-$m$ squares and show that the negative mass fraction
$V_-/A$ tends to $1/2$.

\paragraph{Positivity on degree-$m$ squares.}
To establish non-negativity on degree-$m$ squares, we apply
\cref{lem:lb-reduction} with $g=0$. It suffices to show that,
for every integer $0\le\ell\le m$ and every real polynomial $q$
with $\deg q\le\ell$, the following sum is nonnegative.
\begin{align*}
   \sum_{k=-\ell}^{\ell}
 \binom{2\ell}{\ell+k}u(k)
 q\!\left(k\right)^2 = \sum_{k=-\ell}^{\ell} \frac{(2\ell)!}{ (\ell+k)! (\ell-k)!} \frac{(-1)^k}{ k + \frac{1}{2}} q^2(k)  
\end{align*}

Define $\Phi_\ell (x) = \prod\limits_{i=-\ell}^{\ell} (x - i)$

Then \begin{equation} \label{eq:phi}
    \Phi_\ell^{\prime}(k) = \prod_{-\ell \le i \le \ell, i\neq k} (k-i) = (\ell+k)! (\ell -k)! (-1)^{\ell - k} , \quad for\; k \in \{-\ell, \dots, \ell\} 
\end{equation}

Substituting $(\ell+k)!(\ell - k)!$ by \cref{eq:phi}, we can obtain
\begin{align*}
          \binom{2\ell}{\ell+k}u(k)
 q\!\left(k\right)^2 &= \frac{(2\ell)!}{ (\ell+k)! (\ell-k)!} \frac{(-1)^k}{ k + \frac{1}{2}} q^2(k) \\ &= (2\ell)! \frac{q(k)^2 (-1)^{k}}{
 \Phi_l^{\prime}(k) (-1)^{l-k}(k+\frac{1}{2}) 
 }  \\&= (2\ell)!\frac{q(k)^2 (-1)^\ell}{\Phi_{\ell}^{\prime}(k) (k+\frac{1}{2})}   
\end{align*}

We evaluate the preceding sum using residues of a rational function.
If a rational function $f$ has at most a simple pole at $a$, its
\emph{residue} at $a$ is the coefficient of $(z-a)^{-1}$ in its
expansion near $a$, namely
\[
 \operatorname{Res}_{z=a}f(z)=\lim_{z\to a}(z-a)f(z).
\]
In particular, if $f(z)=N(z)/D(z)$ and $D$ has a simple zero at
$a$, then
\[
 \operatorname{Res}_{z=a}f(z)=\frac{N(a)}{D'(a)}.
\]
If $N(a) = 0$, the singularity is removable and the residue is zero.

For $\deg q\le\ell$, consider
\[
 G_\ell(z)=
 \frac{(2\ell)!(-1)^\ell q(z)^2}
 {(z+\tfrac12)\Phi_\ell(z)}.
\]
Its possible finite poles are $-\tfrac12$ and the integers
$-\ell,\ldots,\ell$, and all are at most simple. Their residues are
\[
 \operatorname{Res}_{z=k}G_\ell(z)
 =\frac{(2\ell)!(-1)^\ell q(k)^2}
 {(k+\tfrac12)\Phi_\ell'(k)}
 =\binom{2\ell}{\ell+k}u(k)q(k)^2
 \quad, k \in \{-\ell, \dots, \ell\} ,
\]

and
\[
 \operatorname{Res}_{z=-\tfrac12}G_\ell(z)
 =\frac{(2\ell)!(-1)^\ell q(-\tfrac12)^2}
 {\Phi_\ell(-\tfrac12)}.
\]

Therefore, the weighted sum we want to prove nonnegative is exactly
the sum of the residues of $G_\ell$ at the integer points.

The numerator of $G_\ell$ has degree at most $2\ell$,
whereas its denominator has degree $2\ell+2$. Hence
$G_\ell(z)=O(z^{-2})$ at infinity.
In its partial-fraction expansion, the coefficient of each
$(z-a)^{-1}$ is the residue at $a$. Since
$(z-a)^{-1}=z^{-1}+O(z^{-2})$ at infinity, the coefficient of
$z^{-1}$ is the sum of all finite residues. The decay
$O(z^{-2})$ forces this coefficient to vanish, so
\[
 \sum_{k=-\ell}^{\ell}
 \operatorname{Res}_{z=k}G_\ell(z)
 +\operatorname{Res}_{z=-\tfrac12}G_\ell(z)=0.
\]
There are exactly $\ell+1$ negative factors in
$\Phi_\ell(-\tfrac12)$, so its sign is $(-1)^{\ell+1}$.
The remaining residue is therefore nonpositive. Substituting its
value gives
\begin{equation}\label{eq:lb-one-pole-positive}
 \begin{aligned}
 \sum_{k=-\ell}^{\ell}
 \binom{2\ell}{\ell+k}u(k)q(k)^2
 &=-\operatorname{Res}_{z=-\tfrac12}G_\ell(z)\\
 &=-\frac{(2\ell)!(-1)^\ell q(-\tfrac12)^2}
 {\Phi_\ell(-\tfrac12)}\\
 &=\frac{(2\ell)!}{|\Phi_\ell(-\tfrac12)|}
 q(-\tfrac12)^2\ge0.
 \end{aligned}
\end{equation}
This identity also holds when $\ell=0$.
Lemma~\ref{lem:lb-reduction}, with $g=0$, now shows that
$\nu(x)=u(|x|-m)$ is non-negative on degree-$m$ squares.

\paragraph{The negative mass fraction.}
Since $V_-=(A-S)/2$, it suffices to show that $S/A=O(1/\log m)$.
Assume $m\ge4$, and put $C_m=\binom{2m}{m}$ and
$J=\lfloor\sqrt m/2\rfloor$.

The signed mass is
\[
 S=\sum_{k=-m}^{m}\binom{2m}{m+k}
   \frac{(-1)^k}{k+\tfrac12}.
\]

First, split the sum defining $S$ into the indices
$k=0,1,\ldots,m$ and $k=-1,-2,\ldots,-m$, in these orders.
Both are alternating sums with decreasing magnitudes and a positive
first term at most $2C_m$. Hence
\begin{equation}\label{eq:lb-one-pole-masses}
 0\le S\le4C_m.
\end{equation}

For the absolute mass, retain only the terms with $0\le k\le J$.
Their binomial coefficients satisfy
\begin{equation}\label{eq:lb-one-pole-binomial}
 \frac{\binom{2m}{m+k}}{C_m}
 =\prod_{i=1}^{k}\left(1-\frac{2i-1}{m+i}\right)
 \ge1-\frac{k^2}{m}\ge\frac34,
\end{equation}
where we used $\prod_i(1-a_i)\ge1-\sum_i a_i$ for
$0\le a_i\le1$, together with
$\sum_{i=1}^{k}(2i-1)/(m+i)\le k^2/m$.

Consequently, comparison with the integral of $1/(t+\tfrac12)$ gives
\[
 A\ge\frac{3C_m}{4}\sum_{k=0}^{J}\frac1{k+\tfrac12}
 \ge\frac{3C_m}{4}\log(2J+3).
\]
Combining the two bounds yields
\begin{equation}\label{eq:lb-one-pole-ratio}
 \frac{V_-}{A}
 =\frac12-\frac{S}{2A}
 \ge\frac12-\frac{8}{3\log(2J+3)}
 =\frac12-O\!\left(\frac1{\log m}\right).
\end{equation}
Applying \cref{lem:lb-obstruction} with $T=m$ proves
\cref{thm:lb-one-extra}, which we restate here.

\lbOneExtraTheorem*

\subsection{The full lower bound}
Next, we are going to show the full lower bound, by extending the construction above.

The residue argument proves non-negativity on degree-$m$
squares, but it does not extend directly to degree-$(m+1)$ squares.
Indeed, Lemma~\ref{lem:lb-reduction} would then require
\[
 \sum_{k=-\ell}^{\ell}\binom{2\ell}{\ell+k}u(k)q(k)^2\ge0
 \qquad\text{for }\deg q\le\min\{\ell+1,2\ell\}.
\]
For $\ell\ge1$, the degree of $q$ can now be $\ell+1$.
If $\deg q=\ell+1$, the numerator and denominator of
\[
 G_\ell(z)=
 \frac{(2\ell)!(-1)^\ell q(z)^2}
 {(z+\tfrac12)\Phi_\ell(z)}
\]
both have degree $2\ell+2$. Thus $G_\ell$ no longer
decays as $O(z^{-2})$: its expansion at infinity can contain a
nonzero $z^{-1}$ term. The sum of its finite residues is therefore
no longer guaranteed to be zero, so the preceding positivity
identity does not follow.

This suggests increasing the degree of the denominator. More
generally, to prove non-negativity on degree-$(m+g)$ squares,
the reduction requires polynomials $q$ of degree up to $\ell+g$.
Replace the single factor $z+\tfrac12$ by a product of $2g+1$
linear factors with distinct noninteger roots. The numerator then has degree at most $2\ell+2g$, while
the denominator has degree$(2g+1)+(2\ell+1)=2\ell+2g+2$
The resulting rational function again decays as $O(z^{-2})$,
so the sum of its finite residues is zero. Thus, in this construction, each additional degree allowed in $q$ requires two additional poles.

With the normalization used in $G_\ell$, the target sum is the
sum of the residues at the integer points. To prove that this sum
is nonnegative, we want every remaining residue to be nonpositive.
The construction below chooses the additional pole locations and
the overall sign to achieve this whenever the window contains all
the poles. We then replace central negative weights by zero to
handle shorter windows.

Fix an integer $g\ge1$. Throughout, $n=2m$ still denotes the input
length, and define
\[
 d=2g-1,\qquad R=g+d^2+1,\qquad
 r_j=R-1-j^2\quad(0\le j\le d).
\]
Thus $r_d=g$. Define the set of $2g+1$ distinct integers
\begin{equation}\label{eq:lb-poles}
 \mathcal P=\{R\}\cup\{r_j:0\le j\le d\}.
\end{equation}
and set
\[
 \Delta(z)=\prod_{\rho\in\mathcal P}(z-\rho),\qquad
 h_g(k)=\frac{(-1)^{g+k}}{\Delta(k+\tfrac12)}
 \quad(k\in\mathbb Z).
\]
Here $\mathcal P$ is the set of roots of $\Delta$; the zeros of
$\Delta(z+\tfrac12)$, which will be the additional poles in the
residue argument, are the half-integers $\rho-\tfrac12$.
Thus $\Delta(k+\tfrac12)\ne0$ at every integer $k$.
The factor $(-1)^g$ in $h_g$ will give the required signs of
the residues.

Remove negative weights in the central interval by setting
\begin{equation}\label{eq:lb-clipped-weight}
 u_g(k)=
 \begin{cases}
  \max\{0,h_g(k)\},&-R\le k<R,\\
  h_g(k),&\text{otherwise}.
 \end{cases}
\end{equation}
Our new cube weighting is
\[
 \nu_g(x)=u_g(|x|-m).
\]
For this weighting, $V_+$, $V_-$, $A$, and $S$ retain exactly
their definitions from the beginning of the section.

\begin{lemma}\label{lem:lb-multipole-positive}
For every integer $\ell\ge0$ and every real polynomial $q$ with
$\deg q\le\min\{\ell+g,2\ell\}$,
\[
 \sum_{k=-\ell}^{\ell}
 \binom{2\ell}{\ell+k}u_g(k)q(k)^2\ge0.
\]
Consequently, $\nu_g$ is non-negative on degree-$(m+g)$ squares.
\end{lemma}

\begin{proof}
If $\ell<R$, every $k\in\{-\ell,\ldots,\ell\}$ lies in the
clipped interval, so all summands are nonnegative.
Suppose now that $\ell\ge R$, so the interval $[-\ell,\ell]$
contains every additional pole $\rho-\tfrac12$.
The roots of $\Delta$ are ordered as
\[
 R>r_0>r_1>\cdots>r_d.
\]
Thus $\Delta'(R)>0$ and
$\operatorname{sgn}\Delta'(r_j)=(-1)^{j+1}$.
Because $d$ is odd,
\[
 R\equiv g\pmod2,\qquad
 r_j\equiv g+1-j\pmod2.
\]
Consequently,
\begin{equation}\label{eq:lb-pole-signs}
 \operatorname{sgn}\Delta'(\rho)=(-1)^{g+\rho}
 \qquad(\rho\in\mathcal P).
\end{equation}

Following the definition of $G_\ell$, put
\[
 G_{\ell,g}(z)=
 \frac{(2\ell)!(-1)^{\ell+g}q(z)^2}
 {\Delta(z+\tfrac12)\Phi_\ell(z)}.
\]
The factor $z+\tfrac12$ has been replaced by
$\Delta(z+\tfrac12)$, and the numerator has gained the sign
$(-1)^g$. At each integer $-\ell\le k\le\ell$,
\begin{align*}
 \operatorname{Res}_{z=k}G_{\ell,g}(z)
 &=\frac{(2\ell)!(-1)^{\ell+g}q(k)^2}
 {\Delta(k+\tfrac12)\Phi_\ell'(k)}\\
 &=\binom{2\ell}{\ell+k}h_g(k)q(k)^2,
\end{align*}
where we used
$\Phi_\ell'(k)=(-1)^{\ell-k}(\ell+k)!(\ell-k)!$.
Thus the weighted sum of squares with the unclipped weights $h_g$ is again
a sum of residues at the integer points.

At each additional pole $\rho-\tfrac12$, exactly
$\ell-\rho+1$ factors in $\Phi_\ell(\rho-\tfrac12)$ are
negative. Hence
\[
 \operatorname{sgn}\Phi_\ell(\rho-\tfrac12)
 =(-1)^{\ell+1+\rho}.
\]
Combining this with \eqref{eq:lb-pole-signs} gives
\begin{align*}
 \operatorname{Res}_{z=\rho-\tfrac12}G_{\ell,g}(z)
 &=\frac{(2\ell)!(-1)^{\ell+g}q(\rho-\tfrac12)^2}
 {\Delta'(\rho)\Phi_\ell(\rho-\tfrac12)}\\
 &=-\frac{(2\ell)!q(\rho-\tfrac12)^2}
 {|\Delta'(\rho)|\,|\Phi_\ell(\rho-\tfrac12)|}\le0.
\end{align*}
Finally, $\deg q\le\ell+g$ implies that the numerator of
$G_{\ell,g}$ has degree at most $2\ell+2g$, while its denominator
has degree $2\ell+2g+2$. Therefore $G_{\ell,g}(z)=O(z^{-2})$,
and the sum of all its finite residues is zero:
\begin{equation}\label{eq:lb-residue-identity}
 \sum_{k=-\ell}^{\ell}\operatorname{Res}_{z=k}G_{\ell,g}(z)
 =-\sum_{\rho\in\mathcal P}
 \operatorname{Res}_{z=\rho-\tfrac12}G_{\ell,g}(z).
\end{equation}
Substituting the residue values proves
\begin{equation}\label{eq:lb-multipole-identity}
 \sum_{k=-\ell}^{\ell}
 \binom{2\ell}{\ell+k}h_g(k)q(k)^2
 =(2\ell)!\sum_{\rho\in\mathcal P}
 \frac{q(\rho-\tfrac12)^2}
 {|\Delta'(\rho)|\,|\Phi_\ell(\rho-\tfrac12)|}
 \ge0.
\end{equation}
Since $u_g(k)\ge h_g(k)$ at every integer $k$, clipping can
only increase this weighted sum of squares. The non-negativity of $\nu_g$ follows
from Lemma~\ref{lem:lb-reduction}.
\end{proof}

\begin{remark}[Why clipping is needed]
\label{rem:lb-clipping}
For $0\le t\le g$, the exact $(m+t)$-query algorithm for
$\operatorname{EXACT}^{2m}_{m-t}$
\cite{AmbainisIraidsSmotrovs2013} yields
\[
 p_t(x)=\mathbf 1_{\{|x|=m-t\}}
 =\sum_\alpha q_\alpha(x)^2,
 \qquad \deg q_\alpha\le m+t \le m+g.
\]
Thus $L_{\nu_g}(p_t) = \sum_{|x| = m-t} \nu(x) = \binom{2 m }{m -t} u_g(- t) \ge 0$, so $u_g(-t) \ge 0$. Similarly, we can argue $u_g(t) \ge 0$
Thus the unclipped weighting fails the required positivity.
\end{remark}

The residue argument establishes the required nonnegativity.
To apply \cref{lem:lb-obstruction} at a fixed positive error, we must also show that
the negative mass is a constant fraction of the absolute mass.
The proof of the next lemma shows that the total magnitude of the negative
weights on one Hamming layer is at least a constant fraction of the total
absolute weight.

\begin{lemma}\label{lem:lb-multipole-mass}
There is an absolute constant $c>0$ such that, whenever $g\ge1$
and $m\ge512g^3$, the weighting $\nu_g$ satisfies
\[
 \frac{V_-}{A}\ge c.
\]
\end{lemma}

The proof is deferred to \Cref{app:lb-multipole-mass}.

Define $F_{m,g}:\{0,1\}^{2m}\to\{0,1\}$ by
\begin{equation}\label{eq:lb-final-function}
 F_{m,g}(x)=
 \begin{cases}
  1,&u_g(|x|-m)<0,\\
  0,&u_g(|x|-m)\ge0.
 \end{cases}
\end{equation}
This is an explicit symmetric total Boolean function.
For clarity, its values can also be written without rational weights.
With $k=|x|-m$ and $j=\max\{k,-k-1\}$, we have
\[
 F_{m,g}(x)=1
 \quad\Longleftrightarrow\quad
 j\ge R+1\ \text{and}\ j-R\ \text{is odd}.
\]
Indeed, inside $-R\le k<R$ all retained weights are nonnegative.
For $k\ge R$, the denominator of $h_g(k)$ is positive, whereas
for $k<-R$ all its $2g+1$ factors are negative.
The displayed characterization follows from $R\equiv g\pmod2$.

We restate \cref{thm:lb-cuberoot} before proving it.

\lbCuberootTheorem*

\begin{proof}
By Lemma~\ref{lem:lb-multipole-positive}, $\nu_g$ is
non-negative on degree-$(m+g)$ squares.
By Lemma~\ref{lem:lb-multipole-mass}, $V_-/A\ge c$ for an
absolute $c>0$. Choose
$\varepsilon_0=\min\{c/2,1/4\}$, so that
$\varepsilon_0<c\le V_-/A$.
Using \cref{lem:lb-obstruction}  we can exclude
every algorithm using
at most $m+g$ queries, proving the first assertion.
The chosen $g_m$ satisfies $m\ge512g_m^3$ and
$g_m=\Theta(m^{1/3})$, proving the second.
\end{proof}

%% file: aimethodology.tex
\section*{AI methodology}

We follow the standardized AI Methodology template proposed by the
AI+TCS working group at the Simons Institute in September 2026.\footnote{See
\emph{AI and TCS: The Next Six Months}, p.~9:
\url{https://simons.berkeley.edu/AITCSreport}.}

\begin{flushleft}
\renewcommand{\arraystretch}{1.3}
\begin{tabular}{@{}l@{\qquad}c@{}}
\textbf{Activity} & \textbf{AI used} \\
Asking the research question & No \\
Coming up with the approach & Yes \\
Proof development & Yes \\
Writing and exposition & Yes \\
Checking for bugs & Yes \\
Other supporting tasks & No \\
\end{tabular}
\end{flushleft}

\begin{sloppypar}
\paragraph{Details.}
This work began with the question of whether van Dam's universal upper bound
on quantum query complexity is optimal. The authors conjectured that Hamming
weight recovery, whose output has $\Theta(\log n)$ bits rather than a single bit,
would require the same additive overhead. Some of the authors posed this
question to OpenAI's GPT 6 Astra Ultra, which found the improved upper bound
presented in this paper but it did not get a matching lower bound. Google's Stellar
Colosseum~\cite{stellar}, consulted next, produced a matching lower bound along with an end-to-end formal verification in Lean.\footnote{\url{https://drive.google.com/file/d/1YByxN8GeiwrGXNzsDj_wpVjKhYLUqT8q/view?usp=sharing}}. The authors then asked for a simpler hard function,
and it supplied a cleaner construction. The authors worked through the argument
in detail until they fully understood it, and then rewrote the analysis
themselves.
\end{sloppypar}

%% file: appendix.tex
\appendix
\crefalias{section}{appendix}
\crefalias{subsection}{appendix}
\section{Proofs for the Lower Bound}

\subsection{Boolean Lattice Reduction}
\label{proof:lb}
We follow the result of \cite{Srinivasan2011} and retain its notation to derive \cref{lem:lb-reduction}.

Let $B(n)$ be the lattice of subsets of $[n]$, and let
$B(n)_i=\{X\subseteq[n]:|X|=i\}$.

Let $V(B(n))$ be
the real vector space with basis $B(n)$.
and let
$V(B(n)_i)$ be the real vector space with basis $B(n)_i$.

For $v=\sum_{X\in B(n)}v_X X$ and
$w=\sum_{X\in B(n)}w_X X$, define the inner product by
\[
 \langle v,w\rangle=\sum_{X\in B(n)}v_Xw_X.
\]
The length of $v$ is$\|v\|=\sqrt{\langle v,v\rangle}
      =\left(\sum_{X\in B(n)}v_X^2\right)^{1/2}.$
      
Define up operator to be
\[
 U(X)=\sum_{j\notin X, j\in [n]}X\cup\{j\}.
\]

\begin{theorem}{\cite[Theorems~1.2]{Srinivasan2011}}
\label{thm:srinivasan}
For every integer $n\ge1$, $V(B(n))$ has an orthogonal
basis $\operatorname{Orth}(n)$ that is a union of symmetric Jordan chains
\[
 (x_k,x_{k+1},\ldots,x_{n-k}),
 \qquad 0\le k\le\lfloor n/2\rfloor.
\]
Here $x_i\in V(B(n)_i)$ is nonzero,
$Ux_i=x_{i+1}$ for $k\le i<n-k$, and $Ux_{n-k}=0$.
Along each chain,
\begin{equation}
 \frac{\|x_{i+1}\|}{\|x_i\|}
 =\sqrt{(i+1-k)(n-k-i)},
 \qquad k\le i<n-k.
 \label{eq:srinivasan-ratio}
\end{equation}
\end{theorem}
We can view a subset $X$ as the monomial
$z_X=\prod_{j\in X}z_j$.Then a vector in $V(B(n)_i)$  can be viewed as a homogeneous multilinear
polynomial of degree $i$.
Thus for $x_i$ denotes a chain vector, it also represents a homogeneous multilinear polynomial  and $x_i(z)$ is its value at a cube point $z \in \{0,1\}^n$.
We will derive orthogonality of these values on each
Hamming layer from the theorem.

\begin{proof}[Proof of \cref{lem:lb-reduction}]
Let $n=2m$ and $T=m+g$. 

Let $\mathcal{A}$ denotes the collection of chains in $\operatorname{Orth}(n)$. And we index the chains in $\operatorname{Orth}(n)$ by $\alpha\in\mathcal A$
 so that
\[
 \operatorname{Orth}(n)
 =
 \bigcup_{\alpha\in\mathcal A}
 \{x^\alpha_{k_\alpha},\ldots,x^\alpha_{n-k_\alpha}\}.
\]

Rescale each entire chain so that
$\|x^\alpha_{k_\alpha}\|=1$.
This preserves both the chain relations and orthogonality.
Multiplying the squared ratios in
\eqref{eq:srinivasan-ratio} gives, for
$k_\alpha\le w\le n-k_\alpha$,
\begin{equation}
 \|x^\alpha_w\|^2
 =\prod_{i = k_\alpha}^{w-1} (i+1-k_\alpha) (n-k_\alpha-i)  =((w-k_\alpha)!)^2
   \binom{n-2k_\alpha}{w-k_\alpha}.
 \label{eq:chain-norm}
\end{equation}

Fix $X\subseteq[n]$ with $|X|=w\ge k_\alpha$.
Every monomial of degree $k_\alpha$ supported inside $X$
reaches $z_X$ under $U^{w-k_\alpha}$ in exactly
$(w-k_\alpha)!$ orders. Therefore
\begin{equation}
 [z_X]\,U^{w-k_\alpha}x^\alpha_{k_\alpha}
 =(w-k_\alpha)!\,x^\alpha_{k_\alpha}(\mathbf1_X),
 \label{eq:layer-incidence-identity}
\end{equation}
where $[z_X]$ denotes a coefficient and $\mathbf1_X \in \{0,1\}^n$
is the Boolean vector with support $X$. Evaluating $x^\alpha_{k_\alpha}$ at $\mathbf1_X$ sums
the coefficients of its monomials $z_S$ with $S\subseteq X$.
Within the chain's range,
$U^{w-k_\alpha}x^\alpha_{k_\alpha}=x^\alpha_w$.
Coefficient orthogonality, \eqref{eq:layer-incidence-identity}  and  \eqref{eq:chain-norm} give
\begin{equation}
 \begin{aligned}
 \sum_{|z|=w}
 x^\alpha_{k_\alpha}(z)x^\beta_{k_\beta}(z) &= \sum_{X\subseteq[n], |X|=w} x^\alpha_{k_\alpha}(\mathbf1_X)x^\beta_{k_\beta}(\mathbf1_X)
 \\&=
 \frac{
\sum\limits_{X\subseteq[n],|X|=w}
   ([z_X]x^\alpha_w)([z_X]x^\beta_w)
 }{(w-k_\alpha)!(w-k_\beta)!}\\
 &=
 \frac{\langle x^\alpha_w,x^\beta_w\rangle}
      {(w-k_\alpha)!(w-k_\beta)!}\\
 &=
 \mathbf1_{\{\alpha = \beta\}}
 \binom{n-2k_\alpha}{w-k_\alpha}.
 \end{aligned}
 \label{eq:layer-orthogonality}
\end{equation}
We interpret out-of-range binomial coefficients
as zero.

For a monomial $z_S$ with $|S|=i$, the definition of $U$ gives
\[
 (Uz_S)(z)
 =\sum_{j\notin S} (z_j \prod_{i \in S} z_i) = (\sum_{j \notin S} z_j) \prod_{i \in S} z_i
 =(|z|-i)z_S(z),\qquad z\in\{0,1\}^n.
\]
By linearity, \[(Uv)(z)=(|z|-i)v(z)\]for every homogeneous
multilinear polynomial $v$ of degree $i$.

Consequently, along a chain,we have
\begin{equation}
\label{eq:lll}
 x^\alpha_i(z)
 =x^\alpha_{k_\alpha}(z)
   \prod_{j=k_\alpha}^{i-1}(|z|-j).
\end{equation}

Let $P$ be a multilinear polynomial with degree at most $T$ and we are going to represent $P$ in the symmetric Jordan chain Basis.
Since the basis $\operatorname{Orth}(n)$ is homogeneous, its expansion
uses only chain vectors of rank $i\le T$.
Collecting the terms from each chain and given \cref{eq:lll} gives
\begin{equation}
 P(z)=\sum_{\alpha\in\mathcal A}
 x^\alpha_{k_\alpha}(z)\,q_\alpha(|z|-m),
 \qquad
 \deg q_\alpha\le
 \min\{T-k_\alpha,n-2k_\alpha\}.
 \label{eq:polynomial-decomposition}
\end{equation}
The two bounds come from $i\le T$ and the chain endpoint
$i\le n-k_\alpha$.
In particular, no restriction $T\le n/2$ is needed.

First, suppose the univariate inequalities in the lemma hold.
On each layer, the factors $q_\alpha(|z|-m)$ are constant,
so \eqref{eq:layer-orthogonality} eliminates all cross
terms in the square of
\eqref{eq:polynomial-decomposition}.
Set $\ell_\alpha=m-k_\alpha$ and $r=|z|-m$.
Summing over the layers gives
\[
 \begin{aligned}
 L_\nu(P^2)
 &=\sum_{w=0}^n u(w-m)\sum_{|z|=w}P(z)^2\\
 &=\sum_{\alpha\in\mathcal A}
   \sum_{w=k_\alpha}^{n-k_\alpha}
   \binom{n-2k_\alpha}{w-k_\alpha}
   u(w-m)q_\alpha(w-m)^2\\
 &=\sum_{\alpha\in\mathcal A}
   \sum_{r=-\ell_\alpha}^{\ell_\alpha}
   \binom{2\ell_\alpha}{\ell_\alpha+r}
   u(r)q_\alpha(r)^2,
 \end{aligned}
\]
Moreover,
\[
 \deg q_\alpha
 \le\min\{T-k_\alpha,n-2k_\alpha\}
 =\min\{\ell_\alpha+g,2\ell_\alpha\}.
\]
Every inner sum is nonnegative by assumption.
Hence $L_\nu(P^2)\ge0$.

Conversely, suppose $\nu$ is non-negative on
degree-$(m+g)$ squares.
Fix an arbitrary integer $0\le\ell\le m$ and an arbitrary
real univariate polynomial $q$ satisfying
$\deg q\le\min\{\ell+g,2\ell\}$.
Put $k=m-\ell$ and take
\[
 F(z)=
 \left(\prod_{i=1}^k(z_{2i-1}-z_{2i})\right)q(|z|-m).
\]
Then
\[
 \deg F\le k+\deg q
 \le(m-\ell)+(\ell+g)=m+g,
\]
so the assumed non-negativity applies to $F^2$.

The squared product $\left(\prod_{i=1}^k(z_{2i-1}-z_{2i})\right)^2$ is $1$ when each pair
contains one $0$ and one $1$, and is $0$ otherwise.
Let $y\in\{0,1\}^{2\ell}$ denote the remaining, unpaired
coordinates. For each fixed $y$, the pairs have $2^k$
orientations with nonzero product, and each contributes
$k$ to $|z|$. Thus $|z|-m=|y| + k- m = |y| - l$ on these assignments,
and
\[
 \begin{aligned}
 0\le L_\nu(F^2)
 &=2^k\sum_{y\in\{0,1\}^{2\ell}}
   u(|y|-\ell)q(|y|-\ell)^2\\
 &=2^k\sum_{r=-\ell}^{\ell}
   \binom{2\ell}{\ell+r}u(r)q(r)^2.
 \end{aligned}
\]
The last equality groups the vectors $y$ by
$|y|=\ell+r$, whose possible values give
$-\ell\le r\le\ell$.
Dividing by $2^k$ proves the required inequality.
\end{proof}

\subsection{A constant negative mass fraction}
\label{app:lb-multipole-mass}
We use the construction in \eqref{eq:lb-poles}--\eqref{eq:lb-clipped-weight}
with $g\ge1$ and $m\ge512g^3$.

\begin{proof}[Proof of \cref{lem:lb-multipole-mass}]
We estimate the negative mass and the absolute mass separately.

\paragraph{Normalization and the proof goal.}
The Hamming layer indexed by $k=|x|-m$ contains
$\binom{2m}{m+k}$ inputs. In particular,
\[
 A=\sum_{k=-m}^{m}\binom{2m}{m+k}|u_g(k)|.
\]
We normalize the rational function so that its residue at the largest
pole is $1$. Set
\[
 \kappa=\Delta'(R)>0,\qquad K(z)=\frac{\kappa}{\Delta(z)}.
\]
For each $\rho\in\mathcal P$, write
\[
 c_\rho=\operatorname{Res}_{z=\rho}K(z)
 =\frac{\kappa}{\Delta'(\rho)}.
\]
In particular, $c_R=1$, and
\[
 \kappa|h_g(k)|=|K(k+\tfrac12)|.
\]
The assumption on $m$ ensures $R+1\le m$. Define
\[
 \mathcal B_R=\binom{2m}{m+R},\qquad
 b(k)=\frac{\binom{2m}{m+k}}{\mathcal B_R}
 \quad(-m\le k\le m),
\]
and the normalized, unclipped absolute mass
\[
 \mathcal M=\sum_{k=-m}^{m}b(k)|K(k+\tfrac12)|.
\]
Since clipping gives $|u_g(k)|\le|h_g(k)|$, we have
\[
 \frac{\kappa A}{\mathcal B_R}\le\mathcal M.
\]
It therefore suffices to prove the two bounds
\[
 \frac{\kappa V_-}{\mathcal B_R}\ge\frac{2e^{-3}}{15},
 \qquad
 \mathcal M\le C_*,
\]
where $C_*>0$ is an absolute constant. The first bound uses only
the layer $k=R+1$; the second requires summing over all layers.

\paragraph{A surviving negative layer.}
The layer $k=R+1$ lies outside the clipped interval. Moreover,
$R\equiv g\pmod2$ and $\Delta(R+\tfrac32)>0$, so
\[
 u_g(R+1)=h_g(R+1)<0.
\]
We now bound the contribution of this layer from below.
Factoring $\Delta(R+t)$ gives, for $t>0$,
\begin{equation}\label{eq:lb-right-product}
 K(R+t)=\frac1{t(t+1)}
 \prod_{j=1}^{d}\left(1+\frac{t}{1+j^2}\right)^{-1}.
\end{equation}
Using $\log(1+s)\le s$ and $\sum_{j\ge1}j^{-2}\le2$, we obtain
\[
 K(R+\tfrac32)
 =\frac4{15}
 \prod_{j=1}^{d}\left(1+\frac{3}{2(1+j^2)}\right)^{-1}
 \ge\frac4{15}e^{-3}.
\]
Also, $R+1\le4g^2$ and $m\ge512g^3$ imply
\[
 b(R+1)=\frac{m-R}{m+R+1}\ge\frac12.
\]
Consequently, the negative mass contributed by this layer satisfies
\begin{equation}\label{eq:lb-surviving-negative}
 \frac{\kappa V_-}{\mathcal B_R}
 \ge b(R+1)K(R+\tfrac32)
 \ge\frac{2e^{-3}}{15}.
\end{equation}
It remains to bound $\mathcal M$. We first estimate the residues,
the values of $K$ between its poles, and the binomial weights.

\paragraph{Residue estimates.}
The quadratic spacing of the poles gives a decay estimate for
$c_{r_j}$. Since $R-r_j=1+j^2$ and $r_0-r_j=j^2$ for $j\ge1$,
the product formula for $\Delta'$ gives $c_R=1$ and
\begin{align*}
 |c_{r_0}|
 &=\frac{\prod_{j=0}^{d}(1+j^2)}{\prod_{j=1}^{d}j^2}\\
 &=\prod_{j=1}^{d}(1+j^{-2})
 \le\exp\!\left(\sum_{j=1}^{d}j^{-2}\right)
 \le e^2.
\end{align*}
For $1\le j\le d$, factoring
$|i^2-j^2|=|i-j|(i+j)$ yields
\[
 \prod_{\substack{0\le i\le d\\i\ne j}}|i^2-j^2|
 =\frac{(d-j)!(d+j)!}{2},
\]
and hence
\[
 \frac{|c_{r_j}|}{|c_{r_0}|}=
 \frac{1}{1+j^2}
 \frac{2 (d!)^2}{(d-j)!(d+j)!}.
\]
The factorial ratio satisfies
\begin{align*}
 \frac{(d!)^2}{(d-j)!(d+j)!}
 &=\prod_{a=0}^{j-1}
 \left(1-\frac{2a+1}{d+a+1}\right)\\
 &\le\exp\!\left(-\frac1{2d}
 \sum_{a=0}^{j-1}(2a+1)\right)
 =e^{-j^2/(2d)},
\end{align*}
where we used $1-t\le e^{-t}$ and $d+a+1\le2d$.
Set $C_0=2e^2$. The preceding estimates give
\begin{equation}\label{eq:lb-residue-decay}
 |c_R|=1,\qquad |c_{r_0}|\le e^2,\qquad
 |c_{r_j}|\le C_0j^{-2}e^{-j^2/(2d)}
 \quad(1\le j\le d).
\end{equation}
In particular, every residue has absolute value at most $C_0$.
This constant is independent of $m$ and $g$.

\paragraph{An estimate inside each gap.}
If $a<b$ are consecutive poles, then
\begin{equation}\label{eq:lb-gap-bound}
 |K(z)|\le\frac{|c_a|}{z-a}+\frac{|c_b|}{b-z}
 \qquad(a<z<b).
\end{equation}
Thus the values of $K$ inside a gap are bounded using only the
residues at its two endpoints.
To prove this, write $\Delta(z)=(z-a)(z-b)Q(z)$ and set
$f(z)=\kappa/|Q(z)|$ on $[a,b]$.
Since the poles are consecutive, $Q$ has no zeros on this interval,
and
\[
 (\log f)''(z)
 =\sum_{\rho\in\mathcal P\setminus\{a,b\}}
 \frac1{(z-\rho)^2}\ge0.
\]
It follows that
$f''= \exp''(\log f) = f((\log f)''+((\log f)')^2)\ge0$, so $f$ is convex.
Its endpoint values are
$f(a)=(b-a)|c_a|$ and $f(b)=(b-a)|c_b|$.
The chord joining these values therefore gives
\[
 f(z)\le(b-z)|c_a|+(z-a)|c_b|.
\]
Dividing by $(z-a)(b-z)$ proves \eqref{eq:lb-gap-bound}.

\paragraph{Binomial weights.}
Since $d+1=2g$, our assumptions imply
\[
 m\ge64(d+1)^3,\qquad R+1\le(d+1)^2,
 \qquad m-R+1\ge m/2.
\]
For $0\le k\le R$, taking ratios of consecutive binomial
coefficients gives
\begin{align}
 \log b(k)
 &=\sum_{a=k+1}^{R}
 \log\!\left(1+\frac{2a-1}{m-a+1}\right)\notag\\
 &\le\frac{R^2-k^2}{m-R+1}
 \le\frac{R-k}{16(d+1)}.
 \label{eq:lb-binomial-growth}
\end{align}
Here the first inequality follows from $\log(1+s)\le s$.
For the second, use $R+k\le2(d+1)^2$ and
$m-R+1\ge32(d+1)^3$.
Setting $k=0$ in the middle bound and using the maximality of the
central binomial coefficient gives
\begin{equation}\label{eq:lb-binomial-center}
 b(k)\le b(0)\le e^{(d+1)/32}
 \qquad(-m\le k\le m).
\end{equation}
On the right tail, we also have $b(k)\le1$ for $k\ge R$.

We now sum the contributions to $\mathcal M$. Each half-integer
sample lies in a gap between consecutive poles or in one of the
two exterior tails.

\paragraph{The two unit gaps.}
Each of $(r_1,r_0)$ and $(r_0,R)$ contains exactly one half-integer
sample. At either sample, $R-k\le2$, so
\eqref{eq:lb-binomial-growth} gives
\[
 b(k)\le e^{1/[8(d+1)]}\le e^{1/16},
\]
where the last inequality uses $d\ge1$.
Each endpoint residue has absolute value at most $C_0$, and the
distance to either endpoint is $1/2$. Thus, by
\eqref{eq:lb-gap-bound}, each gap contributes at most
\[
 e^{1/16}(2C_0+2C_0)=4e^{1/16}C_0.
\]
The two unit gaps therefore contribute at most $8e^{1/16}C_0$
in total.

\paragraph{The remaining gaps.}
Fix $1\le j\le d-1$. The half-integer samples in
$(r_{j+1},r_j)$ are precisely
\[
 k+\tfrac12,\qquad k=r_{j+1},\ldots,r_j-1.
\]
Thus the contribution of this gap to $\mathcal M$ is
\[
 \mathcal M_j
 :=\sum_{k=r_{j+1}}^{r_j-1}b(k)|K(k+\tfrac12)|.
\]

We first bound the binomial weight multiplied by either endpoint
residue. On this range,
\[
 R-k\le R-r_{j+1}=1+(j+1)^2.
\]
Combining \eqref{eq:lb-binomial-growth} with
\eqref{eq:lb-residue-decay}, for $i\in\{j,j+1\}$ we obtain
\begin{align}
 b(k)|c_{r_i}|
 &\le C_0i^{-2}
 \exp\!\left(
   \frac{1+(j+1)^2}{16(d+1)}-\frac{i^2}{2d}
 \right)
 \le C_0i^{-2}.
 \label{eq:lb-binomial-absorption}
\end{align}
Indeed, $(j+1)^2\le4j^2$ and $i\ge j$ imply that the exponent
is at most
\[
 -\frac{j^2}{4d}+\frac1{16(d+1)}<0.
\]
Thus the residue decay compensates for the growth of the
binomial weights.

Applying \eqref{eq:lb-gap-bound} at each sample and summing gives
\begin{align*}
 \mathcal M_j
 &\le
 \sum_{k=r_{j+1}}^{r_j-1}
 b(k)\left(
 \frac{|c_{r_{j+1}}|}{k+\tfrac12-r_{j+1}}
 +\frac{|c_{r_j}|}{r_j-k-\tfrac12}
 \right)\\
 &\le
 \frac{C_0}{(j+1)^2}
 \sum_{k=r_{j+1}}^{r_j-1}
 \frac1{k-r_{j+1}+\tfrac12}
 +
 \frac{C_0}{j^2}
 \sum_{k=r_{j+1}}^{r_j-1}
 \frac1{r_j-k-\tfrac12}.
\end{align*}
The gap has length $r_j-r_{j+1}=2j+1$, so each of the two
reciprocal-distance sums equals
\[
 H_{2j+1},
 \qquad
 H_L:=\sum_{s=0}^{L-1}\frac1{s+\tfrac12}.
\]
By Cauchy--Schwarz and integral comparison,
\[
 H_L^2
 \le L\sum_{s=0}^{\infty}(s+\tfrac12)^{-2}
 \le6L.
\]
Consequently,
\begin{align*}
 \mathcal M_j
 &\le C_0\bigl(j^{-2}+(j+1)^{-2}\bigr)H_{2j+1}\\
 &\le2C_0j^{-2}\sqrt{6(2j+1)}\\
 &\le2\sqrt{18}\,C_0j^{-3/2},
\end{align*}
where the last inequality uses $2j+1\le3j$.

Finally, summing over all remaining gaps yields
\begin{align*}
 \sum_{j=1}^{d-1}
 \sum_{k=r_{j+1}}^{r_j-1}b(k)|K(k+\tfrac12)|
 &=\sum_{j=1}^{d-1}\mathcal M_j\\
 &\le2\sqrt{18}\,C_0\sum_{j=1}^{\infty}j^{-3/2}\\
 &\le6\sqrt{18}\,C_0.
\end{align*}
Here we used
\[
 \sum_{j=1}^{\infty}j^{-3/2}
 \le1+\int_1^\infty t^{-3/2}\,dt=3.
\]
This bound is independent of $d$. When $d=1$, there are no remaining gaps.

\paragraph{The right tail.}
For $k\ge R$, we have $b(k)\le1$.
Dropping the product factors in \eqref{eq:lb-right-product} and
setting $s=k-R$ gives
\[
 \sum_{k=R}^{m}b(k)|K(k+\tfrac12)|
 \le\sum_{s=0}^{\infty}
 \frac1{(s+\tfrac12)(s+\tfrac32)}=2.
\]

\paragraph{The left tail.}
The smallest pole is $r_d=g$. Its nearest neighbor is
$r_{d-1}$, at distance $\delta=2d-1$.
For $t>0$, factoring at $g$ gives
\[
 |K(g-t)|
 =\frac{|c_g|}{t}
 \prod_{\substack{\rho\in\mathcal P\\\rho>g}}
 \frac{\rho-g}{\rho-g+t}
 \le |c_g|\frac{\delta}{t(t+\delta)}.
\]
The inequality retains the nearest-neighbor factor and bounds every
other factor by $1$.
Since $\delta$ is an integer, the identity
\[
 \frac{\delta}{t(t+\delta)}=\frac1t-\frac1{t+\delta}
\]
gives
\[
 \sum_{s=0}^{\infty}
 \frac{\delta}{(s+\tfrac12)(s+\tfrac12+\delta)}=H_\delta.
\]
For the left-tail samples, $k\le g-1$, and
$t=g-(k+\tfrac12)$ ranges over positive half-integers.
Using $b(k)\le b(0)$, \eqref{eq:lb-binomial-center}, and
\eqref{eq:lb-residue-decay} with $j=d$, their total contribution
is at most
\begin{align*}
 b(0)|c_g|H_{2d-1}
 &\le \exp(\frac{d+1}{32}) C_0d^{-2} \exp\!\left(-\frac d2\right)H_{2d-1}
 \\
 &\le C_0d^{-2}\sqrt{12d}\\
 &\le4C_0.
\end{align*}
The second inequality uses $H_{2d-1}\le\sqrt{12d}$ and
$(d+1)/32-d/2=(1-15d)/32\le0$.
The last follows from $d\ge1$ and $\sqrt{12}<4$.

The gaps and the two tails cover all samples in $\mathcal M$.
Adding their contributions gives
\[
 \mathcal M
 \le\bigl(8e^{1/16}+6\sqrt{18}+4\bigr)C_0+2
 =:C_*.
\]
Since $C_0=2e^2$, the constant $C_*$ is independent of $m$ and $g$.
Combining this with \eqref{eq:lb-surviving-negative} gives
\[
 \frac{V_-}{A}
 =\frac{\kappa V_-/\mathcal B_R}{\kappa A/\mathcal B_R}
 \ge\frac{2e^{-3}}{15C_*}>0.
\]
\end{proof}

\begin{remark}[Quadratic spacing]
\label{rem:lb-quadratic-spacing}
The quadratic spacing $r_j=R-1-j^2$ makes the residue decay
compensate for the growth of the binomial coefficients.
Together with the gap length $2j+1$, this gives the summable bound
$2\sqrt{18}\,C_0j^{-3/2}$ for each remaining gap's contribution.
\end{remark}

\section{Proofs for the Quantum Walk Analysis}\label{app:ea-walk}

\subsection{Derivative estimates}\label{app:ea-regularity-proofs}

\paragraph{Coefficient derivatives.}
\label{app:ea-coefficients}

\begin{proof}[Proof of \Cref{clm:ea-coefficients}]
Fix $0<\mathbf{x}<w$, and abbreviate
$\delta=w-\mathbf{x}$ and $r=n-2\mathbf{x}=2\delta+d$.
All derivatives in this proof are with respect to $\mathbf{x}$;
in particular, $\delta'=-1$ and $r'=-2$.
The zeroth-order bounds follow from $\mu^2+\beta^2=1$ and
$\norm{\chi}=\norm{\eta}=1$.

For the derivatives of $\mu$, direct differentiation gives
$\mu^{(q)}=2^q q!b/r^{q+1}$. Since
$\beta=2\sqrt{\delta(\delta+d)}/r\ge2\delta/r$, we obtain
\begin{equation*}
\frac{|\mu^{(q)}|\delta^q}{\beta}
\le 2^{q-1}q!\frac d r
\left(\frac\delta r\right)^{q-1}
\le q!\qquad(q\ge1).
\end{equation*}
The last inequality uses $d/r\le1$ and $\delta/r\le1/2$.

To estimate the derivatives of $\beta$, set $\lambda=\log\beta$.
The formula for $\beta$ gives
$\lambda=\log2+\frac12\log\delta+\frac12\log(\delta+d)-\log r$.
For each $q\ge1$,
\begin{equation*}
\lambda^{(q)}
=-\frac{(q-1)!}{2\delta^q}
-\frac{(q-1)!}{2(\delta+d)^q}
+\frac{2^q(q-1)!}{r^q},
\end{equation*}
and consequently $|\lambda^{(q)}|\le C_q\delta^{-q}$.
For example,
\begin{equation*}
\begin{split}
\beta'&=\beta\lambda',\qquad
\beta''=\beta\bigl((\lambda')^2+\lambda''\bigr),\\
\beta'''&=\beta\bigl((\lambda')^3
+3\lambda'\lambda''+\lambda'''\bigr).
\end{split}
\end{equation*}
The same argument works at every fixed order. Repeatedly applying
the product rule expresses $\beta^{(q)}$ as $\beta$ times a sum
of products $\lambda^{(r_1)}\cdots\lambda^{(r_k)}$ with
$r_1+\cdots+r_k=q$ and $r_i\ge1$. Each product has absolute
value at most $C_q\delta^{-q}$, so
$|\beta^{(q)}|\le C_q\beta\delta^{-q}$.

Suppose next that $d>0$, and write $\sigma=\operatorname{sgn}(b)$.
Differentiating $\kappa=\arcsin\mu$ gives
$\kappa'=\mu'/\beta=\sigma\beta A_0$, where
\begin{equation*}
A_0=\frac d{2\delta(\delta+d)}
=\frac12\left(\frac1\delta-\frac1{\delta+d}\right),\qquad
A_0^{(k)}=\frac{k!}{2}
\left(\delta^{-k-1}-(\delta+d)^{-k-1}\right).
\end{equation*}
Thus $|A_0^{(k)}|\le C_k\delta^{-k-1}$.
Differentiating $\sigma\beta A_0$ another $q-1$ times,
and applying the bounds just proved to each factor, yields
\begin{equation*}
|\kappa^{(q)}|
\le \sum_{k=0}^{q-1}\binom{q-1}{k}
|\beta^{(k)}|\,|A_0^{(q-1-k)}|
\le C_q\beta\delta^{-q}.
\end{equation*}
Here the term $k=0$ uses $|\beta^{(0)}|=\beta$.

Finally, $\chi'=\kappa'\eta/2$ and
$\eta'=-\kappa'\chi/2$. For clarity, the next two derivatives
of $\chi$ are
\begin{equation*}
\begin{split}
\chi''&=\frac{\kappa''}{2}\eta
-\frac{(\kappa')^2}{4}\chi,\\
\chi'''&=\left(\frac{\kappa'''}2
-\frac{(\kappa')^3}{8}\right)\eta
-\frac{3\kappa'\kappa''}{4}\chi.
\end{split}
\end{equation*}
At order $q\ge1$, repeated differentiation produces a fixed
number of terms, each equal to $\chi$ or $\eta$ times a product
of positive-order derivatives of $\kappa$ whose orders add to $q$.
Such a product is bounded by $C_q\beta^k\delta^{-q}$ for some
$k\ge1$, and $\beta^k\le\beta$ because $0<\beta\le1$.
This proves the stated bounds for both $\chi^{(q)}$ and
$\eta^{(q)}$. If $d=0$, then $b=0$, $\mu=0$, $\beta=1$,
and $\chi,\eta$ are constant on the interior. Their positive-order
derivatives are all zero, which also proves the claim in this case.
\end{proof}

\paragraph{Derivatives and samples of the continuous packet.}
\label{app:ea-regularity}

\begin{proof}[Proof of \Cref{lem:ea-regularity}]
We first check that the continuous norm is unchanged. The
characteristic formula and its derivative with respect to the
starting point give
\begin{equation*}
\widetilde f(t,X(t,\xi))
=f_0(\xi)\sqrt{\frac{\beta(\xi)}{\beta(X(t,\xi))}},
\qquad
\partial_\xi X(t,\xi)
=\frac{\beta(X(t,\xi))}{\beta(\xi)}.
\end{equation*}
Both denominators are positive on the transported packet, by
\Cref{clm:ea-interior}. The second identity also shows that
$X(t,\cdot)$ is increasing. We may therefore substitute
$\mathbf{x}=X(t,\xi)$ in the integral defining the norm:
\begin{equation*}
\begin{split}
\norm{\widetilde f_t}^2
&=\int_{X(t,\ell)}^{X(t,2\ell)}
|\widetilde f(t,\mathbf{x})|^2\,d\mathbf{x}\\
&=\int_\ell^{2\ell}|f_0(\xi)|^2
\frac{\beta(\xi)}{\beta(X(t,\xi))}
\frac{\beta(X(t,\xi))}{\beta(\xi)}\,d\xi
=\norm{f_0}^2.
\end{split}
\end{equation*}
The portions outside the displayed intervals contribute zero.
\Cref{clm:ea-initial-derivatives} gives
$\norm{f_0}=\Theta(1)$.

Norm preservation alone does not bound derivatives. We obtain
their bounds by writing the characteristic formula in coordinates
in which the motion is a translation. Put
$y(\mathbf{x})=g(w-\mathbf{x})$, where
$g(q)=\sqrt{q(q+d)}$. Then $y'=-1/\beta$, and
$y(X(t,\xi))=y(\xi)-t$.
Let $\zeta$ be the inverse of $y$ on the initial interval
$[\ell,2\ell]$, and define
$H_0(\xi)=\sqrt{\beta(\xi)}f_0(\xi)$ and
$F_0(z)=H_0(\zeta(z))$ there, setting $F_0$ to zero outside
the image of that interval. The fixed bump is supported strictly
inside $(1,2)$, so this extension is smooth. The characteristic
formula becomes
\begin{equation}\label{eq:app-ea-translated-profile}
\widetilde f_t(\mathbf{x})
=\beta(\mathbf{x})^{-1/2}F_0(y(\mathbf{x})+t)
\end{equation}
where the right-hand side is evaluated on the interior and the
packet is set to zero elsewhere.

We next bound the derivatives of $F_0$. On the initial interval,
$\beta\ge v_\alpha>0$ and $w-\xi\ge c_\alpha n$.
The first assertion follows from the initial speed bound in
\Cref{eq:hd-width-regime}; for the second, use
$w\ge\alpha n/2$ and $2\ell\le\alpha n/6$.
\Cref{clm:ea-coefficients} implies that every derivative of order
$r\ge1$ of $\beta$ is $O_{\alpha,r}(n^{-r})$ on this interval.
Repeated differentiation of $\sqrt\beta$, using
$\beta\ge v_\alpha$, gives the same bound for its derivatives.
Together with \Cref{clm:ea-initial-derivatives}, the product rule
therefore gives, for each fixed $q\ge0$,
\begin{equation*}
|H_0^{(q)}(\xi)|
\le C_q\sum_{k=0}^q n^{-(q-k)}\ell^{-k-1/2}
\le C_q\ell^{-q-1/2}.
\end{equation*}
We used $\ell\le n$ in the last step.

Since $\zeta'(z)=-\beta(\zeta(z))$, differentiation with
respect to $z$ acts as $-\beta\partial_\xi$ on functions of
$\xi=\zeta(z)$. For example,
\begin{equation*}
\begin{split}
F_0'&=-\beta H_0',\\
F_0''&=\beta\beta'H_0'+\beta^2H_0'',\\
F_0'''&=-\beta\bigl((\beta')^2+\beta\beta''\bigr)H_0'
-3\beta^2\beta'H_0''-\beta^3H_0'''.
\end{split}
\end{equation*}
All functions on the right are evaluated at $\zeta(z)$.
More generally, applying $-\beta\partial_\xi$ a total of $q$
times expresses $F_0^{(q)}$ as a sum of terms containing
$H_0^{(k)}$ and derivatives of $\beta$ whose orders add to $q-k$.
The latter factors are bounded by $C_qn^{-(q-k)}$; factors of
$\beta$ without a derivative are at most one. Hence
\begin{equation}\label{eq:app-ea-initial-profile-bound}
|F_0^{(q)}(z)|
\le C_q\sum_{k=0}^q n^{-(q-k)}\ell^{-k-1/2}
\le C_q\ell^{-q-1/2}.
\end{equation}

Fix a time $0\le t\le t_0$ and a point $\mathbf{x}$ in the
transported interval. Write $B(\mathbf{x})=\beta(\mathbf{x})^{-1/2}$,
$Y(\mathbf{x})=y(\mathbf{x})+t$, and $\delta_{\mathbf{x}}=w-\mathbf{x}$.
The coefficient bounds imply
\begin{equation*}
|B^{(r)}|\le C_r\beta^{-1/2}\delta_{\mathbf{x}}^{-r}
\quad(r\ge0),\qquad
|Y^{(r)}|\le C_r\beta^{-1}\delta_{\mathbf{x}}^{-(r-1)}
\quad(r\ge1).
\end{equation*}
For the first inequality, differentiate
$B=\exp(-\frac12\log\beta)$ as in the coefficient proof.
For the second, start from $Y'=-\beta^{-1}$ and apply the same
calculation to the further derivatives of $\beta^{-1}$.

Applying the product and chain rules to $BF_0(Y)$ gives constant
multiples of
\begin{equation*}
B^{(r_0)}F_0^{(k)}(Y)\prod_{i=1}^kY^{(r_i)},
\qquad r_0+\cdots+r_k=q.
\end{equation*}
Here $r_i\ge1$ for $i\ge1$, and the constants depend only on $q$.
By \Cref{eq:app-ea-initial-profile-bound}, the absolute value of
each such term is at most
\begin{equation*}
C_q\ell^{-k-1/2}\beta(\mathbf{x})^{-k-1/2}
\delta_{\mathbf{x}}^{-(q-k)}.
\end{equation*}
\Cref{clm:wb-width,clm:ea-interior} give
$\ell\beta(\mathbf{x})\asymp_\alpha W(t)$ throughout the
packet and $\delta_{\mathbf{x}}\ge\delta(t)$.
Summing the finitely many terms yields
\begin{equation*}
|\partial_{\mathbf{x}}^q\widetilde f_t(\mathbf{x})|
\le C_qW(t)^{-1/2}
\sum_{k=0}^q W(t)^{-k}\delta(t)^{-(q-k)}
\le C_qW(t)^{-q-1/2}.
\end{equation*}
The last inequality uses $\delta(t)\ge c_\alpha W(t)$ from
\Cref{clm:wb-gap}. Outside the transported support all derivatives
are zero. At its ends they agree with these zero values because
$F_0$ is smooth and is zero outside its initial interval. This
proves the pointwise derivative bound everywhere.

It remains to check uniformity over two time units and over shifted
integer samples. Fix $[t,t+2]\subseteq[0,t_0]$.
By \Cref{clm:ea-interior}, $\beta_s\asymp_\alpha\beta_t$
for $t\le s\le t+2$. Applying \Cref{clm:wb-width} at both times
gives $W(s)\asymp_\alpha W(t)$, so the preceding derivative
bound is uniform in $s$ with $W(t)$ on the right.
At each such time, the derivative is supported in an interval
of length $W(s)\le C_\alpha W(t)$.
An interval of length $a$ contains at most $a+1$ points of any
grid $\mathbb Z+z$. Since $W(t)\ge1$, at most
$C_\alpha W(t)$ summands below are nonzero, independently of
$z$. Therefore
\begin{equation*}
\sup_{\substack{t\le s\le t+2\\|z|\le2}}
\sum_{j\in\mathbb Z}
|\partial_{\mathbf{x}}^q\widetilde f_s(j+z)|^2
\le C_qW(t)\bigl(W(t)^{-q-1/2}\bigr)^2
=C_qW(t)^{-2q}.
\end{equation*}
Taking square roots proves the shifted-sample estimate.
The fixed enlargement in \Cref{clm:ea-interior} ensures that
every coefficient evaluation used with these samples in the
Taylor estimates remains inside $(0,w)$.
\end{proof}

\subsection{The three approximation errors}\label{app:ea-error-proofs}

Throughout the following three proofs, fix $0\le t\le t_0-2$ and abbreviate $W=W(t)$, $\delta=\delta(t)$,
$\beta_*=\beta_t$, $f=\widetilde f_t$, and $u=\chi f$.
We use $h=W^{-1}+\delta^{-1}$.
All derivatives below are with respect to the spatial coordinate
$\mathbf{x}$. All coefficient estimates are applied on the interval
containing the packet between times $t$ and $t+2$, together with
the four-site enlargement from \Cref{clm:ea-interior}.
On that interval, the coefficient $\beta$ and the distance from the
wall are within fixed factors of $\beta_*$ and $\delta$.

\paragraph{The second-order error.}
\label{app:ea-second-order-error}
\begin{proof}[Proof of \Cref{clm:ea-second-order-error}]
We first verify the cancellation that produces $E_2$; in particular,
we must keep every term in which a derivative acts on a coefficient.
Set $a=\kappa'/2$.
Differentiating the definitions of $\chi,\eta,\beta,\mu$ gives
$\chi'=a\eta$, $\eta'=-a\chi$, $\beta'=-2a\mu$, and
$\mu'=2a\beta$. Substitution into $D_1$ then yields
\begin{equation*}
D_1(\chi f)=\chi\mathcal Lf-a\eta f,
\qquad
D_1(\eta f)=\eta\mathcal Lf-a\chi f.
\end{equation*}
Since $C\chi=\chi$ and $C\eta=-\eta$, the two first-order
contributions satisfy $(CD_1+D_1C)(\chi f)=2\chi\mathcal Lf$.

To compute the second-order contributions, let
$\mathcal Af=\chi^\dagger D_2(\chi f)$ and put
$Z=\operatorname{diag}(1,-1)$ and
$\Pi_\uparrow=\operatorname{diag}(1,0)$.
The definition of $D_2$ can be written as
$D_2v=(\beta/2)Zv''+\beta'\Pi_\uparrow v'
+(1/2)\Pi_\uparrow C''v$.
The products needed to evaluate its projection onto $\chi$ are
\begin{equation*}
\begin{split}
(\chi f)'&=\chi f'+a\eta f,\qquad
(\chi f)''=\chi(f''-a^2f)+\eta(2af'+a'f),\\
\chi^\dagger Z\chi&=\beta,\qquad
\chi^\dagger Z\eta=-\mu,\qquad
\chi^\dagger\Pi_\uparrow\chi=\frac{1+\beta}{2},\qquad
\chi^\dagger\Pi_\uparrow\eta=-\frac\mu2,\\
C''\chi&=-4a^2\chi+2a'\eta.
\end{split}
\end{equation*}
For the last identity, differentiate $C'\chi=2a\eta$ and use
$C'\eta=2a\chi$. Applying the three summands of $D_2$ separately
therefore gives
\begin{equation*}
\begin{split}
\mathcal Af
&=\frac\beta2\bigl[\beta(f''-a^2f)-\mu(2af'+a'f)\bigr]
+\beta'\left[\frac{1+\beta}{2}f'-\frac{\mu a}{2}f\right]\\
&\quad+\left[-a^2(1+\beta)-\frac{\mu a'}2\right]f\\
&=\frac{\beta^2}{2}f''
+\beta'\left(\beta+\frac12\right)f'
+\left[\frac{(1+\beta)\beta''}{4}
+\frac{(\beta')^2}{8}-\frac{a^2}{2}\right]f.
\end{split}
\end{equation*}
The simplification uses $\beta'=-2a\mu$,
$\beta''=-2a'\mu-4a^2\beta$, and $\mu^2+\beta^2=1$.
On the other hand, applying $\mathcal L=-\beta\partial_{\mathbf{x}}
-\beta'/2$ twice gives
\begin{equation*}
\mathcal L^2f=\beta^2f''+2\beta\beta'f'
+\left[\frac{\beta\beta''}{2}+\frac{(\beta')^2}{4}\right]f.
\end{equation*}
Comparing the coefficients of $f'',f',f$ in these two expressions
shows that
$2\mathcal Af=\mathcal L^2f-a^2f+\beta'f'+\beta''f/2$.

Because $I+C=2\chi\chi^\dagger$ and $C(\chi f)=\chi f$ as
an identity of functions, we have
$(CD_2+D_2C)(\chi f)=2\chi\mathcal Af$.
The two identities for $D_1$ above also give
\begin{equation*}
D_1^2(\chi f)
=\chi(\mathcal L^2f+a^2f)
-\eta\bigl(a\mathcal Lf+\mathcal L(af)\bigr).
\end{equation*}
Here $\mathcal L(af)=a\mathcal Lf-\beta a'f$, so
\begin{equation*}
-a\mathcal Lf-\mathcal L(af)
=2a\beta f'+(a\beta'+\beta a')f
=\mu'f'+\frac12\mu''f.
\end{equation*}
Thus both the $a^2f$ terms and the discrepancy in the $f''$ terms
cancel, leaving
\begin{equation*}
\begin{split}
(CD_2+D_2C+D_1^2)(\chi f)
&=2\chi\mathcal L^2f+E_2f,\\
E_2f
&=\chi\left(\beta'f'+\frac12\beta''f\right)
+\eta\left(\mu'f'+\frac12\mu''f\right).
\end{split}
\end{equation*}

It remains to bound the sampled vector. Multiplication by either
$\chi$ or $\eta$ preserves the norm of a sampled scalar function,
since both have norm one at every point.
By \Cref{clm:ea-coefficients},
$|\beta'|+|\mu'|\le C\beta_*\delta^{-1}$ and
$|\beta''|+|\mu''|\le C\beta_*\delta^{-2}$ on the required
interval. By \Cref{lem:ea-regularity},
$(\sum_j|f'(j)|^2)^{1/2}\le C W^{-1}$ and
$(\sum_j|f(j)|^2)^{1/2}\le C$.
The triangle inequality now gives
\begin{equation*}
\norm{\ket{\mathcal I(E_2f)}}
\le C\beta_*\delta^{-1}W^{-1}
+C\beta_*\delta^{-2},
\end{equation*}
as claimed.
\end{proof}

\paragraph{The spatial remainder.}
\label{app:ea-spatial-error}
\begin{proof}[Proof of \Cref{clm:ea-spatial-error}]
We first derive the exact remainder and then bound its three terms.
Starting from $V=C+D_1+D_2+R_3$, write
$V^2u=V(Cu)+V(D_1u)+V(D_2u)+V(R_3u)$.
In these four summands, use respectively
\begin{equation*}
\begin{split}
V(Cu)&=(C+D_1+D_2)Cu+R_3(Cu),\\
V(D_1u)&=(C+D_1)D_1u+(V-C-D_1)D_1u,\\
V(D_2u)&=CD_2u+(V-C)D_2u,
\end{split}
\end{equation*}
and leave $V(R_3u)$ unchanged.
Since $Cu=u$, collecting terms gives
\begin{equation*}
\begin{split}
V^2u
&=\bigl[I+CD_1+D_1C+CD_2+D_2C+D_1^2\bigr]u
+\mathcal R_3[u],\\
\mathcal R_3[u]
&=(V+I)R_3u+(V-C-D_1)D_1u+(V-C)D_2u.
\end{split}
\end{equation*}
In particular, the proof does not differentiate $R_3u$ or expand
$D_2^2u$. This is why three derivatives of the packet suffice.

For a scalar or two-component function $v$, define
\begin{equation*}
N_q(v):=\sup_{|z|\le2}
\left(\sum_{j\in\mathbb Z}\norm{v^{(q)}(j+z)}^2\right)^{1/2}.
\end{equation*}
For a scalar, the norm inside the sum means absolute value.
These quantities control sampling at the intermediate points that
occur in Taylor's formula. In particular,
$\norm{\ket{\mathcal I(v)}}\le N_0(v)$.
We use them only for functions supported in the interior intervals
specified above; the functions are zero outside their supports.

For $m=0,1,2$, let
$\mathsf T_0=V-C$, $\mathsf T_1=V-C-D_1$, and
$\mathsf T_2=R_3$.
Taylor's formula with its integral remainder gives
\begin{equation}\label{eq:ea-app-spatial-taylor}
\norm{\ket{\mathcal I(\mathsf T_m v)}}
\le C\beta_*\sum_{q=0}^{m+1}
\delta^{-(m+1-q)}N_q(v).
\end{equation}
To verify this estimate, write the scalar remainder as
\begin{equation*}
\mathsf T_\pm h(\mathbf{x})
=\frac{(\pm1)^3}{2}\int_0^1(1-s)^2
h'''(\mathbf{x}\pm s)\,ds.
\end{equation*}
The upper component of $R_3v$ is
$\mathsf T_-(\beta v_\uparrow)
+(\mathsf T_-\mu)v_\downarrow$.
The derivative in its first term expands as
\begin{equation*}
(\beta v_\uparrow)'''
=\beta v_\uparrow'''+3\beta'v_\uparrow''
+3\beta''v_\uparrow'+\beta'''v_\uparrow.
\end{equation*}
By \Cref{clm:ea-coefficients}, the norms after sampling these
four terms are bounded by
$C\beta_*N_3(v)$,
$C\beta_*\delta^{-1}N_2(v)$,
$C\beta_*\delta^{-2}N_1(v)$, and
$C\beta_*\delta^{-3}N_0(v)$, respectively.
The term involving $\mu'''$ has the last of these bounds.
The lower component is $-\beta\mathsf T_+v_\downarrow$ and
has the first bound. Applying the triangle inequality to the
integral proves \Cref{eq:ea-app-spatial-taylor} for $m=2$.
For $m=0,1$, use the same argument with the first- and
second-order integral remainders, whose integrands contain
$(\beta v_\uparrow)'$, $(\beta v_\uparrow)''$,
$\mu'$, and $\mu''$, respectively.
Each contribution retains the factor $\beta_*$.

We next bound the derivatives of the three inputs to these
remainders. The product rule, \Cref{clm:ea-coefficients}, and
\Cref{lem:ea-regularity} give
\begin{equation}\label{eq:ea-app-input-derivatives}
\begin{split}
N_q(u)&\le C h^q \qquad(0\le q\le3),\\
N_q(D_1u)&\le C\beta_*h^{q+1} \qquad(0\le q\le2),\\
N_q(D_2u)&\le C\beta_*h^{q+2} \qquad(0\le q\le1).
\end{split}
\end{equation}
For clarity, the first line follows by expanding
$(\chi f)^{(q)}=\sum_{k=0}^q\binom qk\chi^{(k)}f^{(q-k)}$:
each term is at most $C\delta^{-k}W^{-(q-k)}$ in the shifted
sampled norm. Their sum is at most $Ch^q$.
For the next two lines, write
\begin{equation*}
D_1v=-\beta v'-\Pi_\uparrow C'v,
\qquad
D_2v=\frac\beta2 Zv''+\beta'\Pi_\uparrow v'
+\frac12\Pi_\uparrow C''v,
\end{equation*}
with $Z=\operatorname{diag}(1,-1)$ and
$\Pi_\uparrow=\operatorname{diag}(1,0)$.
The coefficient of $v'$ in $D_1$ has its $k$th derivative bounded
by $C\beta_*\delta^{-k}$; the coefficient of $v$ has the bound
$C\beta_*\delta^{-k-1}$.
Thus every term in $(D_1u)^{(q)}$ is bounded by
$C\beta_*h^{q+1}$.
In $D_2$, the corresponding bounds for the coefficients of
$v''$, $v'$, and $v$ are
$C\beta_*\delta^{-k}$,
$C\beta_*\delta^{-k-1}$, and
$C\beta_*\delta^{-k-2}$.
They give the last line of \Cref{eq:ea-app-input-derivatives}.
No coefficient or packet derivative above has order greater than
three.

Applying \Cref{eq:ea-app-spatial-taylor} to $u$, $D_1u$, and
$D_2u$ at orders $m=2,1,0$, respectively, now yields
\begin{equation*}
\begin{split}
\norm{\ket{\mathcal I(R_3u)}}&\le C\beta_*h^3,\\
\norm{\ket{\mathcal I((V-C-D_1)D_1u)}}&\le C\beta_*^2h^3,\\
\norm{\ket{\mathcal I((V-C)D_2u)}}&\le C\beta_*^2h^3.
\end{split}
\end{equation*}
The support of $R_3u$ is contained in a one-site enlargement of
the support of $u$. Its further image under $V$ is contained in
a two-site enlargement, so all these supports remain strictly
inside the physical interval. Sampling therefore commutes with
the last application of $V$.
Unitarity of the discrete walk gives
\begin{equation*}
\norm{\ket{\mathcal I((V+I)R_3u)}}
=\norm{(V+I)\ket{\mathcal I(R_3u)}}
\le2\norm{\ket{\mathcal I(R_3u)}}.
\end{equation*}
Adding the three bounds and using $\beta_*\le1$ proves
$\norm{\ket{\mathcal I(\mathcal R_3[u])}}\le C\beta_*h^3$.
\end{proof}

\paragraph{The time remainder.}
\label{app:ea-time-error}
\begin{proof}[Proof of \Cref{clm:ea-time-error}]
The operator $\mathcal L$ does not depend on time, so
$\partial_t^q\widetilde f_t=\mathcal L^q\widetilde f_t$.
Taylor's formula on the time interval $[t,t+2]$ gives
\begin{equation*}
\begin{split}
\widetilde u_{t+2}
&=\chi\bigl(\widetilde f_t+2\mathcal L\widetilde f_t
+2\mathcal L^2\widetilde f_t\bigr)+\mathcal R_4(t),\\
\mathcal R_4(t)
&=\frac12\chi\int_0^2(2-s)^2
\mathcal L^3\widetilde f_{t+s}\,ds.
\end{split}
\end{equation*}
To bound every contribution to $\mathcal L^3$, first use
\begin{equation*}
\mathcal L^2f=\beta^2f''+2\beta\beta'f'
+\left(\frac12\beta\beta''+\frac14(\beta')^2\right)f.
\end{equation*}
Apply $-\beta\partial_{\mathbf{x}}-\beta'/2$ once more, including
the derivatives of all coefficients. Collecting terms according
to the derivative of $f$ gives
\begin{equation}\label{eq:ea-app-L-cubed}
\begin{split}
\mathcal L^3f
&=-\beta^3f'''-\frac92\beta^2\beta'f''\\
&\quad-\left[\frac{13}{4}\beta(\beta')^2
+\frac52\beta^2\beta''\right]f'\\
&\quad-\left[\frac12\beta^2\beta'''
+\frac54\beta\beta'\beta''+\frac18(\beta')^3\right]f.
\end{split}
\end{equation}
For example, the coefficient of $f''$ is
$-\beta((\beta^2)'+2\beta\beta')-(\beta'/2)\beta^2
=-(9/2)\beta^2\beta'$.
The other coefficients follow from the same product rule.

Every product in \Cref{eq:ea-app-L-cubed} has three coefficient
factors, counting both $\beta$ and its derivatives as factors.
By \Cref{clm:ea-coefficients}, each such factor contributes
$C\beta_*$, and a total of $k$ coefficient derivatives contributes
the additional factor $\delta^{-k}$.
Together with the sampled derivative bounds in
\Cref{lem:ea-regularity}, this gives, uniformly for $0\le s\le2$,
\begin{equation*}
\begin{split}
\norm{\ket{\mathcal I(\chi\mathcal L^3\widetilde f_{t+s})}}
\le C\beta_*^3\bigl(
W^{-3}+\delta^{-1}W^{-2}
+\delta^{-2}W^{-1}+\delta^{-3}\bigr)
\le C\beta_*^3h^3.
\end{split}
\end{equation*}
The four terms correspond, respectively, to $f'''$, $f''$, $f'$,
and $f$ in \Cref{eq:ea-app-L-cubed}; multiplication by $\chi$
does not change the sampled scalar norm.
Sampling is a finite linear sum, so it commutes with the time
integral. Applying the triangle inequality and using
$\frac12\int_0^2(2-s)^2\,ds=4/3$ yields
\begin{equation*}
\norm{\ket{\mathcal I(\mathcal R_4(t))}}
\le C\beta_*^3h^3\le C\beta_*h^3,
\end{equation*}
as required.
\end{proof}

%% file: main.bbl
\begin{thebibliography}{LPdWY16}
\interlinepenalty=10000

\providecommand{\etalchar}[1]{$^{#1}$}
\providecommand{\natexlab}[1]{#1}
\providecommand{\url}[1]{\texttt{#1}}
\expandafter\ifx\csname urlstyle\endcsname\relax
  \providecommand{\doi}[1]{doi: #1}\else
  \providecommand{\doi}{doi: \begingroup \urlstyle{rm}\Url}\fi

\bibitem[Aar14]{Aaronson2014}
Scott Aaronson.
\newblock The {NEW} ten most annoying questions in quantum computing.
\newblock Shtetl-Optimized, May 2014.
\newblock URL \url{https://scottaaronson.blog/?p=1792}.
\newblock May 13, 2014, Question 5; the question is attributed there to Shelby
  Kimmel.

\bibitem[ABSdW13]{AmbainisEtAl2013}
Andris Ambainis, Art{\=u}rs Ba{\v{c}}kurs, Juris Smotrovs, and Ronald de~Wolf.
\newblock Optimal quantum query bounds for almost all {Boolean} functions.
\newblock In \emph{30th International Symposium on Theoretical Aspects of
  Computer Science (STACS 2013)}, volume~20 of \emph{Leibniz International
  Proceedings in Informatics (LIPIcs)}, pages 446--453, 2013{\natexlab{a}}.
\newblock \doi{10.4230/LIPIcs.STACS.2013.446}.
\newblock URL
  \url{https://drops.dagstuhl.de/entities/document/10.4230/LIPIcs.STACS.2013.446}.

\bibitem[AIS13]{AmbainisIraidsSmotrovs2013}
Andris Ambainis, J{\=a}nis Iraids, and Juris Smotrovs.
\newblock Exact quantum query complexity of {EXACT} and {THRESHOLD}.
\newblock In \emph{8th Conference on the Theory of Quantum Computation,
  Communication and Cryptography (TQC 2013)}, volume~22 of \emph{Leibniz
  International Proceedings in Informatics (LIPIcs)}, pages 263--269,
  2013{\natexlab{b}}.
\newblock \doi{10.4230/LIPIcs.TQC.2013.263}.
\newblock URL
  \url{https://drops.dagstuhl.de/entities/document/10.4230/LIPIcs.TQC.2013.263}.

\bibitem[BBC{\etalchar{+}}01]{BealsEtAl2001}
Robert Beals, Harry Buhrman, Richard Cleve, Michele Mosca, and Ronald de~Wolf.
\newblock Quantum lower bounds by polynomials.
\newblock \emph{Journal of the ACM}, 48\penalty0 (4):\penalty0 778--797, 2001.
\newblock \doi{10.1145/502090.502097}.
\newblock URL \url{https://arxiv.org/abs/quant-ph/9802049}.

\bibitem[BHMT02]{BHMT2002}
Gilles Brassard, Peter H{\o}yer, Michele Mosca, and Alain Tapp.
\newblock Quantum amplitude amplification and estimation.
\newblock In Samuel~J. Lomonaco, Jr., editor, \emph{Quantum Computation and
  Quantum Information}, volume 305 of \emph{Contemporary Mathematics}, pages
  53--74. American Mathematical Society, 2002.
\newblock \doi{10.1090/conm/305/05215}.
\newblock URL \url{https://arxiv.org/abs/quant-ph/0005055}.

\bibitem[dGdW02]{deGraafDeWolf2002}
Mart de~Graaf and Ronald de~Wolf.
\newblock On quantum versions of the {Yao} principle.
\newblock In \emph{STACS 2002}, volume 2285 of \emph{Lecture Notes in
  Computer Science}, pages 347--358. Springer, 2002.
\newblock \doi{10.1007/3-540-45841-7_28}.
\newblock URL \url{https://arxiv.org/abs/quant-ph/0109070}.

\bibitem[FGGS99]{FarhiEtAl1999}
Edward Farhi, Jeffrey Goldstone, Sam Gutmann, and Michael Sipser.
\newblock How many functions can be distinguished with $k$ quantum queries?
\newblock \emph{Physical Review A}, 60:\penalty0 4331--4333, 1999.
\newblock \doi{10.1103/PhysRevA.60.4331}.
\newblock URL \url{https://arxiv.org/abs/quant-ph/9901012}.

\bibitem[Gri01]{Grigoriev2001}
Dima Grigoriev.
\newblock Complexity of {Positivstellensatz} proofs for the knapsack.
\newblock \emph{Computational Complexity}, 10\penalty0 (2):\penalty0 139--154,
  2001.
\newblock \doi{10.1007/s00037-001-8192-0}.
\newblock URL
  \url{https://link.springer.com/article/10.1007/s00037-001-8192-0}.

\bibitem[KLM16]{KurpiszEtAl2016}
Adam Kurpisz, Samuli Lepp\"anen, and Monaldo Mastrolilli.
\newblock Sum-of-squares hierarchy lower bounds for symmetric formulations.
\newblock In \emph{Integer Programming and Combinatorial Optimization
  (IPCO 2016)}, volume 9682 of \emph{Lecture Notes in Computer Science},
  pages 362--374. Springer, 2016.
\newblock \doi{10.1007/978-3-319-33461-5_30}.
\newblock Full version: \url{https://arxiv.org/abs/1407.1746v2}.

\bibitem[LPdWY16]{LeeEtAl2016}
Troy Lee, Anupam Prakash, Ronald de~Wolf, and Henry Yuen.
\newblock On the sum-of-squares degree of symmetric quadratic functions.
\newblock In \emph{31st Conference on Computational Complexity (CCC 2016)},
  volume~50 of \emph{Leibniz International Proceedings in Informatics
  (LIPIcs)}, pages 17:1--17:31, 2016.
\newblock \doi{10.4230/LIPIcs.CCC.2016.17}.
\newblock URL
  \url{https://drops.dagstuhl.de/entities/document/10.4230/LIPIcs.CCC.2016.17}.
\newblock Numbered results and appendices refer to the expanded version,
  arXiv:1601.02311v2.

\bibitem[LWD{\etalchar{+}}26]{stellar}
Honghao Lin, David~P. Woodruff, Yuan Deng, Jieming Mao, Song Zuo, and Vahab Mirrokni.
\newblock Stellar Colosseum: A many-agent harness for long-horizon research in mathematics and theoretical computer science.
\newblock \emph{arXiv preprint arXiv:2609.15983}, 2026.

\bibitem[Pat92]{Paturi1992}
Ramamohan Paturi.
\newblock On the degree of polynomials that approximate symmetric {Boolean}
  functions (preliminary version).
\newblock In \emph{24th Annual ACM Symposium on Theory of Computing (STOC
  1992)}, pages 468--474, 1992.
\newblock \doi{10.1145/129712.129758}.
\newblock URL
  \url{https://cseweb.ucsd.edu/~paturi/myPapers/pubs/Paturi_1992_stoc.pdf}.

\bibitem[Sri11]{Srinivasan2011}
Murali~K. Srinivasan.
\newblock Symmetric chains, {Gelfand--Tsetlin} chains, and the {Terwilliger}
  algebra of the binary {Hamming} scheme.
\newblock \emph{Journal of Algebraic Combinatorics}, 34\penalty0 (2):\penalty0
  301--322, 2011.
\newblock \doi{10.1007/s10801-010-0272-2}.
\newblock URL \url{https://arxiv.org/abs/1001.0280}.
\newblock Result and page locators refer to arXiv:1001.0280v2.

\bibitem[vD98]{VanDam1998}
Wim van Dam.
\newblock Quantum oracle interrogation: Getting all information for almost half
  the price.
\newblock In \emph{39th Annual Symposium on Foundations of Computer Science
  (FOCS 1998)}, pages 362--367, 1998.
\newblock \doi{10.1109/SFCS.1998.743486}.
\newblock URL \url{https://arxiv.org/abs/quant-ph/9805006}.
\newblock Section locators refer to arXiv:quant-ph/9805006v3.

\end{thebibliography}
